\documentclass{jpp}

\usepackage{graphicx}                       
\usepackage[utf8]{inputenc}
\usepackage[T1]{fontenc}
\usepackage{amsmath}

\let\proof\relax

\usepackage{amsthm}

\usepackage[format=plain, font=small, labelfont=bf, justification = raggedright]{caption}
\usepackage{overpic}
\usepackage{empheq} 
\usepackage{hyperref} 
\hypersetup{colorlinks = true,linkcolor = blue, breaklinks = true}

\newcommand{\eq}[1]{(\ref{#1})}

\newcommand{\Fig}[1]{Fig.~\ref{#1}}
\newcommand{\Sec}[1]{Sec.~\ref{#1}}

\newcommand{\App}[1]{Appendix~\ref{#1}}

\newcommand{\ie}{{i.e., }}

\newcommand{\mc}[1]{\mathcal{#1}}

\newcommand{\msf}[1]{\mathsf{#1}}

\newcommand{\mbb}[1]{\mathbb{#1}}

\newcommand{\bra}[1]{\langle#1 |}
\newcommand{\ket}[1]{|#1 \rangle}
\newcommand{\braket}[2]{\langle#1 |  #2 \rangle}

\newcommand{\Vect}[1]{{\boldsymbol{\rm #1}}}

\newcommand{\Mat}[1]{\msf{#1}}
\newcommand{\IMat}[1]{\Mat{I}_{#1}}

\newcommand{\oper}[1]{\smash{\widehat{#1}}}
\newcommand{\unit}[1]{\smash{\check{#1}}}

\newcommand{\hatmap}[1]{\Mat{#1}_\wedge}
\newcommand{\ave}[1]{\left\langle#1 \right\rangle}

\newcommand{\fluct}[1]{\widetilde{#1}}
\newcommand{\mean}[1]{\overline{#1}}

\newcommand{\pd}[1]{\partial_{#1}}
\newcommand{\dd}{\mathrm{d}}

\newcommand{\Symb}[1]{\mc{#1}}
\newcommand{\Weyl}{\mbb{W}}

\renewcommand{\Re}{\textrm{Re}}
\renewcommand{\Im}{\textrm{Im}}

\newcommand{\plasmaF}{\omega_p}
\newcommand{\cycloF}{\Omega}
\newcommand{\collT}{\tau_{ei}}
\newcommand{\larmorR}{r_L}
\newcommand{\thermalV}{v_t}

\newcommand{\nerstV}{v_N}
\newcommand{\groupDISP}{G_d}
\newcommand{\resist}{\eta}
\newcommand{\growthV}{u_\gamma}
\newcommand{\Hall}{\mc{M}}

\newcommand{\inten}{\mc{I}}
\newcommand{\plasmaBETA}{\beta_0}
\newcommand{\betaEFFECT}{\beta_\text{eff}}

\newcommand{\lenT}{L_T}
\newcommand{\lenN}{L_n}
\newcommand{\curveT}{C_T}
\newcommand{\curveN}{C_n}
\newcommand{\shape}{S}
\newcommand{\diffuseT}{\tau_\kappa}
\newcommand{\lenI}{L_I}

\newcommand{\TpsCOEF}{F_{T,1}}
\newcommand{\TppCOEF}{F_{T,2}}
\newcommand{\TNpCOEF}{F_{T,n}}
\newcommand{\NpsCOEF}{F_{n,1}}
\newcommand{\NppCOEF}{F_{n,2}}
\newcommand{\gTpsCOEF}{G_1}
\newcommand{\gTppCOEF}{G_2}

\newcommand{\gFUNC}{g}
\newcommand{\gFUNCnorm}{G}
\newcommand{\paramONE}{A}

\newcommand{\resSOURCE}{\Symb{S}_\alpha}
\newcommand{\advectV}{\Symb{V}}
\newcommand{\modQ}{\Symb{Q}_z}

\newcommand{\fluxREDUCE}{\Gamma}

\newtheorem*{conjecture}{Conjecture}

\newcommand{\nullFrac}{\vphantom{\frac{}{}}}

\newcommand{\Stroke}[1]{\text{\ooalign{ $#1$\cr \hidewidth\raise.225ex \hbox{$-\mkern.5mu$}\cr}}}

\begin{document}
\setlength{\parskip}{0pt}
\setlength{\belowcaptionskip}{0pt}


\shorttitle{Collisional whistler instability}
\shortauthor{N.~A.~Lopez, A.~F.~A.~Bott and A.~A.~Schekochihin}

\title{Collisional whistler instability and electron temperature staircase in inhomogeneous plasma}
\author{N.~A.~Lopez\aff{1}
  \corresp{\email{nicolas.lopez@tokamakenergy.com}},
  A.~F.~A.~Bott\aff{2,3},
 \and A.~A.~Schekochihin\aff{1,4}}

\affiliation{\aff{1}Rudolf Peierls Centre for Theoretical Physics, University of Oxford,
Oxford OX1 3PU, UK
\aff{2}Department of Physics, University of Oxford, Oxford OX1 3PU, UK
\aff{3}Trinity College, Oxford OX1 3BH, UK
\aff{4}Merton College, Oxford OX1 4JD, UK}

\maketitle

\begin{abstract}
    High-beta magnetized plasmas often exhibit anomalously structured temperature profiles, as seen from galaxy cluster observations and recent experiments. It is well known that when such plasmas are collisionless, temperature gradients along the magnetic field can excite whistler waves that efficiently scatter electrons to limit their heat transport. Only recently has it been shown that parallel temperature gradients can excite whistler waves also in collisional plasmas. Here we develop a Wigner--Moyal theory for the collisional whistler instability starting from Braginskii-like fluid equations in a slab geometry. This formalism is necessary because, for a large region in parameter space, the fastest-growing whistler waves have wavelengths comparable to the background temperature gradients. We find additional damping terms in the expression for the instability growth rate involving inhomogeneous Nernst advection and resistivity. They (i) enable whistler waves to re-arrange the electron temperature profile via growth, propagation, and subsequent dissipation, and (ii) allow non-constant temperature profiles to exist stably. For high-beta plasmas, the marginally stable solutions take the form of a temperature staircase along the magnetic field lines. The electron heat flux can also be suppressed by the Ettingshausen effect when the whistler intensity profile is sufficiently peaked and oriented opposite the background temperature gradient. This mechanism allows cold fronts without magnetic draping, might reduce parallel heat losses in inertial fusion experiments, and generally demonstrates that whistler waves can regulate transport even in the collisional limit.

\end{abstract}


\section{Introduction}
\label{sec:intro}

X-ray observations of the diffuse plasma residing within galaxy clusters have revealed intricately structured temperature fields whose sharp gradients are inferred to have persisted far longer than classical transport theory predicts~\citep{Peterson06,Markevitch07}. Recent experiments on NIF with laser-produced hot, magnetized, high-beta plasmas also feature such anomalously structured temperature fields~\citep{Meinecke22}. Such structure would require a significantly reduced level of electron heat transport, which might be due to plasma microinstabilities. Indeed, it is now well-established~\citep{Kunz22} that the macroscopic transport properties of high-beta magnetized plasmas are significantly modified by small-scale instabilities such as firehose~\citep{Rosin11,Kunz14}, mirror~\citep{Kunz14,Komarov16}, or heat-flux whistler instabilities~\citep{Levinson92,Pistinner98,Komarov18,RobergClark18a,Drake21,Yerger24}. 

The heat-flux whistler instability is a particularly promising candidate because it is the fastest of all possible instabilities in the relevant parameter regime~\citep{Bott24} and quickly limits the parallel heat flux to a marginal value via electron scattering. However, this mechanism requires a resonant interaction with heat-carrying electrons that can be disrupted by collisions. This is not a problem for astrophysical plasmas whose magnetization is typically of order $\Hall \doteq \cycloF \collT \sim 10^{12}$, where $\cycloF$ is the electron cyclotron frequency and $\collT$ is the electron-ion collision time. However, it is not understood whether such an instability could persist in the more collisional laboratory analogues, whose magnetization range is $\Hall \sim 10^{-2} - 10^{2}$~\citep{Meinecke22}.

Recently a collisional mechanism for exciting whistler waves via anisotropic friction forces was identified~\citep{Bell20}. However, this initial analysis was restricted to the short-wavelength geometrical-optics limit, which prevented it from correctly describing all aspects of the long-wavelength fluid limit. This is problematic because in a typical laser-plasma experiment~\citep{Meinecke22}, the long-wavelength modes will be (i) the most easily observable in diagnostics, and (ii) the first modes excited as the plasma heats up, and therefore the modes most capable of subsequently manipulating the plasma. 

Here we remove this shortcoming by deriving the Wigner--Moyal equations that govern the collisional whistler instability in a slab geometry, similar to what has been done in modelling drift-wave turbulence beyond the geometrical-optics approximation~\citep{Ruiz16,Zhu18a,Tsiolis20}. We find additional terms in the instability dispersion relation and growth rate due to gradients in the background plasma. We proceed to show that these additional terms can actually stabilize a non-constant temperature profile along the magnetic field, which is not possible if only the geometrical-optics approximation is used. Equivalently, these additional terms cause the instability to be damped at low temperatures, providing a mechanism for the instability to re-arrange the temperature profile into a marginally stable state via excitation at high temperature, propagation down the temperature gradient via Nernst advection, and subsequent damping at low temperature. We proceed to show that the marginally stable temperature profile generically takes the form of a staircase where isothermal regions are insulated from each other by abrupt jumps in the temperature, which occur at the zeros of a certain function comprised of an intricate combination of magnetic transport coefficients. These staircases can be in pressure balance, resembling the ubiquitous cold fronts in galaxy clusters~\citep{Markevitch07} but with the magnetic field no longer required to drape the front. 

We then derive the back-reaction of the instability on the background temperature profile and show that, in the initial stages of the instability, the frictional work done by the instability actually cools the background plasma instead of heating it (as demanded by energy conservation). Moreover, when the gradient of the unstable whistlers' amplitude is anti-aligned with the background temperature gradient along the magnetic field, the parallel heat flux can be reduced via the Ettingshausen effect, although this is more difficult to achieve in high-beta plasmas. If this mechanism can be reliably engineered, however, it might allow higher hotspot temperatures to be achieved in magnetized inertial fusion, since the parallel heat flux is the present limiting factor~\citep{Walsh22}.

\section{Summary}
Here we first provide an executive summary that highlights the main definitions, discussions, and results for each section, serving as an overall roadmap of the paper that can be consulted later for easy reference.

In \Sec{sec:Equations}, the governing electron MHD equations and slab geometry are introduced, leading to the set \eq{eq:MHDslab}. Fundamentally, the electron-MHD limit in slab geometry results in the simplification that only the electron temperature and the perpendicular components of the magnetic field (with respect to the single direction of inhomogeneity) have non-trivial time evolution; the electron density and the parallel magnetic field both remain constant in time. The perpendicular magnetic field is then expressed in the diagonalizing eigenbasis for its evolution equation, leading to the simplified description in terms of mode amplitudes given in \eq{eq:psiEVO} and \eq{eq:Tevo}. This is all for a general, unspecified friction and heat flux; the remaining parts of this section (and the remainder of the paper) specialize to when the friction and heat flux are determined by the standard Chapman--Enskog expressions \eq{eq:CEfrictionQ}. The nine transport coefficients ($\alpha_\parallel$, $\alpha_\perp$, $\alpha_\wedge$, $\beta_\parallel$, $\beta_\perp$, $\beta_\wedge$, $\kappa_\parallel$, $\kappa_\perp$, $\kappa_\wedge$) are all generally functions of the dimensionless magnetization parameter $\Hall$ defined in \eq{eq:Mdef}. Ultimately, after inserting the same eigenmode decomposition into the expressions for the friction and heat flux, one arrives at the final set of working equations: the magnetic-field mode amplitudes are governed by \eq{eq:psiNONLIN} and the temperature evolution is governed by \eq{eq:Tnonlin}. Up to that point, all manipulations of the initial equations are exact, and all nonlinearities are retained.

In \Sec{sec:main}, the dynamical equations are linearized about the small transverse field amplitude in order to obtain their dispersion relation and growth rate; since the temperature evolution involves terms quadratic or more in the mode amplitude, the two dynamical equations decouple in this limit. The temperature profiles are treated as fixed in time from the perspective of the magnetic field fluctuations, even though heat conduction is still present in the lowest-order temperature evolution equation; the validity of this approximation is outlined in \Sec{sec:dynamREL}. To maintain an exact treatment (within the linear regime), the dispersion relation and growth rate are obtained as the Hamiltonian and non-Hamiltonian components of the generalized wave-kinetic equation \eq{eq:wignerEVO} that governs the magnetic fluctuations. Specifically, the Hermitian and anti-Hermitian parts of the dispersion relation are given in \eq{eq:hamiltonian}, with the constituent terms defined in \eq{eq:hamCONSTIT}. This method of deriving the dispersion relation is chosen because it does not rely on making a short-wavelength approximation. The remainder of \Sec{sec:main} is therefore dedicated to defining the parameter space in which such a general treatment in necessary; it is found that the wavelength for the maximally growing mode becomes comparable to the background medium inhomogeneity lengthscale in the regime $\Hall^3 \plasmaBETA \ll 1$, where $\plasmaBETA$ is the plasma beta defined in \eq{eq:betaDEF}.

In \Sec{sec:stable}, the expression for the growth rate \eq{eq:Daham} is examined in more detail to understand the condition for instability \eq{eq:linSTAB}. Allowing for wavelengths comparable to inhomogeneity scale of the medium introduces two additional stabilizing mechanisms: one related to the gradient of the Nernst velocity that advects the perturbations, and one to the curvature of the plasma resistivity that diffuses the perturbations. Simple heuristic descriptions of these two stabilizing mechanisms are then presented briefly. The section concludes with a specific discussion for the simple case \eq{eq:linTstable} when the plasma is in pressure balance with a linear temperature profile; in this case, it is shown that the additional stabilizing terms cause a transition from unstable to stable behaviour analogous to the transition of the short-wavelength approximation from being valid to being violated discussed in \Sec{sec:main}.

In \Sec{sec:dynamREL}, the validity for the approximation that the temperature profile remain fixed in time for the linear stability analysis is assessed by comparing the instability growth rate \eq{eq:peakGROW} with the diffusion time \eq{eq:diffuseT} of the background temperature profile. This analysis accommodates arbitrarily shaped plasma profiles, with local lengthscales and curvatures defined in \eq{eq:lengths}. For convenience, the growth rate \eq{eq:peakGROW} is also split into terms driven by temperature gradients, temperature curvatures, mixed temperature and density gradients, density gradients, and density curvatures; the coefficient functions, defined in \eq{eq:auxFfuncs}, are grouped together respectively as $\TpsCOEF$, $\TppCOEF$, $\TNpCOEF$, $\NpsCOEF$, and $\NppCOEF$. Generally, it is found that the instability's growth rate \eq{eq:whistREL} normalised by the dynamical evolution time is only greater than unity in a sub-region of parameter space where both $\Hall$ and $\plasmaBETA$ are large; this is in contrast to the short-wavelength approximation that predicts the instability growth to be dynamically relevant everywhere [cf. \eq{eq:WKBrel}]. However, this result is nuanced because only regions where the instability is present are considered, and in fact, in much of the excluded parameter space there is no instability and perturbations are instead strongly damped (at a rate much larger than the diffusion time). The strong damping is due to the additional stabilizing mechanisms that occur when the wavelengths become comparable to the inhomogeneity scale discussed in \Sec{sec:stable}.

In \Sec{sec:global_stable}, the marginally stable temperature profile (satisfying the condition for zero growth rate everywhere) is derived. Due to the additional stabilizing terms in the expression for the growth rate, the solution involves non-trivial global structure. As discussed first in the general case, the governing equation \eq{eq:marginalMeqNOAPPROX} for the marginally stable $\Hall$ profile (a proxy for temperature) has the general structure of a nonlinear boundary-layer differential equation \eq{eq:boundaryEQ}, due to the singular nature as $\plasmaBETA$ becomes large of the coefficient $\gFUNC(\Hall)$ defined in \eq{eq:auxG} and \eq{eq:auxG12}. A variable transformation is performed to recast the nonlinear differential equation into a linear one that is then readily solved by successive integrations, yielding the solution \eq{eq:marginalZ} for the inverse function of the spatial profile $\Hall(z)$. This solution is then shown to exhibit a staircase structure as a generic feature due to the boundary layers. The remainder of the section specializes the general theory to two situations: plasmas with constant pressure and plasmas with constant density. Both cases are qualitatively similar when viewed in a parameter space consisting of $\Hall$ and some measure of the plasma beta; for the constant-pressure case, this is simply $\plasmaBETA$, but for the constant density case, an effective plasma beta is defined in \eq{eq:betaEFF}. The staircase is shown to have a single step, in rough correspondence to the transition between (dynamically relevant) instability growth and damping outlined in the previous section \ref{sec:dynamREL}. The predicted staircase feature of the solution is then verified with direct numerical simulation of the governing differential equation along with a closed-form analytical approximation described in \eq{eq:exampleT}.

In \Sec{sec:quasilinear}, the linear analysis is extended to a quasilinear study of the electron temperature response by restoring the lowest-order (quadratic) fluctuation terms in the temperature-evolution equation \eq{eq:TevoFRIC}. Two different groupings of the various terms are presented depending on the physics to be emphasized, with definitions provided in \eq{eq:perturbQUANT}. First, it is discussed how (as required by energy conservation), the growth of unstable perturbations also corresponds to a net cooling effect of the friction forces when the condition \eq{eq:negFRICcond} is satisfied. Next, the total heat flux (including Ettingshausen terms generated by the instability) is analyzed, and found to be reduced compared to the standard conductive heat flux when the condition \eq{eq:negETTINGS} is satisfied, although the reduction is generally modest. Lastly, the heat-flux reduction is analyzed when the temperature has the marginally stable staircase profile derived in the previous section \ref{sec:global_stable}. The extreme situation in which the heat flux is completely suppressed for the marginal-stability profiles is then considered to derive required conditions on the instability intensity profiles; the resulting expression \eq{eq:boundFLUXreduce} suggests that such high degrees of heat-flux suppression are unlikely to occur for high-beta plasmas within the validity of the quasilinear fluid model adopted here.

Finally, in \Sec{sec:concl}, the main results are summarized and directions for future work are outlined. Auxiliary calculations and review material are presented in appendices.


\section{Governing fluid equations}
\label{sec:Equations}


\subsection{Extended electron MHD equations in slab geometry}
\label{sec:MHDslab}

We are interested in the dynamics of electromagnetic oscillations in the whistler-frequency range in a collisional plasma. To allow an analytical description, let us consider for simplicity the extended electron MHD equations for a Lorentz plasma with stationary ions and isotropic pressure tensor:
\begin{subequations}
    \label{eq:mhd}%
	\begin{align}
        \label{eq:densMHD}
        \pd{t} n &= 0
		, \\
        \label{eq:BMHD}
		\pd{t} \Vect{B} &= - c \nabla \times 
		\left[
			\frac{ (\nabla \times \Vect{B}) \times \Vect{B}}{4 \pi e n} 
			- \frac{\nabla(nT)}{e n } 
			+ \frac{\Vect{R}}{e n}
		\right]
		, \\
        \label{eq:tempMHD}
		\frac{3}{2} n \, \pd{t} T
		&= 
		\frac{c}{4 \pi n e} (\nabla \times \Vect{B}) \cdot 
		\left(
			\frac{3}{2} n \nabla T 
			- T \nabla n
			+ \Vect{R}
		\right)
		- \nabla \cdot \Vect{q}
        .
	\end{align}
\end{subequations}

\noindent Here all symbols have their usual meaning: $n$ and $T$ are the electron density and temperature, respectively, $\Vect{B}$ is the magnetic field, $c$ is the speed of light in vacuum, $e > 0$ is the absolute value of the electron charge, $\Vect{R}$ is the frictional force experienced by the electron fluid due to pitch-angle-scattering collisions with ions, and $\Vect{q}$ is the electron heat flux. Expressions for $\Vect{R}$ and $\Vect{q}$ will be provided later in this section. Physically, the term within parenthesis in \eq{eq:tempMHD} is comprised of three distinct contributions: in order of appearance, they are (i) advection of temperature with the electron flow, (ii) compressional heating, and (iii) frictional heating that can be related to Joule heating by replacing $\Vect{R}$ with $\Vect{E}$ via Ohm's law -- indeed, the terms inside square brackets in \eq{eq:BMHD} are precisely $\Vect{E}$.

The simplest setup exhibiting the collisional whistler instability has a temperature gradient, density gradient, and wavevector of unstable perturbations all aligned with a mean background magnetic field~\citep{Bell20}. Hence, it can be adequately described by a $1$-D slab model in which the total magnetic field is given by 
\begin{equation}
    \Vect{B}(t, z) = \begin{pmatrix}
		\fluct{B}_x(t, z), \fluct{B}_y(t, z), B_z(t)
	\end{pmatrix}^\intercal
    ,
\end{equation}

\noindent with $B_z > 0$ being the mean field and $\fluct{B}_x$ and $\fluct{B}_y$ being fluctuating quantities associated with the instability. We also take $n$ and $T$ to be functions of $z$ and $t$ only. As discussed in \App{app:decouple}, this constraint means that the collisional whistler instability has no associated density or temperature fluctuations at the fundamental frequency. Since $B_z$ is independent of $z$, one has automatically
\begin{equation}
    \nabla \cdot \Vect{B}(t, z) = 0
    .
\end{equation}

\noindent One also has
\begin{equation}
    \nabla \times \Vect{B}(t, z) = 
	\begin{pmatrix}
		- \pd{z} \fluct{B}_y(t, z) \\
		\pd{z} \fluct{B}_x(t, z) \\
		0
	\end{pmatrix}
	= \begin{pmatrix}
		- \Mat{J} & 0 \\
		0 & 0
	\end{pmatrix}
	\pd{z} \Vect{B}
	,
\end{equation}

\noindent where we have introduced the skew-symmetric matrix
\begin{equation}
    \Mat{J}
	= \begin{pmatrix}
		0 & 1 \\
		-1 & 0
	\end{pmatrix}
    .
\end{equation}

Using the assumed form for the dynamical variables, the extended electron MHD equations \eq{eq:mhd} become
\begin{subequations}
    \label{eq:MHDslab}
	\begin{align}
		\pd{t} n &= 0
		, \\
		\pd{t} B_z &= 0
		, \\
        \label{eq:fluctBeq0}
		\pd{t} \fluct{\Vect{B}}_\perp
		&=
		\pd{z}
		\left(
			\frac{ \cycloF c^2}{\plasmaF^2 } 
			\Mat{J} \, \pd{z} \fluct{\Vect{B}}_\perp
			+ \frac{c}{e n}
			\Mat{J} \, \Vect{R}_\perp
		\right)
		, \\
        \label{eq:Teq0}
		\frac{3}{2} n \, \pd{t} T
		&= 
		- \frac{\cycloF c^2}{\plasmaF^2 B_z} \Vect{R}_\perp \cdot \Mat{J} \cdot \pd{z} \fluct{\Vect{B}}_\perp
		- \pd{z} q_z
        ,
	\end{align}
\end{subequations}

\noindent where we have defined the local plasma frequency and the mean electron cyclotron frequency respectively as
\begin{equation}
	\plasmaF^2(z) = \frac{4 \pi e^2}{m} n(z)
	, \quad
	\cycloF
	= \frac{e B_z}{m c}
    .
\end{equation}

\noindent Since the evolution equations for $n$ and $B_z$ are trivial, we shall omit them in the following analysis.


\subsection{Eigenbasis projection}
\label{sec:projection}

Further simplifications can be obtained by expanding $\fluct{\Vect{B}}_\perp$ and $\Vect{R}_\perp$ onto the eigenbasis of $\Mat{J}$, viz., the circular polarization vectors. These eigenvectors and their eigenvalues are given by
\begin{equation}
    \unit{e}_\pm = \frac{1}{\sqrt{2} } 
	\begin{pmatrix}
		1 \\
		\pm i
	\end{pmatrix}
	, \quad
	\Mat{J} \, \unit{e}_\pm
	= 
	\pm i
	\, \unit{e}_\pm
    ,
\end{equation}

\noindent and satisfy the orthogonality condition
\begin{equation}
    \unit{e}_\pm^* \cdot
	\unit{e}_\pm
	= 1
	, \quad
	\unit{e}_\mp^* \cdot
	\unit{e}_\pm
	= 0
	.
\end{equation}

\noindent Since $\fluct{\Vect{B}}_\perp$ and $\Vect{R}_\perp$ are both real-valued vectors and since $\unit{e}_+^* = \unit{e}_-$, the eigenbasis expansion takes the form
\begin{equation}
    \fluct{\Vect{B}}_\perp
    = \epsilon B_z \frac{ \psi \, \unit{e}_+ + \psi^* \, \unit{e}_+^*}{2}
    , \quad
    \Vect{R}_\perp
    =
    \frac{ \xi \, \unit{e}_+ + \xi^* \, \unit{e}_+^*}{2}
	,
    \label{eq:eigenEXPAND}
\end{equation}

\noindent where $\psi$ and $\xi$ are the complex scalar wavefunctions of $\fluct{\Vect{B}}_\perp$ and $\Vect{R}_\perp$, respectively, and $\epsilon > 0$ is a constant (assumed small) that parameterizes the relative size of the fluctuations. Later, it will be shown that $\xi$ is of order $\epsilon$ as well. Note that since
\begin{equation}
    \begin{pmatrix}
		\fluct{B}_x \\
		\fluct{B}_y
	\end{pmatrix}
	=
	\frac{\epsilon B_z}{\sqrt{2}}
	\begin{pmatrix}
		\Re \, \psi \\
		- \Im \, \psi
	\end{pmatrix}
	, \quad
	\begin{pmatrix}
		R_x \\
		R_y
	\end{pmatrix}
	=
	\frac{1}{\sqrt{2}}
	\begin{pmatrix}
		\Re \, \xi \\
		- \Im \, \xi
	\end{pmatrix}
    ,
\end{equation}

\noindent by introducing $\psi$ and $\xi$ we have essentially traded two real-valued degrees of freedom for a single complex-valued degree of freedom. 

Since $\unit{e}_+$ is independent of $t$ and $z$, one can readily show using orthogonality that the real-vector-valued evolution equation \eq{eq:fluctBeq0} for $\fluct{\Vect{B}}_\perp$ is equivalent to the following complex-scalar-valued evolution equation for $\psi$:
\begin{equation}
	i \pd{t} \psi
	= - \pd{z}
	\left(
		\frac{ \cycloF c^2}{\plasmaF^2} 
		\pd{z} \psi
		+ \frac{c}{e n} \frac{\xi}{\epsilon B_z}
	\right)
    .
	\label{eq:psiEVO}
\end{equation}

\noindent Similarly, the temperature equation \eq{eq:Teq0} takes the form
\begin{equation}
	\frac{3}{2} n \, \pd{t} T
	= 
	\epsilon \frac{\cycloF c^2}{\plasmaF^2} \frac{\Im\left( \xi^* \pd{z} \psi \right)}{2 }
	- \pd{z} q_z
    .
	\label{eq:Tevo}
\end{equation}


\subsection{Chapman--Enskog friction coefficients}
\label{sec:friction}

Equations \eq{eq:psiEVO} and \eq{eq:Tevo} are valid for any friction force, allowing one to study driven systems. For undriven systems, the friction force is determined by the plasma fluid variables themselves according to some closure. A common closure that we shall adopt here is provided by the Chapman--Enskog method~\citep{Helander02,Bott24}, which for the Lorentz collision operator yields the following expressions for the friction force and for the heat flux~\citep{Epperlein84}:
\begin{equation}
	\Vect{R} = \frac{n e c }{ \plasmaF^2 \collT} \Mat{M}_\alpha \cdot \nabla \times \Vect{B} - n \Mat{M}_\beta \cdot \nabla T
	, \quad
	\Vect{q}
	= 
	- n \collT \thermalV^2
	\Mat{M}_\kappa \cdot \nabla T
	- \frac{\cycloF c^2}{\plasmaF^2} \frac{n T}{B_z} \, \Mat{M}_\beta \cdot \nabla \times \Vect{B}
	,
    \label{eq:CEfrictionQ}
\end{equation}

\noindent where $\collT$ is the electron-ion collision time~\citep{Helander02, Epperlein86} and $\thermalV$ is the thermal speed, defined respectively as
\begin{equation}
    \collT
	=
	\frac{12 \pi^2}{ \sqrt{2 \pi} }
	\frac{
		n \thermalV^3
	}{
		Z \plasmaF^4 \log \Lambda
	}
    , \quad
    \thermalV = \sqrt{\frac{T}{m}}
    .
\end{equation}

\noindent The dimensionless resistivity ($\alpha$), thermoelectric ($\beta$), and conductivity ($\kappa$) matrices are anisotropic with respect to the magnetic field, taking the form
\begin{equation}
    \Mat{M}_\sigma
    = 
    \sigma_\perp(\Hall)\IMat{3}
    + \Delta_\sigma(\Hall) \frac{\Vect{B} \Vect{B}}{|B|^2}
    \pm \sigma_\wedge(\Hall) \frac{\hatmap{B} }{|B|}
    , \quad
    \sigma = \alpha, \beta, \kappa
    ,
\end{equation}

\noindent where $\hatmap{B}$ denotes the skew-symmetric hat-map matrix that enacts the cross-product $\hatmap{B} \cdot \Vect{v} = \Vect{B} \times \Vect{v}$~\citep{Zhang20} and $\Delta_\sigma \doteq \sigma_\parallel - \sigma_\perp$ is the anisotropy measure. In the last term, the minus sign applies to $\alpha_\wedge$ only. Also note that $\sigma_\perp$ and $\sigma_\wedge$ are both positive, but $\Delta_\alpha \le 0$, while $\Delta_\beta \ge 0$ and $\Delta_\kappa \ge 0$. The various transport coefficients are functions purely of the magnetization parameter 
\begin{align}
    \Hall \doteq \frac{|B|}{B_z} \cycloF \collT
    &=
    \cycloF \collT
    \sqrt{
        1 + \frac{\epsilon^2}{2} |\psi|^2
    }
    \nonumber\\
    &\equiv
    \frac{3}{ 4 \sqrt{2 \pi} }
    \frac{
        \cycloF \sqrt{m}
    }{
        Z e^4 \log \Lambda
    }
    \frac{T^{3/2} }{n }
    \sqrt{
        1 + \frac{\epsilon^2}{2} |\psi|^2
    }.
    \label{eq:Mdef}
\end{align}

\noindent In the remainder of the analysis, we shall use the rational interpolants for the various transport coefficients as functions of $\Hall$ developed in \citet{Lopez24a}. Their limiting forms as $\Hall \to 0$ and $\Hall \to \infty$ are listed in \App{app:transport}.

Lastly, note that the slab geometry allows the relevant matrices to be constructed explicitly with a relatively simple form. Since
\begin{equation}
    \hatmap{B}
	= \begin{pmatrix}
		0 & - B_z & \fluct{B}_y \\[2mm]
		B_z & 0 & - \fluct{B}_x \\[2mm]
		- \fluct{B}_y & \fluct{B}_x & 0
	\end{pmatrix}
    ,
\end{equation}

\noindent the anisotropic transport matrices are given as
\begin{equation}
    \Mat{M}_\sigma
	=
	\frac{1}{|B|^2}
	\begin{pmatrix}
		\Delta_\sigma \fluct{B}_x^2 + \sigma_\perp |B|^2
        & \Delta_\sigma \fluct{B}_x \fluct{B}_y \mp \sigma_\wedge |B| B_z
        & \Delta_\sigma \fluct{B}_x B_z \pm \sigma_\wedge |B| \fluct{B}_y 
        \\[2mm]
		\Delta_\sigma \fluct{B}_x \fluct{B}_y \pm \sigma_\wedge |B| B_z 
        & \Delta_\sigma \fluct{B}_y^2 + \sigma_\perp |B|^2
        & \Delta_\sigma \fluct{B}_y B_z \mp \sigma_\wedge |B| \fluct{B}_x
        \\[2mm]
		\Delta_\sigma \fluct{B}_x B_z \mp \sigma_\wedge |B| \fluct{B}_y
        & \Delta_\sigma \fluct{B}_y B_z \pm \sigma_\wedge |B| \fluct{B}_x
        & \Delta_\sigma B_z^2 + \sigma_\perp |B|^2
	\end{pmatrix}
    .
\end{equation}


\subsection{Resulting equations for Chapman--Enskog friction forces}

From \eq{eq:CEfrictionQ}, the perpendicular component of the Chapman--Enskog friction force in slab geometry is
\begin{equation}
	\Vect{R}_\perp
	=
	\frac{n e c}{\plasmaF^2 \collT}
	\left(
		\frac{\alpha_\wedge B_z}{|B|} \IMat{2}
		- \alpha_\perp \Mat{J}
		- \frac{\Delta_\alpha}{|B|^2} \fluct{\Vect{B}}_\perp \fluct{\Vect{B}}_\perp^\intercal \Mat{J}
	\right)
	\, \pd{z} \fluct{\Vect{B}}_\perp
	-\frac{n \, \pd{z}T}{|B|}
	\left(
		\frac{\Delta_\beta B_z}{|B|} \IMat{2}
		+ \beta_\wedge \Mat{J}
	\right)
	\fluct{\Vect{B}}_\perp
    .
\end{equation}

\noindent Hence, the complex amplitude $\xi$ is given in terms of $\psi$ as
\begin{align}
	\xi
	&=
	\epsilon B_z
	\frac{n e c}{\plasmaF^2 \collT}
	\left(
		\frac{\alpha_\wedge B_z}{|B|}
		- i \alpha_\perp 
	\right)
	\pd{z} \psi
    \nonumber\\
    &\hspace{4mm}-
    \epsilon B_z
    \left[
		\frac{n \, \pd{z}T}{|B|}
		\left(
			\frac{\Delta_\beta B_z}{|B|}
			+ i \beta_\wedge
		\right)
		- \frac{\epsilon^2 B_z^2}{2} \frac{n e c}{\plasmaF^2 \collT} \frac{\Delta_\alpha}{|B|^2}
		\Im \left( \psi^* \pd{z}\psi \right)
	\right]
	\psi
    .
\end{align}

\noindent Thus, $\xi = O(\epsilon)$, as promised following \eq{eq:eigenEXPAND}. Hence, \eq{eq:psiEVO} becomes
\begin{align}
    i \pd{t} \psi
    &= - \pd{z}
    \left[
        \left(
            \groupDISP
            - i \resist
        \right)
        \pd{z} \psi
        \nullFrac
    \right]
    + \pd{z}
    \left[
        \left(
            \growthV
            - i \nerstV
        \right)
        \psi
        \nullFrac
	\right]
    .
	\label{eq:psiNONLIN}
\end{align}

\noindent where we have defined the following quantities:
\begin{subequations}
    \label{eq:dispFUNCnonlin}
    \begin{align}
        \groupDISP
        &= \frac{ \cycloF c^2}{\plasmaF^2}
        \left(
            1 
            + \frac{\alpha_\wedge }{\Hall }
        \right)
        , \\
        \resist
        &= \frac{ \cycloF c^2}{\plasmaF^2} \frac{|B|}{B_z}\frac{\alpha_\perp }{ \Hall }
        , \\
        \growthV &=
        \frac{\Delta_\beta }{m \cycloF}
        \left( \frac{B_z}{|B|}\right)^2 \pd{z}T
        - \frac{\epsilon^2}{2} \frac{\cycloF c^2}{\plasmaF^2} 
        \frac{B_z}{|B|} \frac{\Delta_\alpha }{\Hall}
        \Im \left( \psi^* \pd{z}\psi \right)
        , \\
        \nerstV &=
        - \frac{\beta_\wedge}{m \cycloF} \frac{B_z}{|B|} \pd{z}T
        .
    \end{align}
\end{subequations}

\noindent Recall that $|B|$ and $\Hall$ depend on $|\psi|^2$, so \eq{eq:psiNONLIN} still contains nonlinear effects.

The $z$-component of Chapman--Enskog heat flux \eq{eq:CEfrictionQ}, $q_z$, that appears in \eq{eq:Tevo} can be shown to take the form
\begin{align}
	q_z
	&=
	- \frac{n T \collT}{m}
	\left[
		\kappa_\text{eff} \, \pd{z} T
        + \epsilon^2 \frac{\cycloF^2 m c^2}{2 \plasmaF^2}
		\left(
			\frac{\beta_\wedge}{2 \Hall} \, \pd{z} |\psi|^2
			+ \frac{B_z}{|B|} \frac{\Delta_\beta}{\Hall}
            \Im \left( \psi^* \pd{z}\psi \right)
		\right)
	\right]
    ,
    \label{eq:qzDEF}
\end{align}

\noindent where the effective conductivity is given as
\begin{equation}
    \kappa_\text{eff}
    =
    \frac{2 \kappa_\parallel
		+ \epsilon^2 \kappa_\perp |\psi|^2 
	}{
        2 + \epsilon^2 |\psi|^2
    }
    .
\end{equation}

\noindent Hence, \eq{eq:Tevo} becomes
\begin{align}
    \frac{3}{2} n \, \pd{t} T
    &= 
    \epsilon^2 \frac{B_z^2}{8\pi} 
    \left[
        \resist |\pd{z} \psi|^2
        - \growthV \Im\left( \psi^*\pd{z} \psi \right)
        - \frac{\nerstV}{2} \pd{z} |\psi|^2
    \right]
    - \pd{z} q_z
    ,
    \label{eq:Tnonlin}
\end{align}

\noindent with $q_z$ given in \eq{eq:qzDEF}. 

In summary, the two main equations we shall be working with in the remainder of this paper are the evolution of the transverse field perturbations (describing the linear instability dynamics) governed by \eq{eq:psiNONLIN}, and the evolution of the temperature profile (describing the back-reaction of the instability on the equilibrium dynamics) governed by \eq{eq:Tnonlin}. More will be said about the physical content of \eq{eq:Tnonlin} in \Sec{sec:quasilinear}, but as a brief preview, it is worthwhile to highlight now that only the first term within square brackets is manifestly positive (corresponding to resistive heating). The remaining terms, although still related to frictional forces, can take either sign depending on whether the instability is growing or damping, allowing the instability to re-distribute the temperature profile. As we shall discuss in later sections (beginning in the following section), this effect can be particularly prominent in the high-beta, low-magnetization regime.


\section{Dispersion relation, growth rate, and breakdown of the short-wavelength approximation}
\label{sec:main}

Let us now restrict attention to the linear limit when $\epsilon \to 0$. To lowest order, \eq{eq:Tnonlin} is decoupled from \eq{eq:psiNONLIN}, so we shall just consider the dynamics of \eq{eq:psiNONLIN} with prescribed stationary temperature and density profiles (the back-reaction of the instability on the temperature profile will be considered in \Sec{sec:quasilinear}). Moreover, \eq{eq:psiNONLIN} maintains the same form when $\epsilon \to 0$, but the constituent functions \eq{eq:dispFUNCnonlin} become simply
\begin{subequations}
    \begin{align}
        \label{eq:group}
        \groupDISP
        &= \frac{ \cycloF c^2}{\plasmaF^2}
        \left(
            1 
            + \frac{\alpha_\wedge }{\Hall }
        \right)
        , \\
        \label{eq:resist}
        \resist
        &= \frac{ \cycloF c^2}{\plasmaF^2} \frac{\alpha_\perp }{ \Hall }
        , \\
        \label{eq:growthV}
        \growthV &=
        \frac{\Delta_\beta }{m \cycloF} \pd{z}T
        , \\
        \label{eq:nerstV}
        \nerstV &=
        - \frac{\beta_\wedge}{m \cycloF} \pd{z}T
        ,
    \end{align}
    \label{eq:hamCONSTIT}
\end{subequations}

\noindent where $\Hall = \cycloF \collT$. In this limit, the physical interpretation of these quantities becomes clearer: $\groupDISP$ represents the familiar group-velocity dispersion for whistler waves but modified by friction-induced Hall effect~\citep{Davies21}, $\resist$ governs the resistive diffusion of magnetic-field perturbations, $\growthV$ is the cross-gradient Nernst advection velocity, and $\nerstV$ is the standard Nernst advection velocity. In particular, the presence of $\groupDISP$ explains our choice to call this instability the `collisional whistler instability'; in the absence of friction terms (\ie with all $\alpha$ and $\beta$ coefficients set to zero), the dispersion relation \eq{eq:Dham}, which we shall derive shortly, would be identical to the standard whistler dispersion relation in the electron-MHD limit [see, for example, \citet{Komarov18}].

\begin{figure}
   \centering
   \includegraphics[width=0.45\linewidth,trim={4mm 4mm 3mm 4mm},clip]{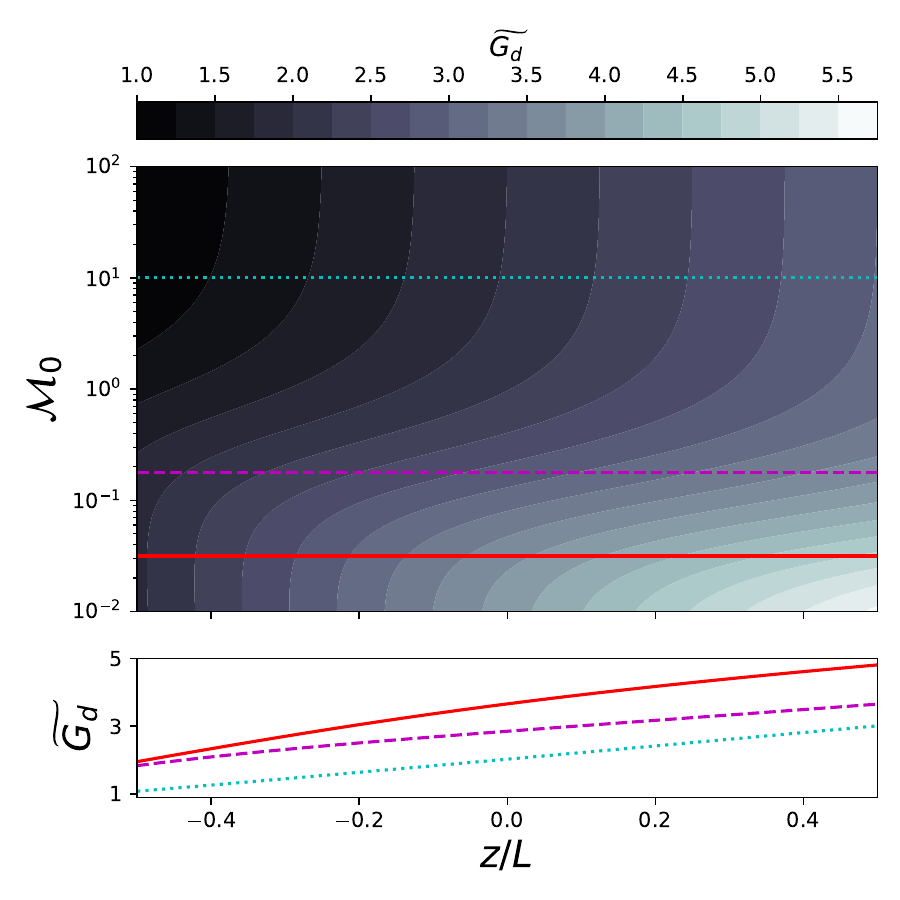}
   \includegraphics[width=0.45\linewidth,trim={4mm 4mm 3mm 4mm},clip]{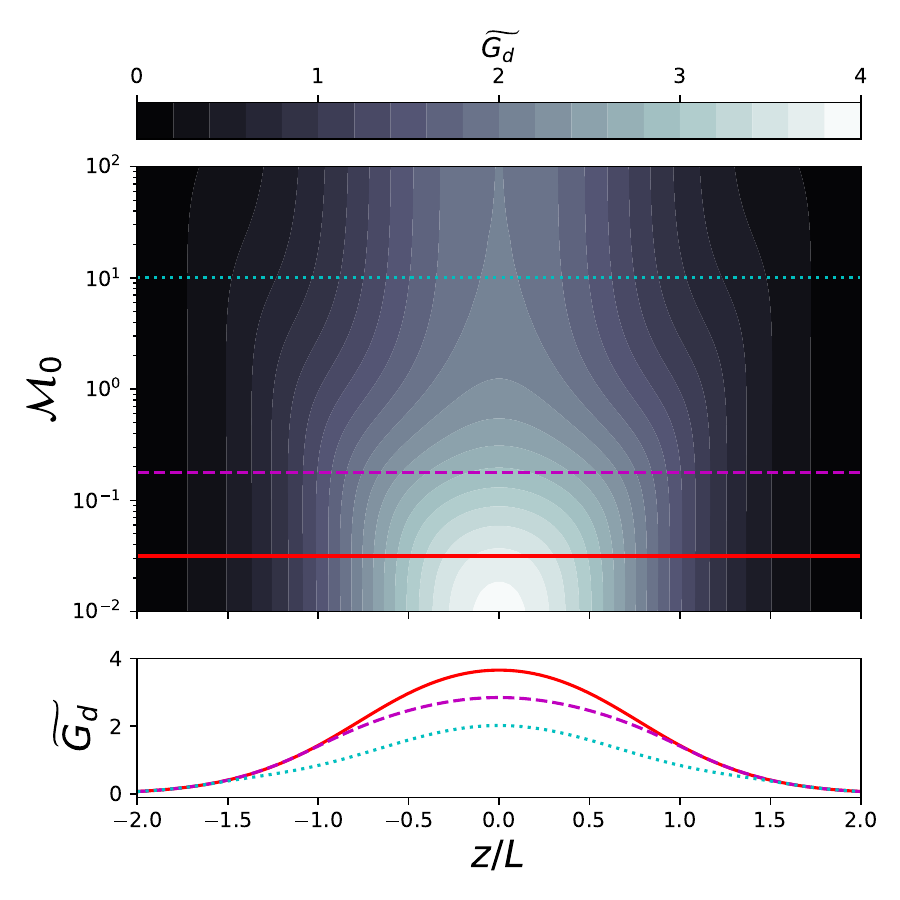}

   \caption{Plots of the normalized group-velocity dispersion $\widetilde{\groupDISP} = \plasmaBETA\groupDISP/\thermalV \larmorR$ [see \eq{eq:group}] for a linear temperature profile \eq{eq:linT} (left) and a Gaussian temperature profile \eq{eq:gaussT} (right). All normalization quantities are defined with respect to $T_0$, and $\Hall_0 = \Hall(T_0)$.}
   \label{fig:group}
\end{figure}

\begin{figure}
   \centering
   \includegraphics[width=0.45\linewidth,trim={4mm 4mm 3mm 4mm},clip]{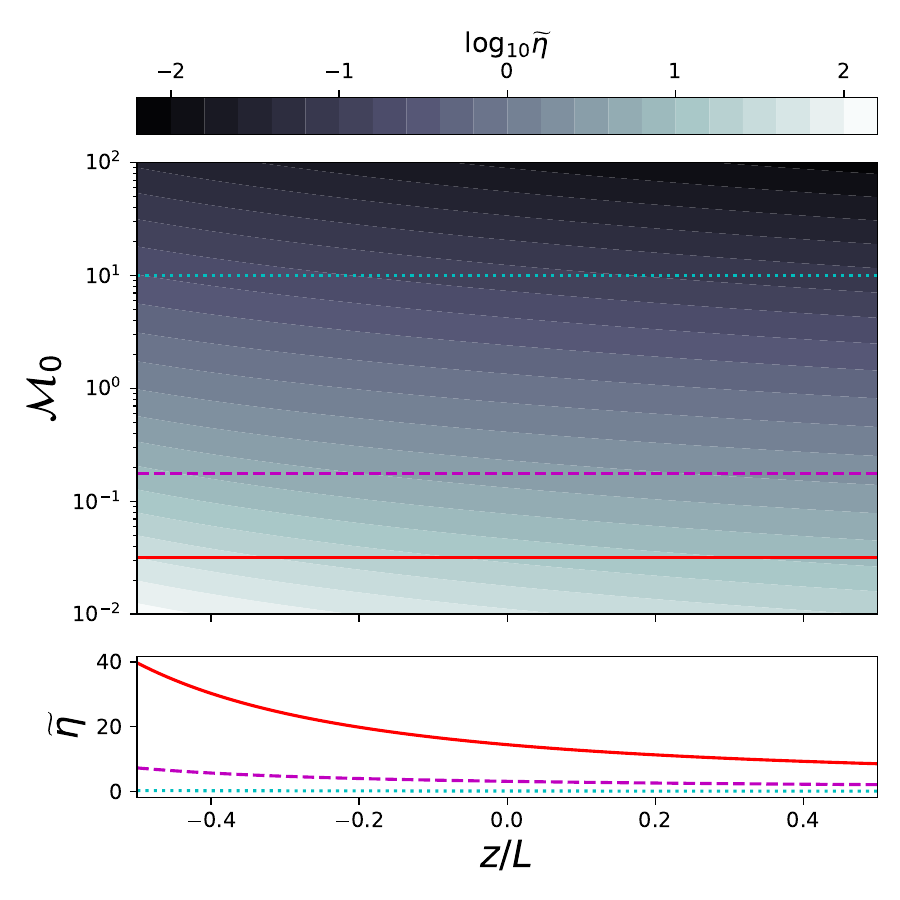}
   \includegraphics[width=0.45\linewidth,trim={4mm 4mm 3mm 4mm},clip]{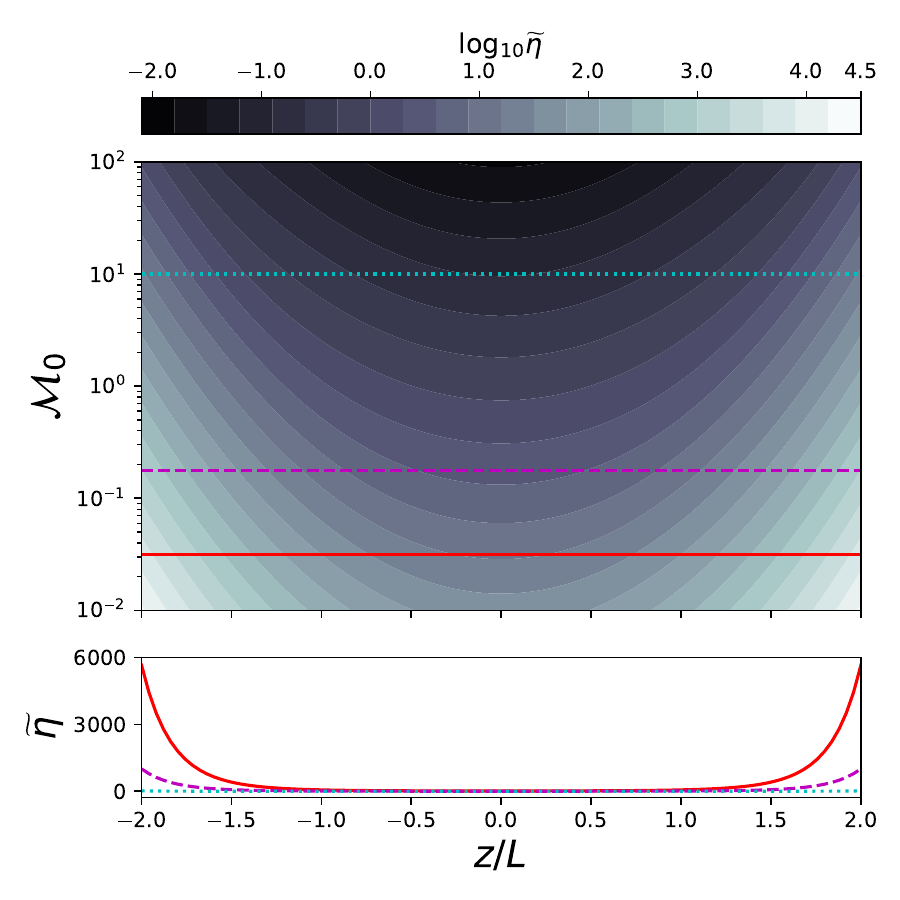}

   \caption{Same as \Fig{fig:group} but for the normalized resistivity $\widetilde{\resist} = \plasmaBETA\resist/\thermalV \larmorR$ [see \eq{eq:resist}].}
   \label{fig:resist}
\end{figure}

\begin{figure}
   \centering
   \includegraphics[width=0.45\linewidth,trim={4mm 4mm 4mm 4mm},clip]{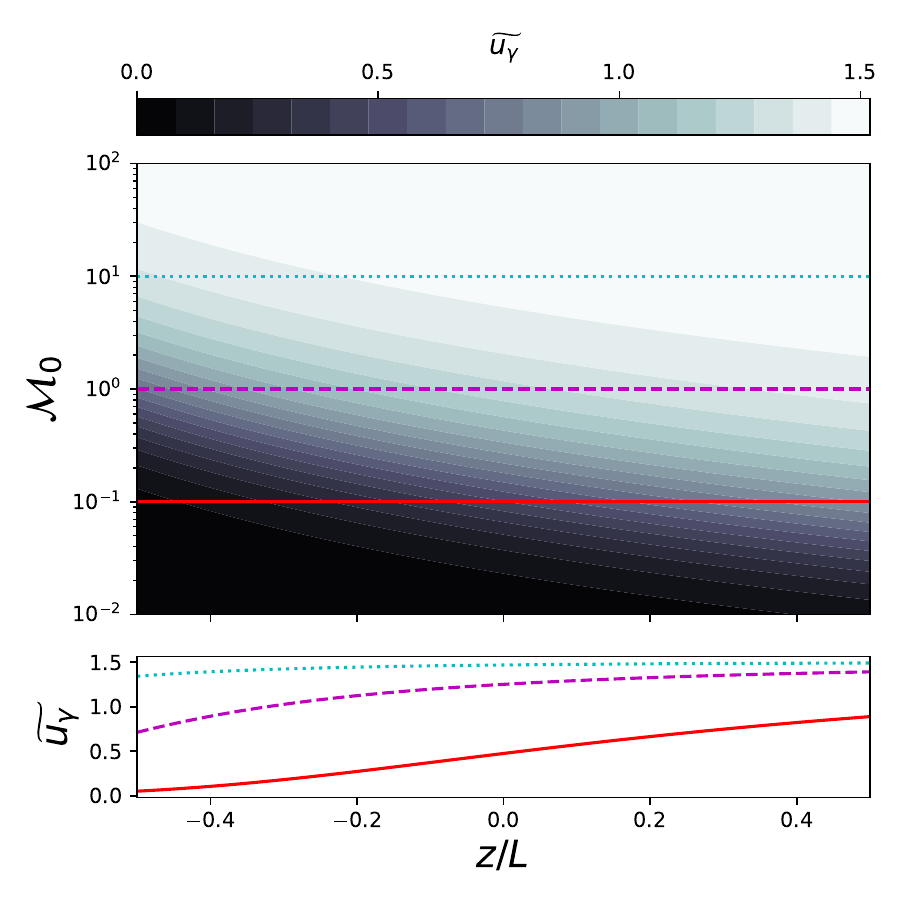}
   \includegraphics[width=0.45\linewidth,trim={4mm 4mm 4mm 4mm},clip]{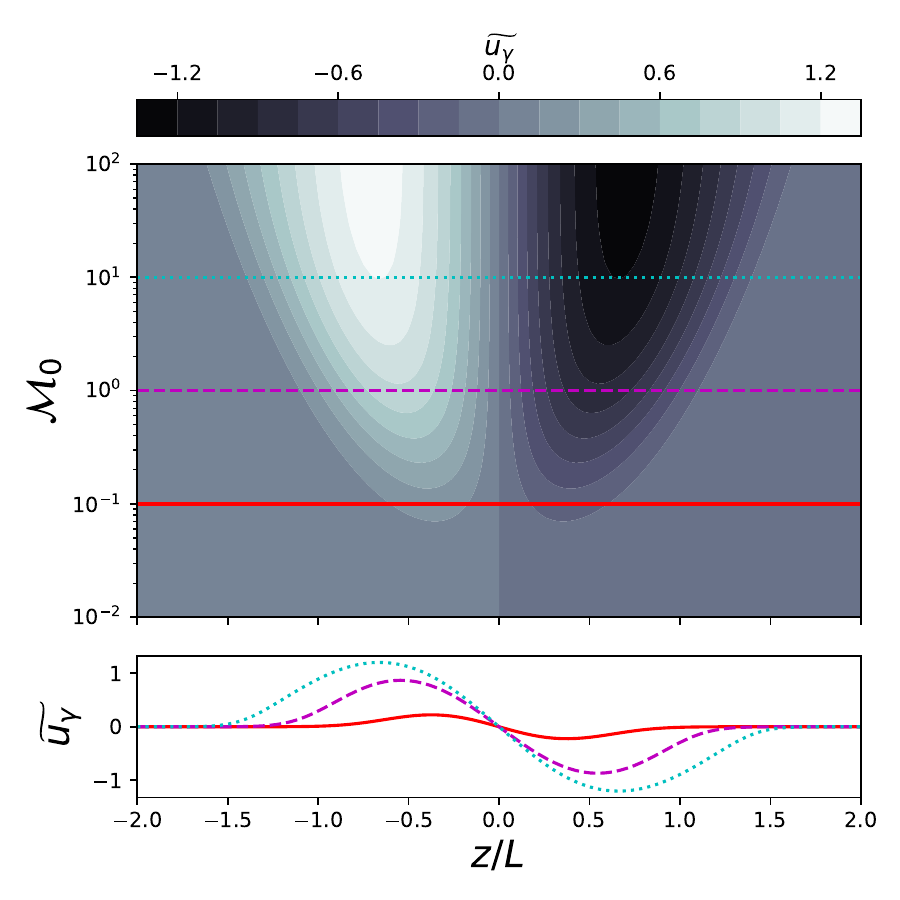}

   \caption{Same as \Fig{fig:group} but for the normalized cross-gradient Nernst velocity $\widetilde{\growthV} = L\growthV/\larmorR \thermalV$ [see \eq{eq:growthV}].}
   \label{fig:crossnernst}
\end{figure}

\begin{figure}
   \centering
   \includegraphics[width=0.45\linewidth,trim={4mm 4mm 4mm 4mm},clip]{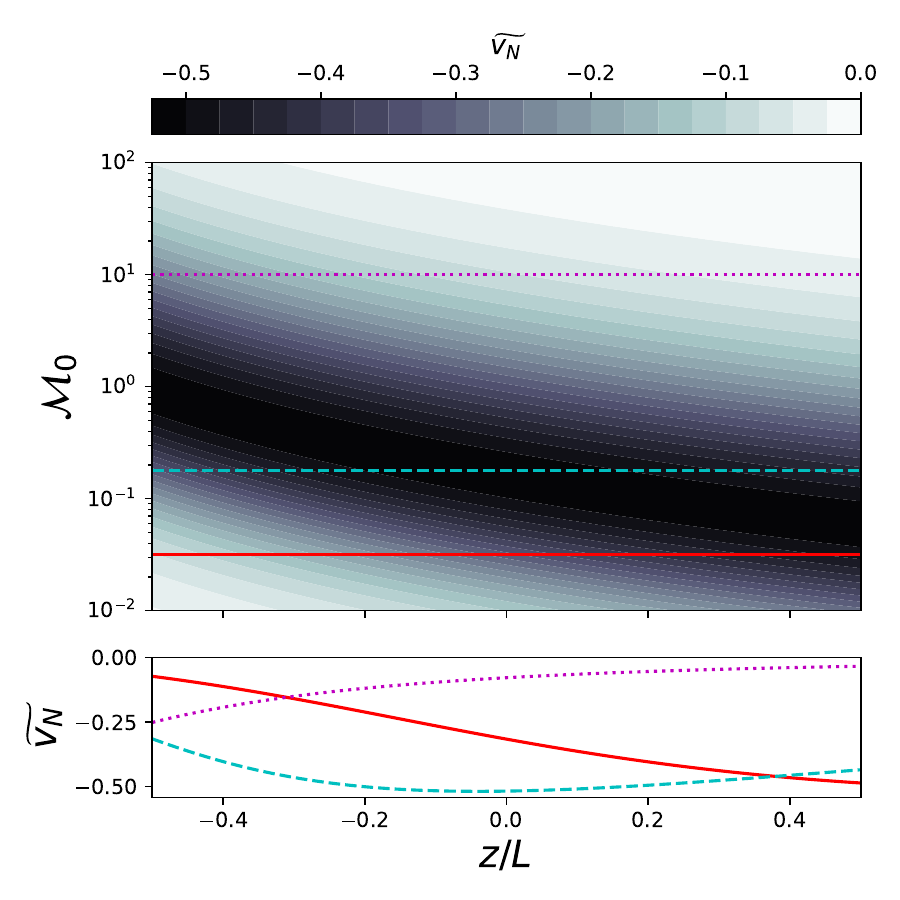}
   \includegraphics[width=0.45\linewidth,trim={4mm 4mm 4mm 4mm},clip]{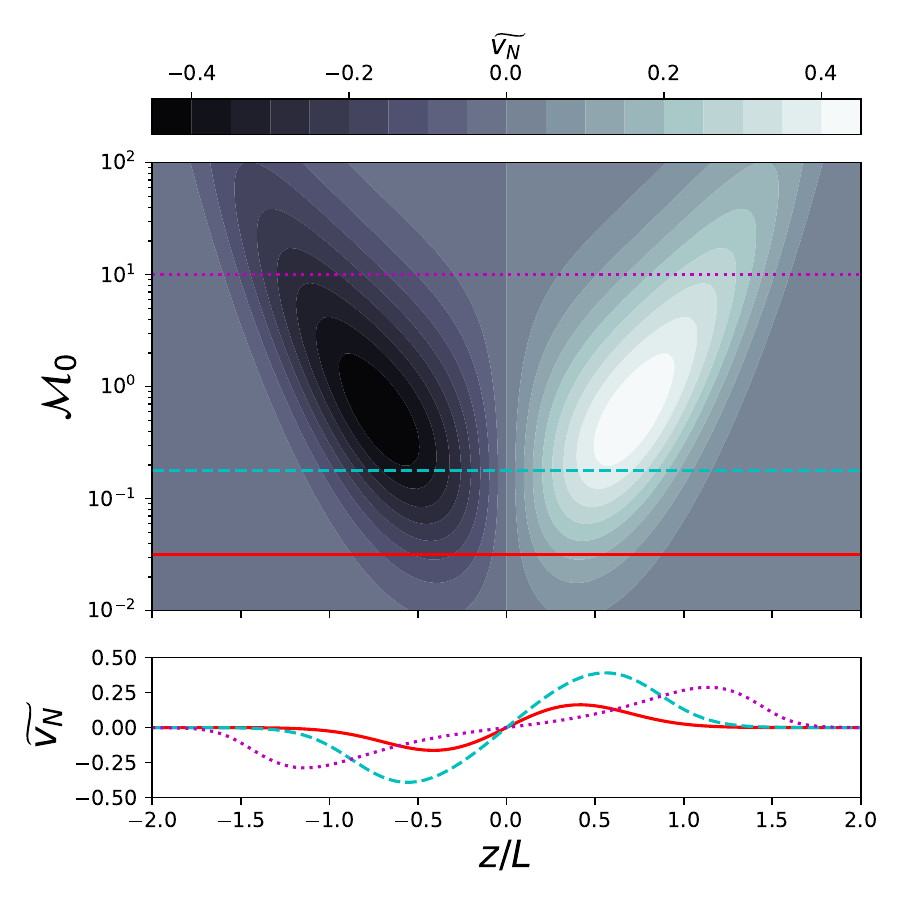}

   \caption{Same as \Fig{fig:group} but for the normalized Nernst velocity $\widetilde{\nerstV} = L\nerstV/\larmorR \thermalV$ [see \eq{eq:nerstV}].}
   \label{fig:nernst}
\end{figure}

For illustration purposes, Figs.~\ref{fig:group} -- \ref{fig:nernst} show how $\groupDISP$, $\resist$, $\growthV$, and $\nerstV$ vary in space for a linear temperature profile
\begin{equation}
   T(z) = T_0 (1 + z/L)
   ,
   \label{eq:linT}
\end{equation}

\noindent and for a Gaussian temperature profile
\begin{equation}
   T(z) = T_0 \exp\left[-(z/L)^2\right]
   , 
   \label{eq:gaussT}
\end{equation}

\noindent both with an isobaric density profile $n \propto 1/T$. 

We shall now derive the linear dispersion relation and growth rate for the collisional whistler instability using two approaches -- a phase-space-based approach and a more traditional configuration-space-based approach.


\subsection{Derivation via Wigner--Moyal phase-space formulation}
\label{sec:wigner}

Here we shall derive the phase-space analog of \eq{eq:psiNONLIN} that governs the Wigner function of the complex mode amplitude $\psi$, defined as 
\begin{equation}
	\Symb{W}(z, k_z, t)
	= \int \dd s \, 
	\psi^*\left(z + \frac{s}{2}, t \right) \psi \left(z - \frac{s}{2}, t \right)
	\exp(i k_z s)
    .
    \label{eq:wignerDEF}
\end{equation}

\noindent This will bring the Hamiltonian structure of \eq{eq:psiNONLIN} to light, allowing us then to extract the dispersion relation and growth rate immediately. The following derivation closely follows the presentation of \citet{Ruiz16}, so the reader is invited to consult that reference along with the brief summary provided in \App{app:Weyl} for more detailed explanations of the steps involved. A more conventional derivation of the same dispersion relation and growth rate, based on a polar decomposition of the wavefield into an amplitude and phase, is presented in \App{app:polar}.

To begin, let us introduce the state vector $\ket{\psi}$ whose spatial projection is given as $\braket{z}{\psi} = \psi(z)$. Let us also introduce the operators $\oper{z}$ and $\oper{k}_z$ whose action on state vectors is given respectively as $\bra{z} \oper{z} \ket{\psi} = z \psi(z)$ and $\bra{z} \oper{k}_z \ket{\psi} = - i \pd{z} \psi(z)$. Then, \eq{eq:psiNONLIN} can be viewed as the spatial projection of the Schr\"odinger equation
\begin{equation}
    i \pd{t} \ket{\psi}
    =
    \oper{D}
    \ket{\psi}
    \label{eq:Schrodinger}
    ,
\end{equation}

\noindent where the non-Hermitian Hamiltonian operator $\oper{D}$ has the form
\begin{equation}
    \oper{D}
    = \oper{k}_z
    \left[
        \groupDISP(\oper{z}) 
        - i \resist(\oper{z})
    \right]\oper{k}_z
    + \oper{k}_z
    \left[
        \nerstV(\oper{z})
        + i \growthV(\oper{z})
    \right]
    .
\end{equation}

\noindent By right-multiplying \eq{eq:Schrodinger} by $\bra{\psi}$ and subtracting its adjoint equation, one arrives at the evolution equation for the density operator $\oper{W} \doteq \ket{\psi}\bra{\psi}$:
\begin{equation}
    i \pd{t} \oper{W}
    =
    \oper{D} \oper{W}
    - \oper{W} \oper{D}^\dagger
    .
\end{equation}

\noindent Finally, applying the Wigner--Weyl transform (WWT, see \App{app:Weyl}) yields the Wigner--Moyal kinetic equation that governs the Wigner function \eq{eq:wignerDEF} of the fluctuations:
\begin{equation}
	\pd{t} \Symb{W}
    = i \Symb{W} \star \Symb{D}^*
    - i \Symb{D} \star \Symb{W}
	\equiv 2 \, \Im \left( \Symb{D}_H \star \Symb{W} \right)
	+ 2 \, \Re \left(\Symb{D}_A\star \Symb{W} \right)
	,
    \label{eq:wignerEVO}
\end{equation}

\noindent where the Moyal product $\star$ is defined in \App{app:Weyl} and $\Symb{D}$ is the dispersion function whose Hermitian and anti-Hermitian parts are
\begin{subequations}
    \label{eq:hamiltonian}
    \begin{align}
        \label{eq:Dham}
        \Symb{D}_H
        &=
        k_z \nerstV(z)
        + k_z^2 \groupDISP(z)
        + \frac{1}{2} \pd{z} \growthV(z)
        + \frac{1}{4} \pd{z}^2 \groupDISP(z)
        , \\
        \Symb{D}_A
        &=
        k_z \growthV(z)
        - k_z^2 \resist(z)
        - \frac{1}{2} \pd{z}\nerstV(z)
        - \frac{1}{4} \pd{z}^2 \resist(z)
        .
        \label{eq:Daham}
    \end{align}
\end{subequations}

Equation \eq{eq:wignerEVO} states that $\Symb{D}_H$ governs the Hamiltonian dynamics of the collisional whistler instability in phase space while $\Symb{D}_A$ acts as the growth rate. This latter point is seen more easily by integrating \eq{eq:wignerEVO} over $k_z$ (\ie taking the lowest-order `fluid' moment). This gives
\begin{equation}
    \pd{t} \inten
    = 2 \ave{\Symb{D}_A} \inten
    - \pd{z}
    \left[ 
        \ave{\pd{k_z} \Symb{D}_H} \inten
        - \frac{1}{2} \pd{z} (\resist \inten)
    \right]
    ,
    \label{eq:intenEQ}
\end{equation}

\noindent where moments of $\Symb{W}$ have been defined as follows:
\begin{equation}
    \inten(z) \doteq \int \frac{\dd k_z}{2\pi} \Symb{W}(z, k_z)
    \equiv |\psi(z)|^2
	, \quad
    \ave{f(z)} \doteq
	\frac{
        \int \dd k_z f(z, k_z) \Symb{W}(z, k_z)
    }{2\pi \inten(z)}
    .
    \label{eq:aveDEF}
\end{equation}

\noindent Hence, if the flux at the boundary is negligible, then the total amount of energy contained within the fluctuations remains constant if $\ave{\Symb{D}_A} = 0$.

Both the Hermitian and anti-Hermitian parts \eq{eq:hamiltonian} of the instability dynamics contain additional terms (the final two gradient terms) that are absent from previous treatments performed in \citet{Bell20} based on the short-wavelength approximation. These terms arise because of the spatial variation in the plasma profiles and the consequent non-commutation with the differential operator $\pd{z}$, similar to how additional non-Hermitian terms arise when studying zonal flows~\citep{Ruiz16}. More will be said about the gradient terms in \Sec{sec:stable}.


\subsection{Insufficiency of short-wavelength approximation}
\label{sec:noWKB}

One might be tempted to drop the gradient drives when the equilibrium lengthscales are sufficiently long (\ie to apply the short-wavelength approximation), but this not always valid. From \eq{eq:Daham}, we see that the wavevector for the fastest-growing mode at a given point $z$ is given by
\begin{equation}
    k_{z,\text{max}} = 
    \frac{\growthV(z)}{2\resist(z)}
    .
    \label{eq:kzmax}
\end{equation}

\begin{figure}
    \centering
    \includegraphics[width=0.6\linewidth,trim={16mm 6mm 30mm 2mm},clip]{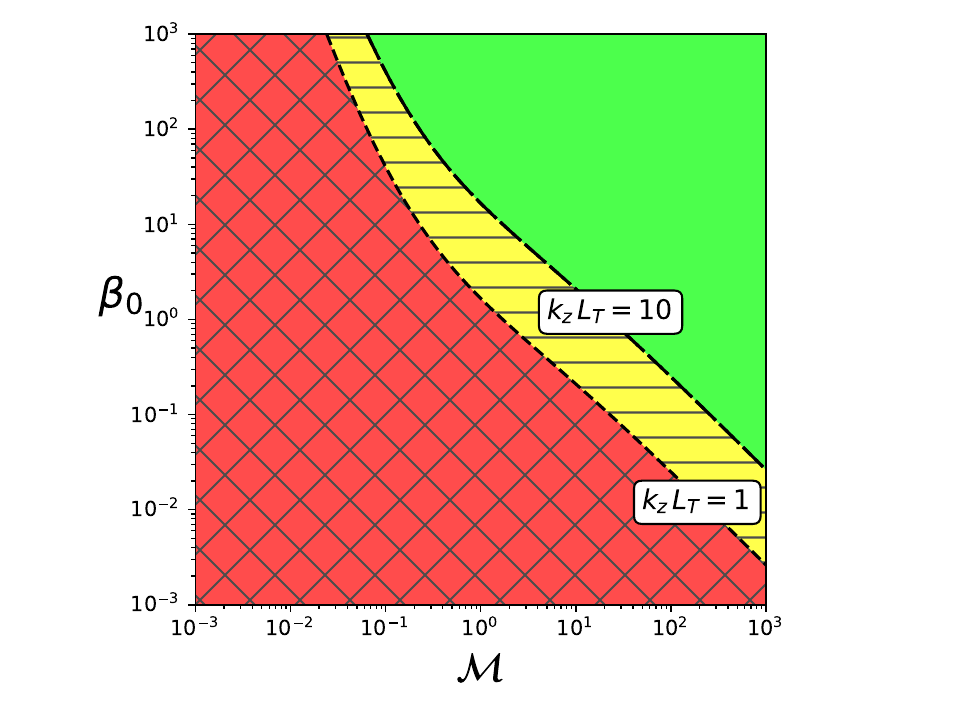}
    \caption{Region of parameter space where a geometrical-optics description of the collisional whistler instability is valid (green, $k_z \lenT \gg 1$), questionable (yellow lined region, $k_z \lenT \gtrsim 1$), and not valid (red crossed region, $k_z \lenT < 1$). The regions are determined by the expression for $k_z \lenT$ given by \eq{eq:WKBvalid} using the transport coefficients of \citet{Lopez24a}.}
    \label{fig:WKBvalid}
\end{figure}

\noindent Since $\growthV \propto \pd{z} T$, \eq{eq:kzmax} shows that the fluctuation wavelength may be comparable to, or even larger than, the temperature lengthscale $\lenT = (\pd{z} \log T)^{-1}$, depending on the prefactor. Indeed, one has
\begin{equation}
    k_{z,\text{max}} \lenT
    = \frac{ \Hall \Delta_\beta}{4 \alpha_\perp} \plasmaBETA
    \sim \Hall^3 \plasmaBETA
    ,
    \label{eq:WKBvalid}
\end{equation}

\noindent where the plasma beta is defined as
\begin{equation}
    \plasmaBETA = \frac{8\pi n T}{B_z^2}
    ,
    \label{eq:betaDEF}
\end{equation}

\noindent and the final expression in \eq{eq:WKBvalid} holds in the weakly magnetized limit $\Hall \ll 1$. If $k_{z,\text{max}} \lenT \gg 1$, the additional gradient terms in $\Symb{D}_A$ can be neglected and one recovers the growth rates of \citet{Bell20}. However, as shown in \Fig{fig:WKBvalid}, this condition is not satisfied for a weakly magnetized plasma. In this parameter regime, the fastest-growing modes will have wavelengths comparable to the equilibrium scale, so describing them requires the Wigner--Moyal formalism employed here. Also, as we shall show in \Sec{sec:global_stable}, it is only by retaining the additional gradient terms in the growth rate that one can obtain non-trivial temperature profiles that are stable to the collisional whistler instability.

Let us conclude this section with a brief discussion regarding the relevance of our analysis to current laser-plasma experiments. Consider the initial stages of an experiment such as that performed by \citet{Meinecke22}. In such an experiment, pressure balance is quickly established before self-generated magnetic fields have time to grow to appreciable strength (there are no imposed zeroth-order fields). Thus, we can view the initial phase of the experiment as residing within the upper-left corner of \Fig{fig:WKBvalid} (low $\Hall$ and high $\plasmaBETA$). As time progresses, dynamo action causes magnetic fields to grow while maintaining constant pressure, so the system evolves to a higher $\Hall$ state along the trajectory $\plasmaBETA \sim \Hall^{-2}$ (if temperature is not constant during this time, then the evolution of $\Hall$ is even faster, following the shallower trajectory $\plasmaBETA \sim T^5 \Hall^{-2}$). Along such a trajectory, $k_{z,\text{max}} L$ will be an increasing function since the contours go as $\plasmaBETA \sim \Hall^{-3}$~\eq{eq:WKBvalid}. There is a subtlety, however, in that the maximum growth rate for the collisional whistler instability is negative when $k_{z,\text{max}} L \lesssim 1$, as will be discussed later (see \Sec{sec:stable} and also \Fig{fig:drives}). This means that whistler waves will not be excited in the experiment until the plasma is magnetized enough so that $k_{z,\text{max}} L \sim 1$, at which point modes whose wavelengths are comparable with the gradient lengthscale will appear. Due to their early excitation, these modes will have the most time subsequently to manipulate the plasma evolution%
\footnote{This assumes the experiment progresses slowly enough for the time difference to be dynamically meaningful.}
(see \Sec{sec:quasilinear}), but due to their long wavelengths, they can only be accurately described by the Wigner--Moyal analysis performed here.


\section{Linear stability condition}
\label{sec:stable}

Let us consider the case when the whistler-intensity profile has some infinitesimally small (noise-level) initial value that is constant over space. Then, by integrating \eq{eq:polarIeq} over all space, one can readily see that the growth rate for whistlers with a given $k_z = \pd{z} \theta$ is governed by $\Symb{D}_A(k_z, z)$. Hence, the whistlers will be linearly unstable if the maximum growth rate is positive. Since \eq{eq:kzmax} implies that
\begin{equation}
    \Symb{D}_A\left[ k_{z,\text{max}}, z \right]
    =
    \frac{\growthV^2}{4\resist}
    - \frac{1}{2} \pd{z}\nerstV
    - \frac{1}{4} \pd{z}^2 \resist
    ,
    \label{eq:maxDaham}
\end{equation}

\noindent linear instability requires that
\begin{equation}
    \frac{\growthV^2}{\resist}
    \ge
    2 \pd{z}\nerstV
    + \pd{z}^2 \resist
    .
    \label{eq:linSTAB}
\end{equation}

\noindent Note that the left-hand side of \eq{eq:linSTAB} is always positive; therefore, if gradients were neglected in \eq{eq:Daham} under the assumption that $k_z \lenT \gg 1$, corresponding to setting the right-hand side of \eq{eq:linSTAB} to zero, one would erroneously conclude that any non-constant temperature profile would be unstable, \ie that $\pd{z} T = 0$ is the only stable profile.

\begin{figure}
    \centering
    \includegraphics[width=0.48\textwidth,trim={4mm 32mm 2mm 32mm},clip]{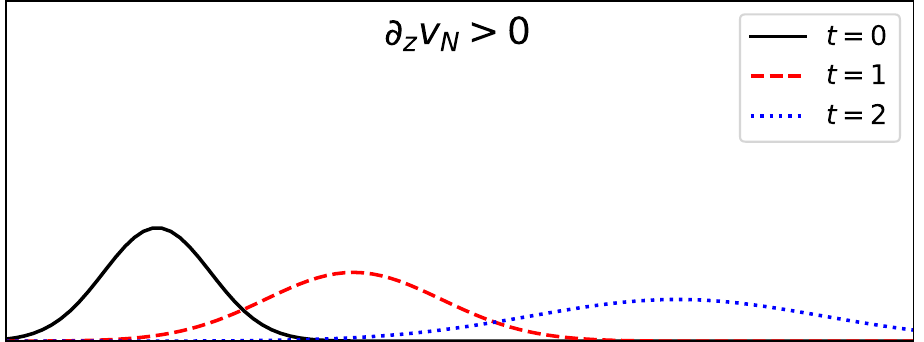}
    \hspace{2mm}\includegraphics[width=0.48\textwidth,trim={4mm 32mm 2mm 32mm},clip]{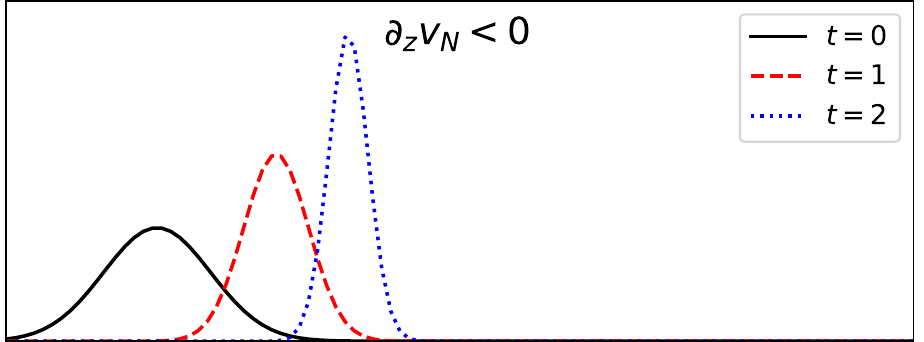}
    \caption{Evolution of a Gaussian pulse advected by an inhomogeneous velocity field $\nerstV(z) > 0$ (\ie directed towards the right). The pulse spreads when $\pd{z} \nerstV(z) > 0$, and compresses when $\pd{z} \nerstV(z) < 0$.}
    \label{fig:compress}
\end{figure}

Instead, we see that two gradient-driven stabilization mechanisms are present. The first term is the well-known compressional amplification that can result from a perturbation being advected by an inhomogeneous flow. As illustrated in \Fig{fig:compress}, if the Nernst advection velocity is a decreasing function of the propagation direction ($\pd{z} \nerstV < 0$) then the flow can pile up and amplify the initial perturbation, otherwise, when $\pd{z} \nerstV > 0$, an initial perturbation will be spread out and stabilized. In the specific context of Nernst advection, this is a well-known mechanism for amplifying magnetic fields near the ablation fronts of laser-compressed fuel pellets~\citep{Nishiguchi84,Nishiguchi85}. As seen in \Fig{fig:nernst}, $\pd{z} \nerstV < 0$ tends to occur when the plasma is weakly magnetized, \ie at small values of $\Hall$.

\begin{figure}
    \centering
    \includegraphics[width=0.7\textwidth,trim={4mm 4mm 4mm 2mm},clip]{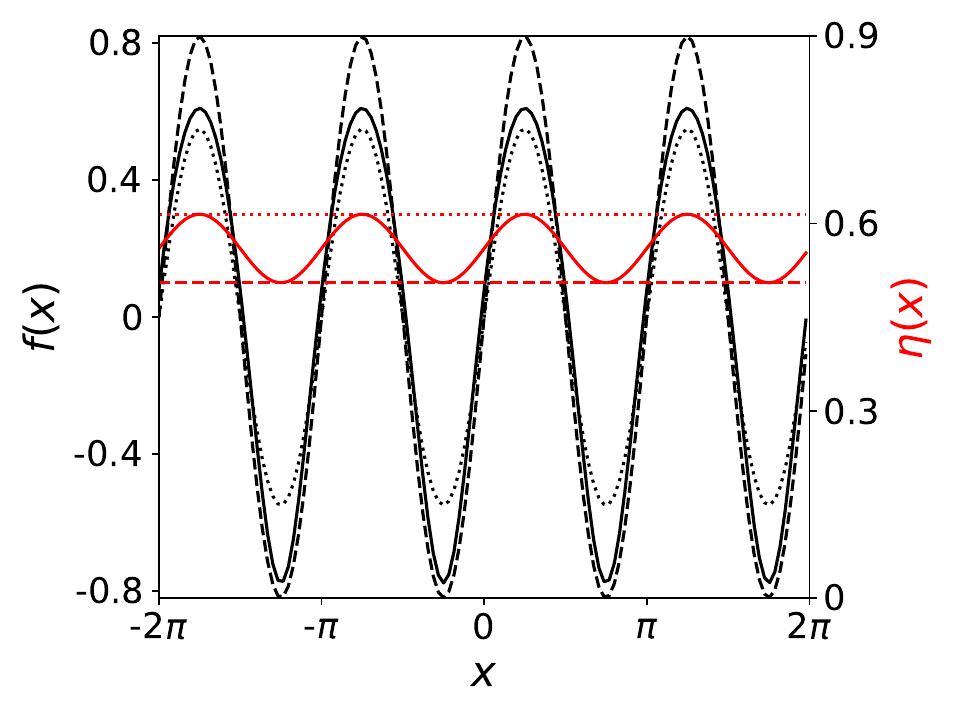}
    \caption{Diffusion of a sinusoidal perturbation $f(x) = \sin(2x)$ by either a sinusoidal diffusion coefficient $\resist(x) = [2 + \sin(2x)]/10$ (solid) or a constant diffusion coefficient given by the maximum (dotted) or minimum (dashed) value of $\resist(x)$.}
    \label{fig:diffuse}
\end{figure}

Less well-understood is the second stabilization term due to `resistivity curvature'. As shown in \Fig{fig:diffuse}, the diffusion of a perturbation is faster when $\pd{z}^2 \resist > 0$ than homogeneous theory would predict (thus increasing the stability of the system against the perturbation), and the diffusion is slower when $\pd{z}^2 \eta < 0$ (decreasing the stability of the system). Per \Fig{fig:resist}, one generally has $\pd{z}^2 \eta > 0$ in the vicinity of a hotspot, with $\pd{z}^2 \eta$ becoming increasingly larger as the magnetization level decreases.  

In the limit when the geometrical-optics approximation is only weakly violated, this effect can be understood as the result of using the wavelength-averaged resistivity in place of the resistivity when determining the damping of a wave. Indeed, if we define the effective resistivity as
\begin{equation}
    \resist_\text{eff}(z) = \frac{1}{2}
    \left[ 
        \resist\left(z - \frac{1}{2k} \right)
        + \resist\left(z + \frac{1}{2k} \right)
    \right]
    ,
\end{equation}

\noindent then in the limit that $k$ is still sufficiently large, a simple Taylor expansion yields
\begin{equation}
    \resist_\text{eff}(z) \approx
    \resist(z )
    + \frac{1}{4 k^2} \pd{z}^2 \resist(z)
    .
\end{equation}

\noindent Thus, including the resistivity-curvature correction in \eq{eq:Daham} is equivalent to using $k^2 \resist_\text{eff}$ as the dissipation term in growth rate of \citet{Bell20}.

\begin{figure}
    \centering
    \includegraphics[width=0.14\textwidth,trim={34mm 4mm 44mm 2mm},clip]{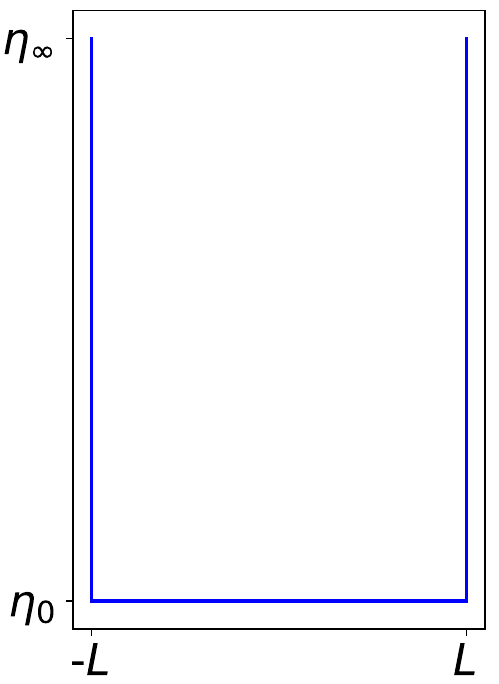}
    \includegraphics[width=0.27\textwidth,trim={0mm 0mm 0mm 0mm},clip]{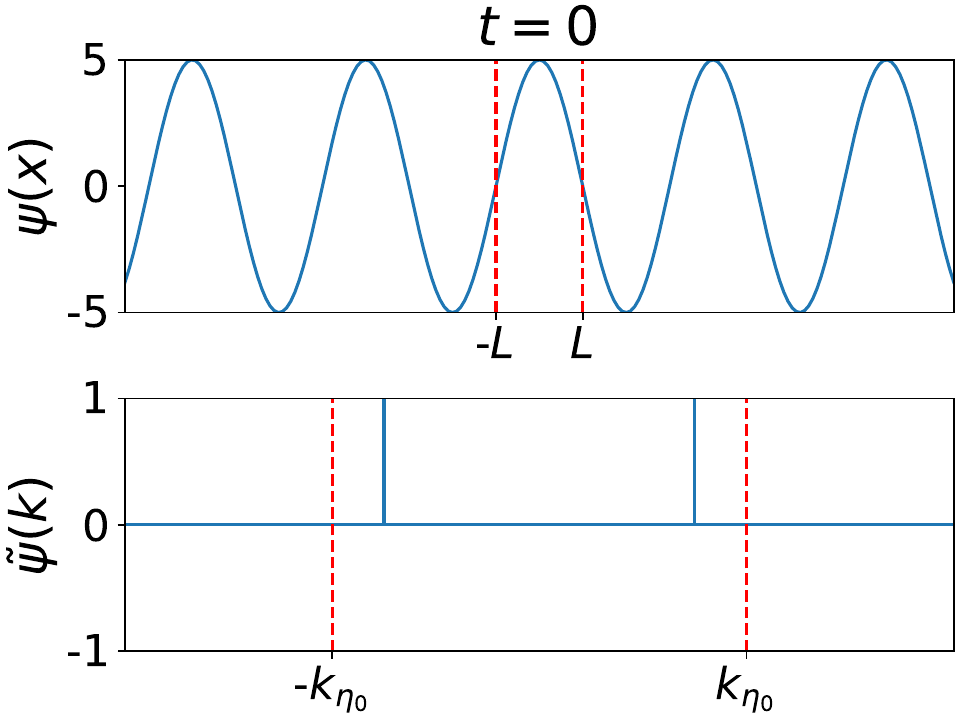}
    \includegraphics[width=0.27\textwidth,trim={0mm 0mm 0mm 0mm},clip]{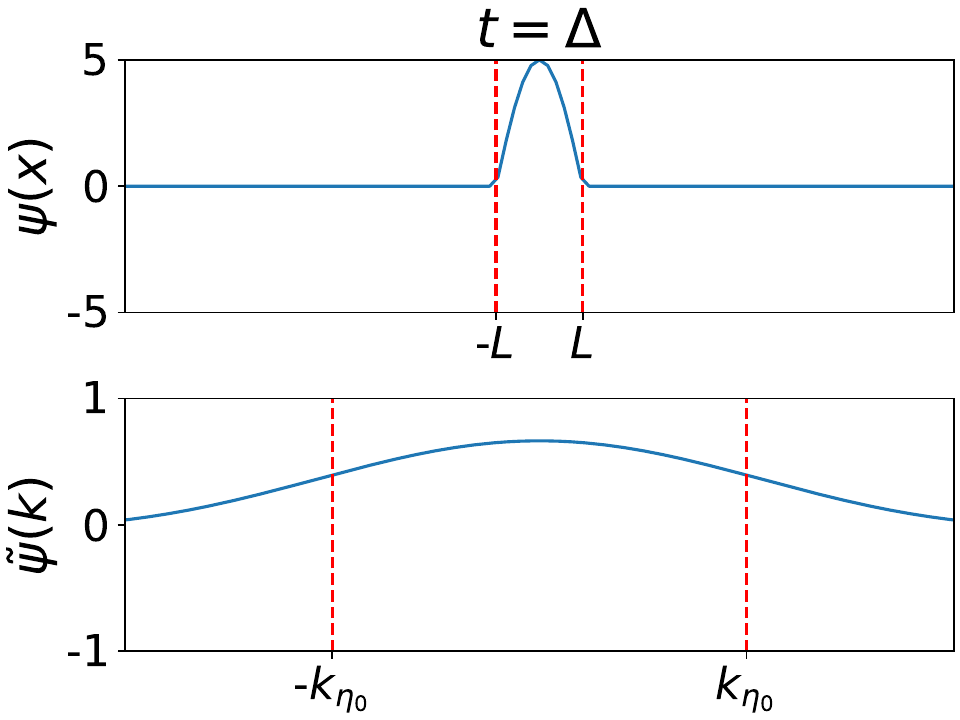}
    \includegraphics[width=0.27\textwidth,trim={0mm 0mm 0mm 0mm},clip]{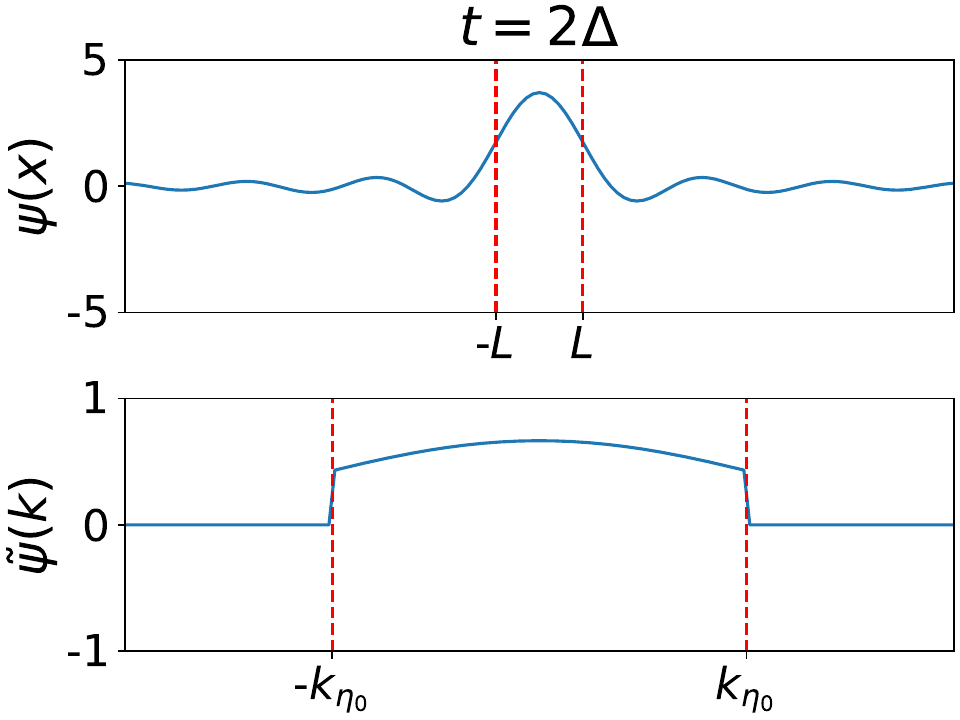}

    \vspace{2mm}
    \includegraphics[width=0.14\textwidth,trim={34mm 4mm 44mm 2mm},clip]{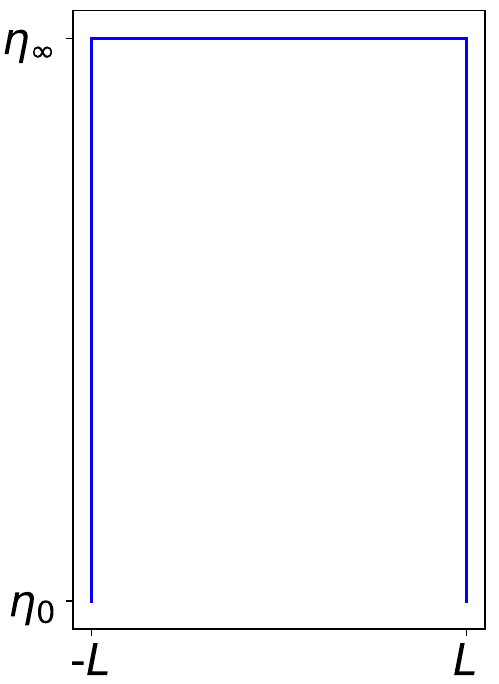}
    \includegraphics[width=0.27\textwidth,trim={0mm 0mm 0mm 0mm},clip]{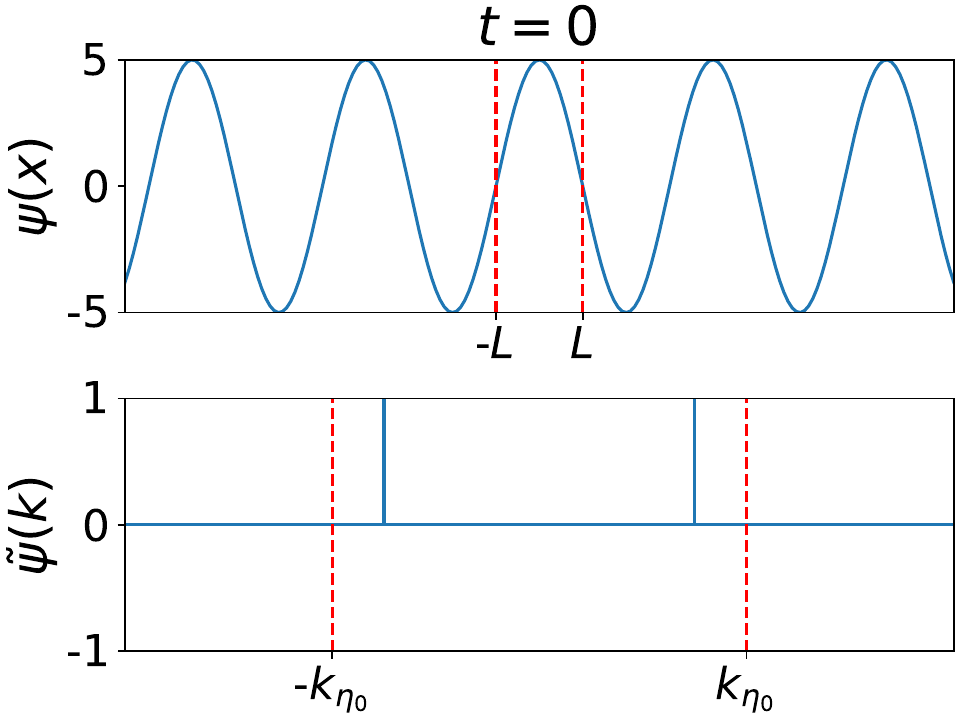}
    \includegraphics[width=0.27\textwidth,trim={0mm 0mm 0mm 0mm},clip]{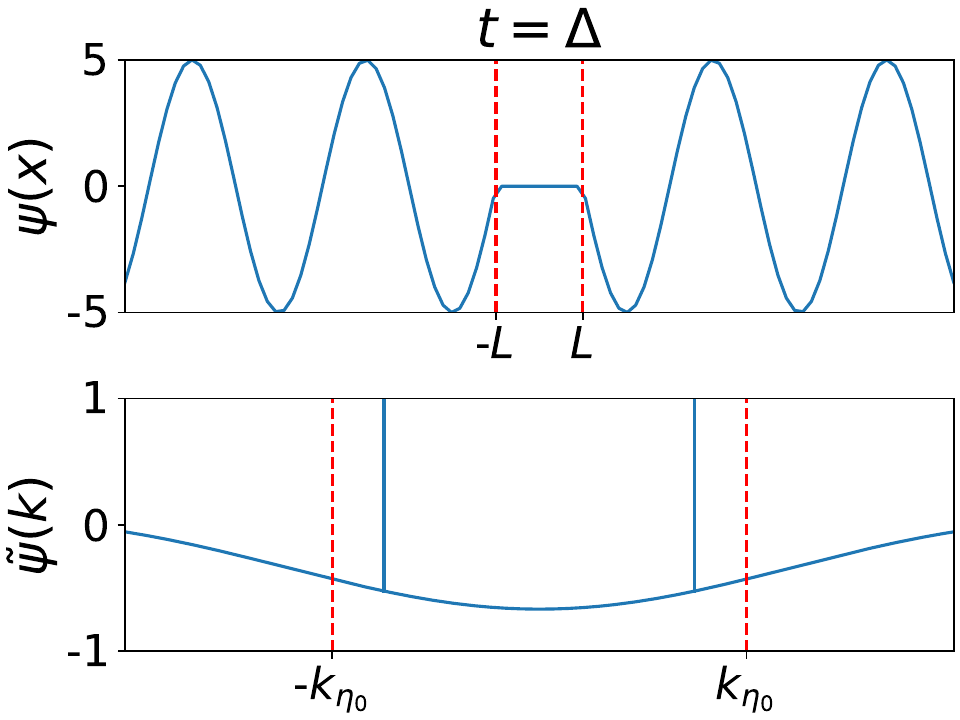}
    \includegraphics[width=0.27\textwidth,trim={0mm 0mm 0mm 0mm},clip]{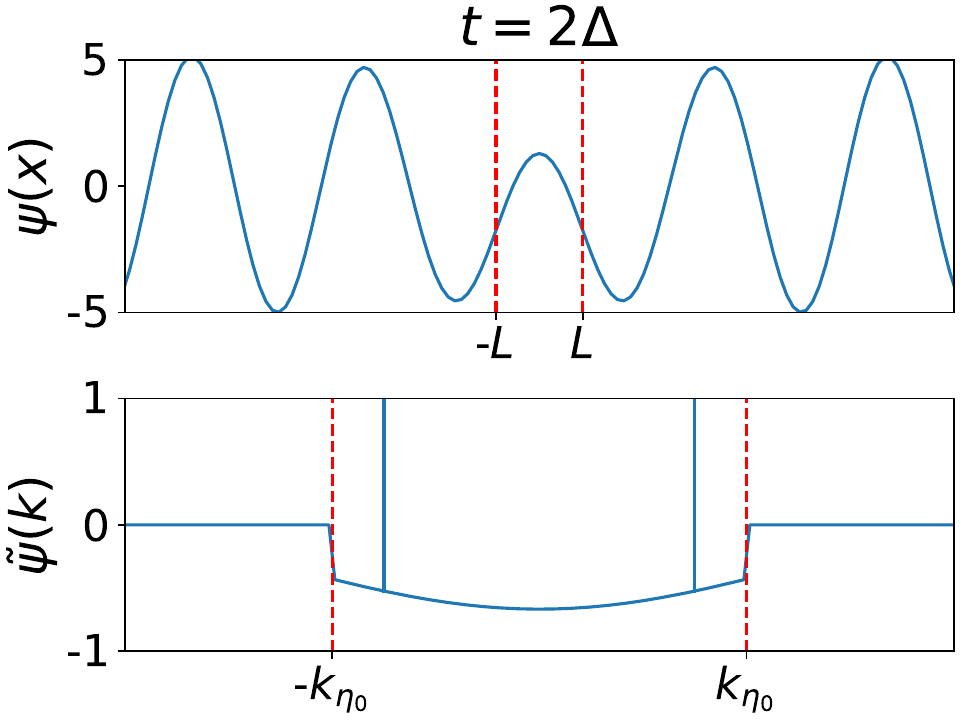}
    \caption{Evolution of a sinusoidal pulse subject to a discrete-time diffusion model (with step size $\Delta$) in which the diffusion acts as a low-pass filter with respect to wavevectors larger than $k_\resist \sim 1/\sqrt{\resist}$; accordingly, the spectral filter becomes a spatial filter where $\resist = \resist_\infty \to \infty$. }
    \label{fig:leakage}
\end{figure}

An alternative explanation for the stabilization by resistivity curvature can be formulated based on spectral leakage (\ie the uncertainty principle), as depicted in \Fig{fig:leakage}. This figure shows the evolution of an initially sinusoidal perturbation as a heuristic diffusion operator is repeatedly applied. This heuristic diffusion operator acts as a low-pass filter for wavevectors larger than the diffusion scale $k_\resist \sim 1/\sqrt{\resist}$; accordingly, if a bilevel diffusion coefficient is used where one value of $\resist$ is much larger than the other, then the heuristic diffusion operator acts as a combined spectral and spatial filter with respect to the diffusive scale of the smaller value $k_{\resist_0}$ and the spatial domain of the larger value. Spectral leakage then enables the entire perturbation to decay away eventually, but at different rates when $\resist$ is a local minimum compared to a local maximum. Indeed, as seen by comparing the central peak at $t = \Delta$ and $t = 2\Delta$ in the figure, diffusion is increased when $\resist$ is concave up and diffusion is decreased when $\resist$ is concave down compared to the nominal diffusion one would expect if only the local value of $\resist$ was considered.

To illustrate the impact of these additional stabilization mechanisms, let us consider a plasma with a linear temperature profile $T \sim z/\lenT$ and in pressure balance, so that $n \sim 1/T$. Then it can be shown [see \eq{eq:peakGROW}] that \eq{eq:linSTAB} becomes
\begin{equation}
    \frac{ \plasmaBETA \Delta_\beta^2(\Hall) }{\alpha_\perp(\Hall) }
    \ge
    \frac{15 \alpha_\perp(\Hall) }{\plasmaBETA \Hall^2}
    - \frac{15 \alpha_\perp'(\Hall) }{\plasmaBETA \Hall}
    + \frac{25 \alpha''_\perp(\Hall) }{\plasmaBETA}
    - 10 \beta_\wedge'(\Hall)
    ,
    \label{eq:linTstable}
\end{equation}

\noindent where as a reminder, $\Hall$ is defined in \eq{eq:Mdef} with $\epsilon = 0$. Note that \eq{eq:linTstable} is actually independent of the temperature gradient. Hence, when the right-hand side is sufficiently positive there will be no unstable temperature gradients. It is clear that this will happen for weakly magnetized plasmas (small $\Hall$) or for low-$\plasmaBETA$ plasmas due to the divergent denominator%
\footnote{The plot of \eq{eq:linTstable} as a function of $\Hall$ and $\plasmaBETA$ is nearly identical to the plot of $\TpsCOEF$ in \Fig{fig:drives} for reasons that will be discussed in \Sec{sec:dynamREL}, with the unstable region colored in red and the stable region colored in blue.}; %
this weakly-magnetized regime will be discussed further in Secs.~\ref{sec:global_stable} and \ref{sec:quasilinear}.

At fixed $\plasmaBETA$, the same divergent denominator that ensures stability at low $\Hall$ causes the system to become unstable at high $\Hall$. This implies the existence of a critical value $\Hall_\text{crit}$ across which the transition from stable to unstable behavior occurs. Hence, if one were to set up a simulation similar to \citet{Komarov18}, in which a linear temperature gradient is initialized across a plasma of length $L$ with $T(0)$ held fixed and $T(L)$ allowed to vary between simulations, one would see whistler waves beginning to grow once $T(L)$ exceeds a critical value corresponding to $\Hall_\text{crit}$. It might be tempting to conclude that the destabilization is due to the temperature gradient exceeding a critical value, but subsequent simulations with increased box size $L$ and the same temperature difference should feature whistler waves continuing to be excited despite the temperature gradient being reduced.


\section{Dynamical relevance of collisional whistler instability}
\label{sec:dynamREL}

The collisional whistler instability will grow on time-scales determined by the maximum growth rate, denoted $\gamma_\text{whist}$. This is given by \eq{eq:maxDaham}, which can be re-written as
\begin{align}
    \gamma_\text{whist}
    =
    \frac{3}{4}
    \left[
        \TpsCOEF
        + \TppCOEF \frac{\lenT^2}{\curveT}
        + \TNpCOEF \frac{ \lenT }{ \lenN }
        + \NpsCOEF \left( \frac{ \lenT }{ \lenN } \right)^2
        + \NppCOEF \frac{\lenT^2 }{ \curveN}
    \right]
    \frac{\cycloF \larmorR^2}{\lenT^2}
    ,
    \label{eq:peakGROW}
\end{align}

\noindent where we have introduced the temperature lengthscale $\lenT$, the `temperature curvature' $\curveT$, the density lengthscale $\lenN$, and the `density curvature' $\curveN$, as follows:
\begin{align}
    \lenT
    &= \frac{T}{\pd{z} T}
    , \quad
    \curveT
    = \frac{T}{\pd{z}^2 T}
    , \quad
    \lenN
    = \frac{n}{\pd{z} n}
    , \quad
    \curveN
    = \frac{n}{\pd{z}^2 n}
    ,
    \label{eq:lengths}
\end{align}

\noindent and we have also introduced the auxiliary functions
\begin{subequations}
    \label{eq:auxFfuncs}
    \begin{align}
        \label{eq:FT1}
        \TpsCOEF
        &= 
        \frac{\plasmaBETA \Hall \Delta_\beta^2 }{6 \alpha_\perp} 
        + \plasmaBETA \, \pd{\plasmaBETA} \TppCOEF
        + \frac{3}{2} \Hall \, \pd{\Hall} \TppCOEF
        , \\
        \TppCOEF
        &=
        \frac{2 \beta_\wedge}{3 } 
        + \frac{\alpha_\perp - \Hall \alpha'_\perp}{\plasmaBETA \Hall}
        , \\
        \TNpCOEF
        &=
        \plasmaBETA \, \pd{\plasmaBETA} \TppCOEF
        - \Hall \, \pd{\Hall} \TppCOEF
        + \frac{3}{2} \Hall \, \pd{\Hall} \NppCOEF
        , \\
        \NpsCOEF
        &= 
        - \Hall \, \pd{\Hall} \NppCOEF
        - 2 \NppCOEF
        , \\
        \NppCOEF
        &= \frac{2 \alpha'_\perp}{3 \plasmaBETA }
        .
    \end{align}
\end{subequations}

\noindent Note that here and in what follows, we use $'$ to denote $\pd{\Hall}$ for univariate functions of $\Hall$. Also note that all the length scales in \eq{eq:lengths} are signed quantities.

\begin{figure}
    \centering
    \includegraphics[width = 0.48\linewidth, trim = {20mm 6mm 6mm 2mm}, clip]{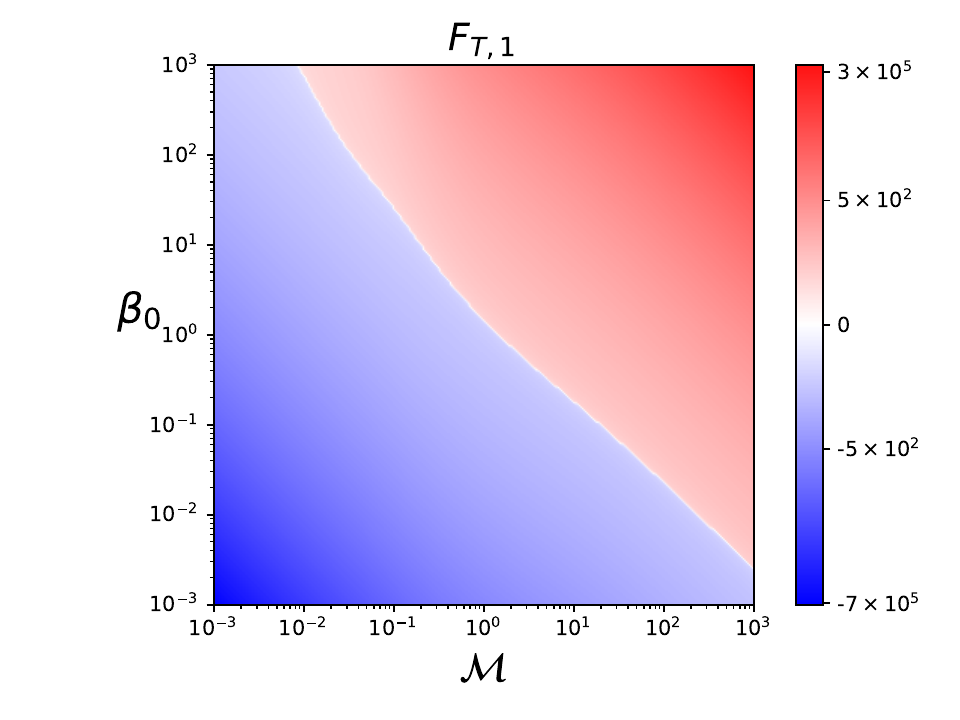}
    \includegraphics[width = 0.48\linewidth, trim = {20mm 6mm 6mm 2mm}, clip]{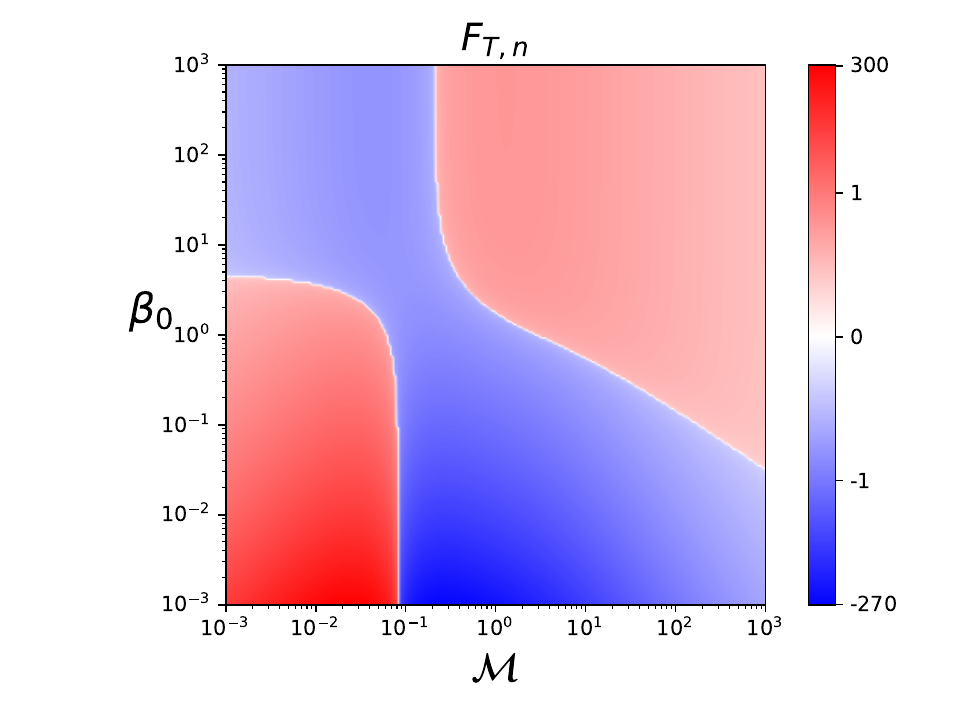}
    
    \includegraphics[width = 0.48\linewidth, trim = {20mm 6mm 6mm 2mm}, clip]{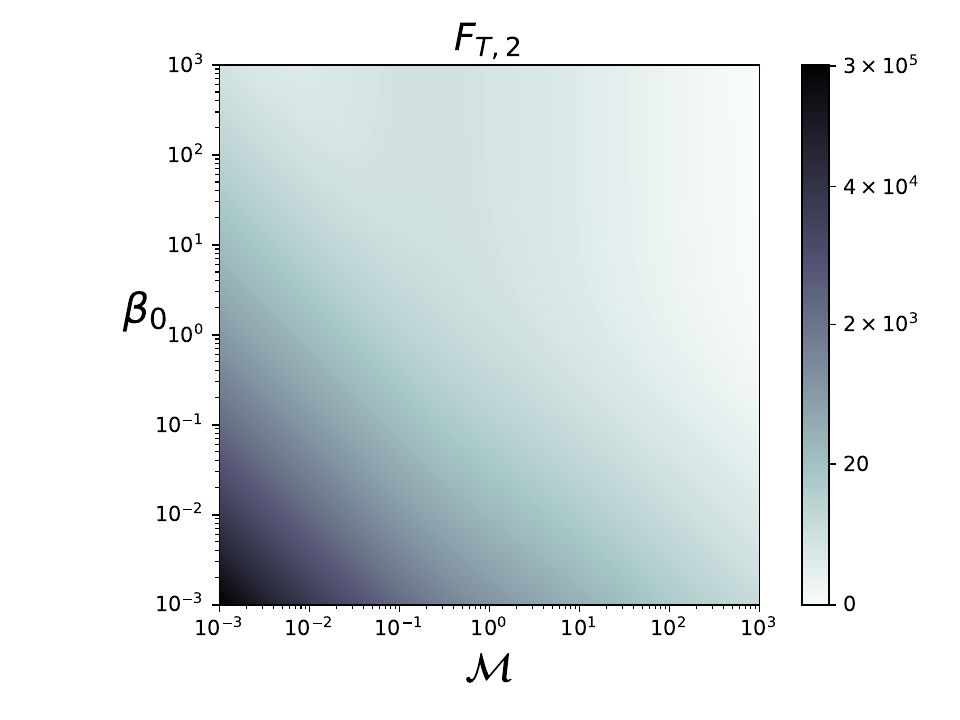}
    \includegraphics[width = 0.48\linewidth, trim = {20mm 6mm 6mm 2mm}, clip]{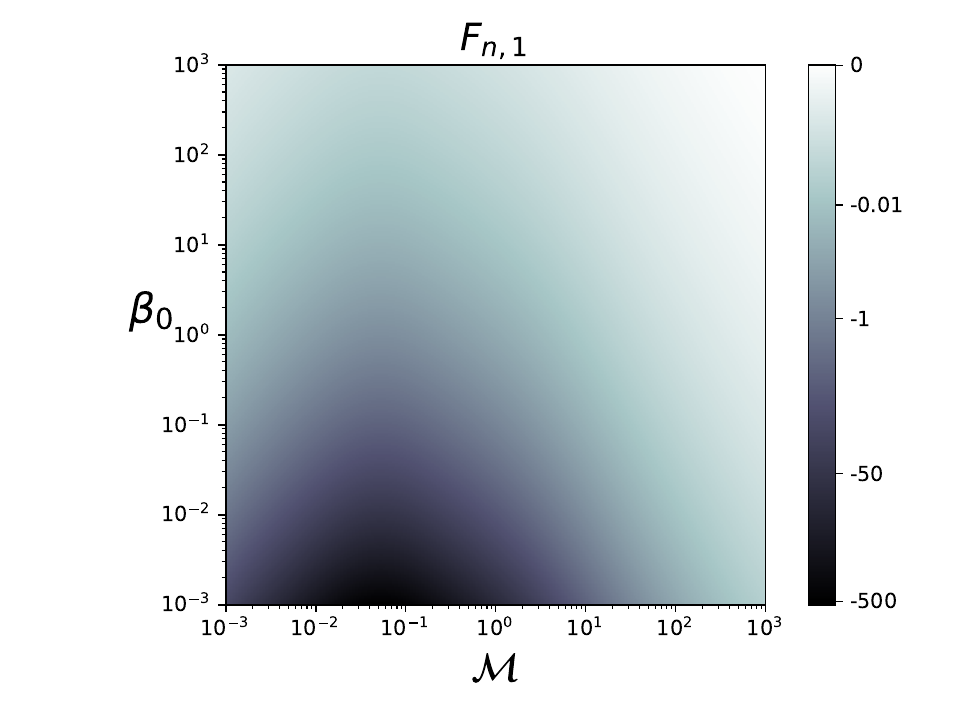}

    \includegraphics[width = 0.48\linewidth, trim = {20mm 6mm 6mm 2mm}, clip]{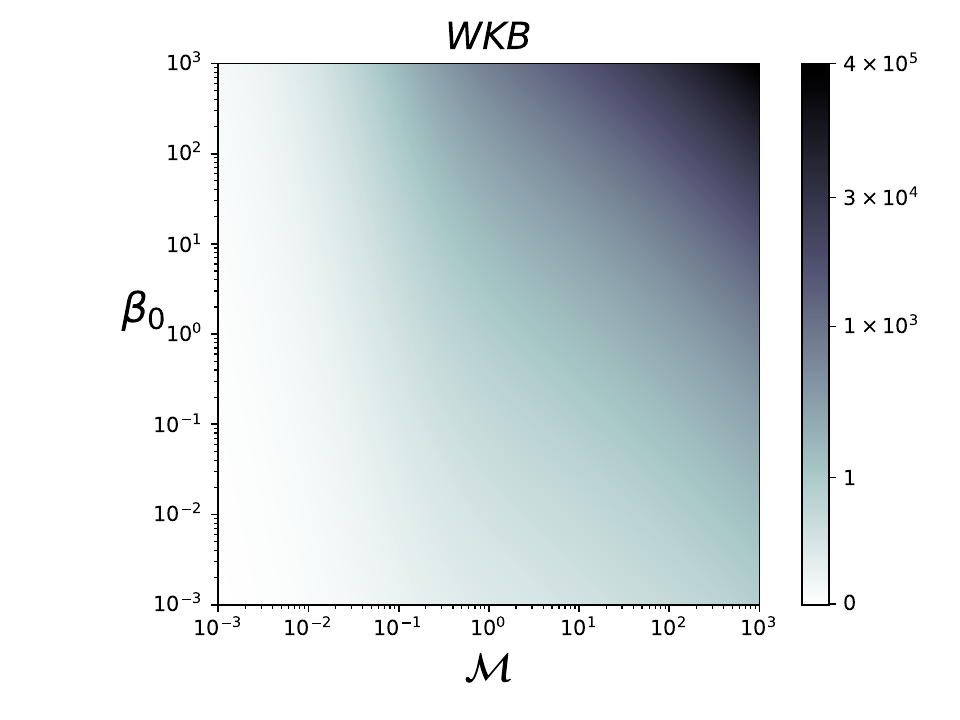}
    \includegraphics[width = 0.48\linewidth, trim = {20mm 6mm 6mm 2mm}, clip]{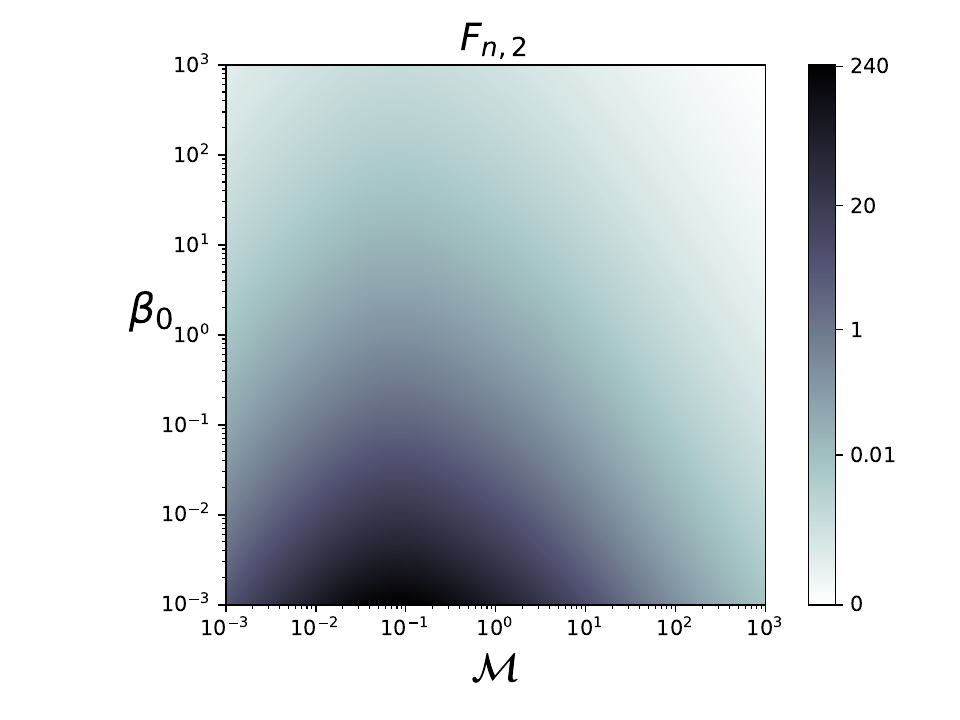}

    \caption{Plots of the various drive terms for the collisional whistler instability, as defined in~\eq{eq:peakGROW}. Here `WKB' refers to the only drive term that survives the short-wavelength approximation: see \eq{eq:wkbGROW}. Importantly, note that the color-bar axis can differ by orders of magnitude between plots.}
    \label{fig:drives}
\end{figure}

Figure \ref{fig:drives} shows the various drive terms as functions of $\Hall$ and $\plasmaBETA$. First, one notes that terms corresponding to the density-gradient drives ($\TNpCOEF$, $\NpsCOEF$, and $\NppCOEF$) are generally smaller (by about two orders of magnitude) than the terms corresponding to temperature-gradient drives ($\TpsCOEF$, and $\TppCOEF$). The temperature-gradient-drive terms are larger because they either contain a factor $\plasmaBETA \Hall$ that grows unbounded towards the upper right corner of parameter space, or a divergent factor $1/\plasmaBETA \Hall$ that grows unbounded towards the lower left corner. The former corresponds to the `WKB' term $\plasmaBETA \Hall \Delta_\beta^2/6 \alpha_\perp$ in \eq{eq:peakGROW}, which is proportional to the maximum growth rate when no additional gradient terms are included in \eq{eq:Daham} [see \eq{eq:wkbGROW}], while the latter is associated with the resistivity curvature. Thus, we expect that the instability dynamics will be largely independent of the density profile when a non-uniform temperature profile is present. 

For the collisional whistler instability to be dynamically relevant, the whistler waves must grow faster than the time that it takes for the driving temperature inhomogeneity to diffuse away. The diffusion time $\diffuseT = (\pd{t} \log T)^{-1}$ can be calculated from \eq{eq:qzDEF}--\eq{eq:Tnonlin} with $\epsilon = 0$:
\begin{equation}
    \diffuseT
    = 
    \frac{\lenT^2}{ \cycloF \larmorR^2}
    \frac{
        3
    }{
        \Hall \kappa_\parallel
        \left| 
            5 
            + 2 \shape
        \right| 
    } 
    ,
    \label{eq:diffuseT}
\end{equation}

\noindent where $\shape = \lenT^2/\curveT$ describes the shape of the temperature profile. Hence, one has
\begin{equation}
    \gamma_\text{whist} \diffuseT
    = 
    \frac{9}{4}
    \frac{
        \TpsCOEF
        + \TppCOEF \shape
        + \TNpCOEF \lenT \lenN^{-1}
        + \NpsCOEF \left( \lenT \lenN^{-1} \right)^2
        + \NppCOEF \lenT^2 \curveN^{-1}
    }{
        \Hall \kappa_\parallel
        \left| 
            5 
            + 2 \shape
        \right| 
    } 
    ,
    \label{eq:whistREL}
\end{equation}

\noindent with dynamical relevance requiring $\gamma_\text{whist} \diffuseT > 1$. Note that when $\shape = - 5/2$ the diffusion time becomes infinite ($T \propto z^{2/7}$ yields a spatially constant heat flux), so the whistlers will be dynamically relevant anywhere there is a positive growth rate.

For simplicity, let us restrict attention to when the density profile is either isobaric or constant:
\begin{equation}
    \lenN^{-1}
    = 
    \begin{cases}
        - \lenT^{-1} & \text{isobaric} \\
        0 & \text{constant}
    \end{cases}
    , \quad
    \curveN^{-1}
    = 
    \begin{cases}
        2 \lenT^{-2} - \curveT^{-1} & \text{isobaric} \\
        0 & \text{constant}
    \end{cases}
    .
\end{equation}

\noindent Hence, one has
\begin{equation}
    \gamma_\text{whist} \diffuseT
    =
    \frac{9}{4 \Hall \kappa_\parallel
        \left| 
            5
            + 2 \shape
        \right| }
    \times
    \begin{cases}
        \TpsCOEF
        - \TNpCOEF
        + \NpsCOEF
        + 2 \NppCOEF
        + (\TppCOEF - \NppCOEF) \shape
        & \text{isobaric} \\
        \TpsCOEF
        + \TppCOEF \shape
        & \text{constant}
    \end{cases}
    .
\end{equation}

\begin{figure}
    \centering
    \begin{overpic}[width=0.49\linewidth,trim={18mm 6mm 32mm 4mm}, clip]{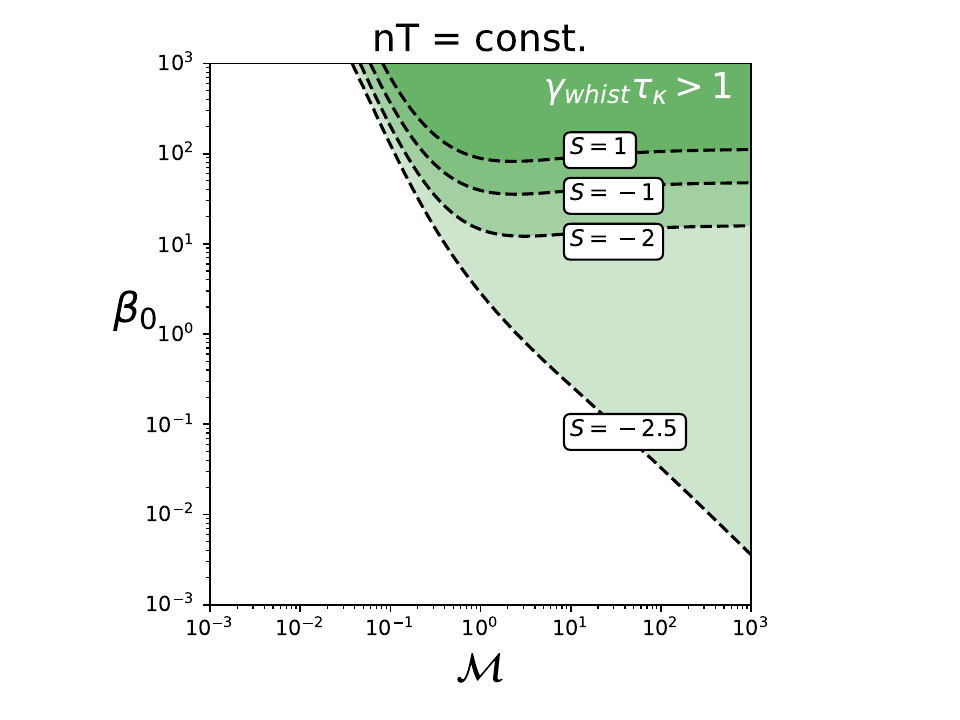}
        \put(17,88){(a)}
    \end{overpic}
    \begin{overpic}[width=0.49\linewidth,trim={18mm 6mm 32mm 4mm}, clip]{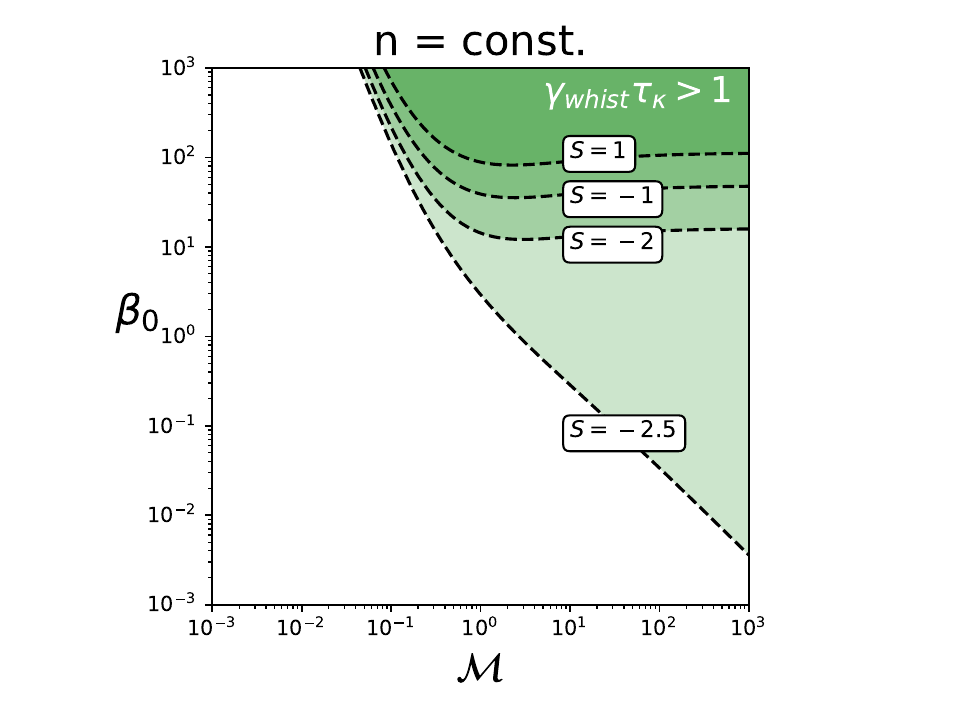}
        \put(17,88){(b)}
    \end{overpic}

    \begin{overpic}[width=0.49\linewidth,trim={18mm 6mm 32mm 4mm}, clip]{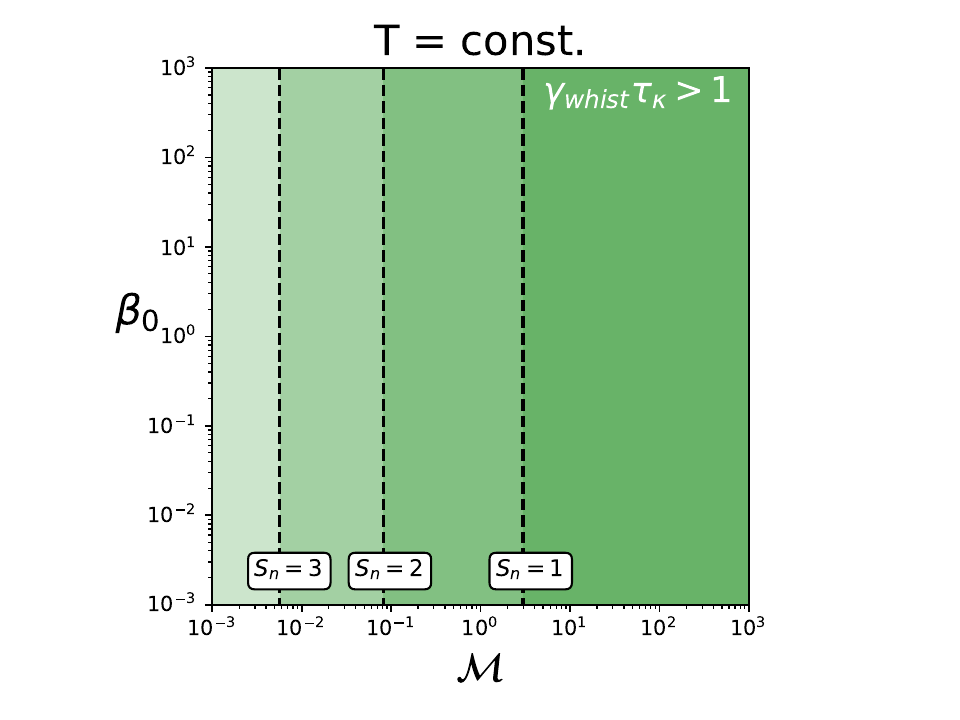}
        \put(17,88){(c)}
    \end{overpic}
    \begin{overpic}[width=0.49\linewidth,trim={18mm 6mm 32mm 4mm}, clip]{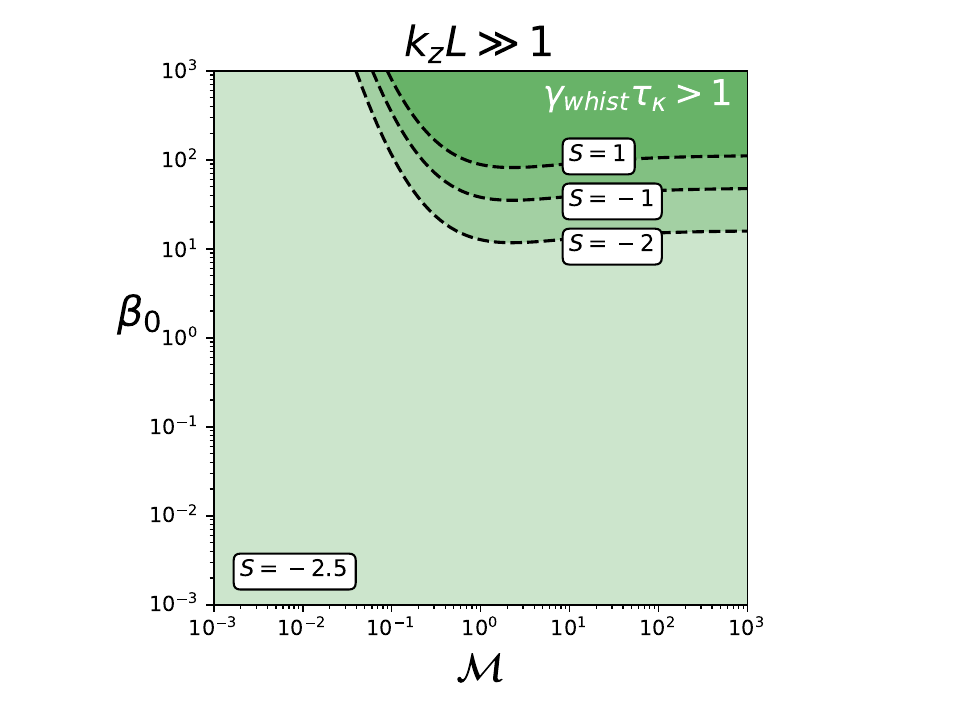}
        \put(17,88){(d)}
    \end{overpic}
    \caption{\textbf{(a)-(b)} Parameter space where the collisional whistler instability is dynamically relevant (green) for the specified density profile, as determined by comparing the peak growth rate $\gamma_\text{whist}$ with the diffusion time $\diffuseT$ associated with standard parallel conduction [see \eq{eq:whistREL}]. The boundary of this region depends on the shape factor $S \doteq T T''/(T')^2$, and is roughly symmetric about $S = -2.5$ (\eg the boundaries for $S = -3$ and $S = -2$ are approximately the same). \textbf{(c)} Same, but when the temperature profile is constant and the instability is instead driven by a density inhomogeneity with shape factor $S_n$ defined analogously to $S$. \textbf{(d)} Same, but for the growth rate provided in \eq{eq:wkbGROW}, which is valid in the short-wavelength limit and is independent of the density profile.}
    \label{fig:whistleREL}
\end{figure}

\noindent The regions of parameter space where whistlers are dynamically relevant for isobaric or constant density profiles are shown in \Fig{fig:whistleREL}(a,b). 

First of all, there is no visible difference between the results for an isobaric versus a constant density profile. This is because the density-drive terms in \eq{eq:peakGROW}, $\NpsCOEF$ and $\NppCOEF$, are generally smaller than the principal temperature-drive term $\TpsCOEF$ (\Fig{fig:drives}). Secondly, not all of the parameter space is susceptible to whistlers even when the initial profile is diffusion-free ($\shape = -2.5$). This is because the instability actually disappears for sufficiently low $\Hall$ and all whistler waves are instead strongly damped. This is in stark contrast to the prediction made with the short-wavelength asymptotic growth rate obtained in \citet{Bell20}:
\begin{equation}
    \gamma_{\text{wkb}}
    =
    \frac{\plasmaBETA \Hall \Delta_\beta^2 }{8 \alpha_\perp}
    \frac{\cycloF \larmorR^2}{\lenT^2}
    .
    \label{eq:wkbGROW}
\end{equation}

\noindent The region of the instability's dynamical relevance in this limit, which is shown in \Fig{fig:whistleREL}(d), is determined by the quantity
\begin{equation}
    \gamma_{\text{wkb}}\diffuseT
    = 
    \frac{3 \plasmaBETA\Delta_\beta^2 }{8 \kappa_\parallel \alpha_\perp}
    \frac{
        1
    }{
        \left| 
            5 
            + 2 \shape
        \right| 
    } 
    .
    \label{eq:WKBrel}
\end{equation}

\noindent Since $\gamma_{\text{wkb}} \ge 0$, this approximation does not capture the strong damping that occurs at low magnetization, instead predicting that all of the parameter space is susceptible to the collisional whistler instability.

Finally, one should note that there actually exists an instability even when the temperature is constant, driven instead by a density gradient. Indeed, setting $\pd{z} T$ and $\pd{z}^2 T$ both equal to zero in \eq{eq:peakGROW} yields
\begin{equation}
    \gamma_\text{whist}
    =
    \frac{3}{4}
    \left(
        \NpsCOEF
        + \NppCOEF \frac{\lenN^2 }{ \curveN}
    \right)
    \frac{\cycloF \larmorR^2}{\lenN^2}
    .
    \label{eq:densGROWTH}
\end{equation}

\noindent Furthermore, since $T$ is constant, there is no diffusion so $\diffuseT$ is infinite; the density-gradient-driven collisional whistler instability will be dynamically relevant whenever $\gamma_\text{whist} \ge 0$, with $\gamma_\text{whist}$ given now by \eq{eq:densGROWTH}. This region is shown in \Fig{fig:whistleREL}(c). Since $\NppCOEF > 0$ while $\NpsCOEF$ has no definite sign, the region of dynamical relevance increases as $\lenN^2/ \curveN$ becomes increasingly positive. Eventually, as $\lenN^2/ \curveN \to \infty$, the entire parameter space is susceptible to the density-gradient-driven instability. 

That said, we should again emphasize that the density-gradient-driven instability has much smaller growth rates than the temperature-gradient-driven instability. Also, by using \Fig{fig:drives} to estimate $\TpsCOEF \sim 10^5$ and $\NpsCOEF \sim 10^2$, \eq{eq:peakGROW} shows that only when $\lenN^{-1} \gtrsim 30 \lenT^{-1}$ will the density-gradient drives be important for the collisional whistler instability. We shall defer the study of such isothermal plasmas to future work, and instead consider either isobaric or constant-density plasmas to facilitate comparisons with \citet{Meinecke22} or \citet{Bell20}, respectively. In these plasmas, the density-gradient drives play a negligibly small role.


\section{Global marginally stable temperature profiles}
\label{sec:global_stable}

It is interesting to consider what plasma profiles are marginally stable over an arbitrarily large spatial domain, since these can potentially correspond to the final states obtained after the collisional whistler instability has saturated quasilinearly. We shall obtain these global marginally stable states by considering the condition
\begin{equation}
    \gamma_\text{whist}
    = 0
    ,
    \label{eq:marginalCOND}
\end{equation}

\noindent with $\gamma_\text{whist}$ given by \eq{eq:peakGROW}, as a differential equation governing $T(z)$ for a prescribed $n(z)$, since $n$ does not evolve in time. Note that non-trivial (\ie inhomogeneous) marginally stable profiles are only possible when the gradient terms are included in $\Symb{D}_A$; if instead one were to consider marginally stable states with respect to $\gamma_\text{wkb}$ given by \eq{eq:wkbGROW}, the answer would be simply a uniform temperature profile, regardless of the density profile.

We shall first discuss the general case before considering two special cases in detail. The first special case has the plasma density constrained by pressure balance, as occurs in astrophysical and recent experimental contexts~\citep{Markevitch07,Meinecke22}. The second special case will be the simpler situation in which the density is constant, corresponding to the analysis performed in \citet{Bell20}.

\subsection{General theory}

Let us consider monotonic profiles such that $\pd{z} \Hall \neq 0$ everywhere, $\Hall$ having been defined in \eq{eq:Mdef} but with $\epsilon = 0$. Then one can formally parameterize the inverse function $z(\Hall)$ so that all functions can be considered functions of $\Hall$. Suppose further that
\begin{equation}
    \pd{\Hall} n[z(\Hall)] \neq -\frac{n}{\Hall}
    .
\end{equation}

\noindent Then one has $\pd{\Hall} T \neq 0$ everywhere, so the composite map $z[\Hall(T)]$ can be formally constructed and all functions can be parameterized by $T$ instead of $z$. One then has
\begin{equation}
    \pd{z} n
    = n'
    (\pd{T} \Hall) 
    \pd{z} T
    , \quad
    \pd{z}^2 n
    = n'' ( \pd{T} \Hall)^2 (\pd{z} T)^2
    + n' (\pd{T}^2 \Hall) (\pd{z} T)^2
    + n' (\pd{T} \Hall) \pd{z}^2 T
    ,
\end{equation}

\noindent where $'$ again denotes $\pd{\Hall}$. Then, \eq{eq:marginalCOND} can be recast in the form
\begin{align}
    \gTpsCOEF(\Hall) (\pd{z} T)^2
    + \gTppCOEF(\Hall) \pd{z}^2 T
    = 0
    ,
    \label{eq:marginalTeqNOAPPROX}
\end{align}

\noindent where the two new auxiliary functions are defined as follows:
\begin{subequations}
    \label{eq:auxG12}
    \begin{align}
        \gTpsCOEF(\Hall)
        &=
        \frac{2\pi n(\Hall) }{B_z^2}
        \Hall
        \frac{
            [\Delta_\beta(\Hall)]^2
        }{
            \alpha_\perp(\Hall)
        }
        + \gTppCOEF'(\Hall) \pd{T} \Hall
        , \\
        \gTppCOEF(\Hall)
        &=
        \beta_\wedge(\Hall) 
        + \frac{B_z^2}{8\pi}
        \left\{
            \alpha_\perp(\Hall) 
            \left[ 
                1
                + \Hall \frac{n'(\Hall)}{n(\Hall)}
            \right]
            - \Hall \alpha_\perp'(\Hall)
        \right\}
        \frac{\pd{T} \Hall}{n(\Hall) \Hall^2}
        .
    \end{align}
\end{subequations}

Using the chain rule,
\begin{equation}
    \pd{z} T
    = (\pd{\Hall} T) \pd{z} \Hall
    , \quad
    \pd{z}^2 T
    = (\pd{\Hall}^2 T) (\pd{z} \Hall)^2
    + (\pd{\Hall} T) \pd{z}^2 \Hall
    ,
\end{equation}

\noindent along with the standard relations between the derivatives of inverse functions, viz.,
\begin{equation}
    \pd{x} y
    = \frac{1}{\pd{y} x}
    , \quad
    \pd{x}^2 y
    =
    - \frac{\pd{y}^2 x}{(\pd{y}x)^3}
    ,
    \label{eq:inverseDER}
\end{equation}

\noindent we deduce that \eq{eq:marginalTeqNOAPPROX} can be written as a differential equation for $\Hall$:
\begin{equation}
    \pd{z}^2 \Hall
    = 
    \gFUNC(\Hall)
    (\pd{z} \Hall)^2
    ,
    \label{eq:marginalMeqNOAPPROX}
\end{equation}

\noindent where we have defined
\begin{equation}
    \gFUNC(\Hall)
    =
    \frac{
        \gTppCOEF(\Hall) \pd{T}^2 \Hall
        - \gTpsCOEF(\Hall) \pd{T} \Hall
    }{\gTppCOEF(\Hall) (\pd{T} \Hall)^2}
    .
    \label{eq:auxG}
\end{equation}

Clearly, \eq{eq:marginalMeqNOAPPROX} can be trivially satisfied by $\Hall$ = constant. To obtain a nontrivial solution, note that \eq{eq:marginalMeqNOAPPROX} possesses affine symmetry with respect to $z$, \ie $z \mapsto c_1 + c_2 z$; hence any solution will have the general form $\Hall(c_1 + c_2 z)$ with $c_1$ and $c_2$ being the two integration constants. The fact that the two integration constants enter in this manner suggests that it will be simpler to solve for the inverse function $z(\Hall)$, since one expects $\log \pd{\Hall} z$ to satisfy an equation of the form $\pd{\Hall} \log \pd{\Hall} z = f(\Hall)$. Indeed, by making use again of \eq{eq:inverseDER}, the \textit{nonlinear} differential equation \eq{eq:marginalMeqNOAPPROX} is recast as a \textit{linear} differential equation
\begin{equation}
    z''
    = 
    - \gFUNC(\Hall)
    z'
    .
    \label{eq:marginalZeqNOAPPROX}
\end{equation}

\noindent As this is now a first-order differential equation with respect to $z'$, the solution to \eq{eq:marginalZeqNOAPPROX} can be obtained directly via two successive integrations as
\begin{equation}
    z(\Hall) 
    = 
    z(\Hall_1)
    + z'(\Hall_2)
    \int_{\Hall_1}^\Hall
    \dd \mu \,
    \exp\left[
        -
        \int_{\Hall_2}^\mu
        \dd m \, \gFUNC(m)
    \right]
    ,
    \label{eq:marginalZ}
\end{equation}

\noindent where $\Hall_1$ and $\Hall_2$ are arbitrary values at which boundary conditions can be applied. One notes that \eq{eq:marginalZ} manifestly respects the affine symmetry of the original equation. For \eq{eq:marginalZ} to be physically relevant, though, it must be the case that $\Hall(z) \ge 0$; a sufficient condition to ensure positivity is that $\gFUNC \sim A/\Hall$ with $A > 1$ as $\Hall \to 0$, as shown in \App{app:positivity}. Realistic friction coefficients do indeed have this property, as we shall see in \Sec{sec:isoCHAP} and \Sec{sec:consCHAP}.

\subsection{Magnetization staircases as a general class of solutions}

Let us now consider a high-$\plasmaBETA$ plasma. Although not obvious from \eq{eq:marginalZ}, in this limit, $\Hall(z)$, and thus $T(z)$, naturally forms a staircase structure. To see this more easily, note that $\gTpsCOEF \sim O(\delta^{-1})$ and $\gTppCOEF \sim O(1)$ with respect to the small parameter $\delta \sim 1/\plasmaBETA$ [see \eq{eq:auxG12}, or more simply, see that $\TpsCOEF$ is significantly larger than any other coefficient in \Fig{fig:drives} when $\plasmaBETA$ is large]. Hence, \eq{eq:marginalMeqNOAPPROX} has the general abstract form
\begin{equation}
    \delta \, y''(z)
    =
    \gFUNCnorm(y)
    \left[ y'(z) \right]^2
    ,
    \label{eq:boundaryEQ}
\end{equation}

\noindent where $\gFUNCnorm$ is a nominally $O(1)$ function. It is well established~\citep{Bender78} that the solutions to such equations can exhibit boundary layers when $\delta \to 0$; for \eq{eq:boundaryEQ} such boundary layers will occur where $\gFUNCnorm(y) = 0$. 

Away from the boundary layers, the `outer' solution of \eq{eq:boundaryEQ} is approximately constant, \ie $y \approx y_j$ for some $y_j$. However, the small gradient of $y$ will eventually bring $\gFUNCnorm(y)$ sufficiently close to zero to trigger a rapid change in $y$ across the boundary layer to reach the next plateau region where $y \approx y_{j+1}$. A staircase pattern thereby emerges whose steps are dictated by the root structure of $\gFUNCnorm$ (equivalently, the inflection points of $y$), with the widths $W$ of the steps set by $\delta$ as
\begin{equation}
    W
    \propto
    a^{\gFUNCnorm'(y_*)/\delta}
    ,
    \label{eq:stairWIDTH}
\end{equation}

\noindent where $y_*$ is a root of $\gFUNCnorm(y) = 0$ and $a > 1$ is a constant that depends on boundary conditions. This behavior is summarized as follows:

\vspace{3mm}
\begin{conjecture}
    A staircase step forms in the solution of \eq{eq:boundaryEQ} for $|\delta| \ll 1$ when $\gFUNCnorm(y)$ traverses a root $y_*$ where $\gFUNCnorm(y_*) = 0$ and $\gFUNCnorm'(y_*)/\delta < 1$. Therefore, a multi-step staircase forms when $\gFUNCnorm(y)$ has multiple roots.
\end{conjecture}
\vspace{3mm}

\begin{figure}
    \centering
    \begin{overpic}[width=0.7\linewidth, trim={4mm 4mm 4mm 4mm}, clip]{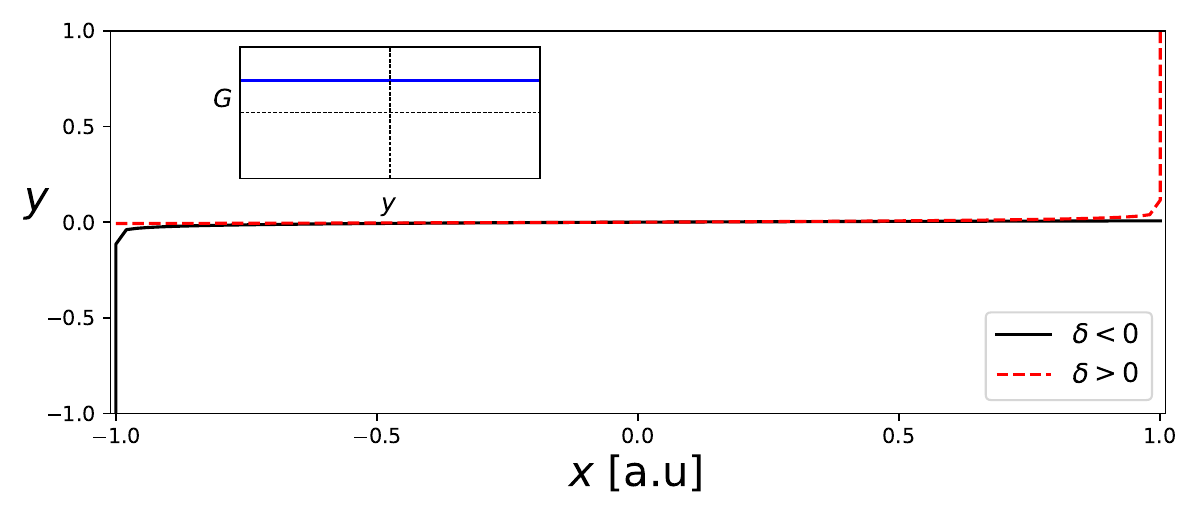}
        \put(9,9){\textbf{\small(a)}}
    \end{overpic}

    \begin{overpic}[width=0.7\linewidth, trim={4mm 4mm 4mm 4mm}, clip]{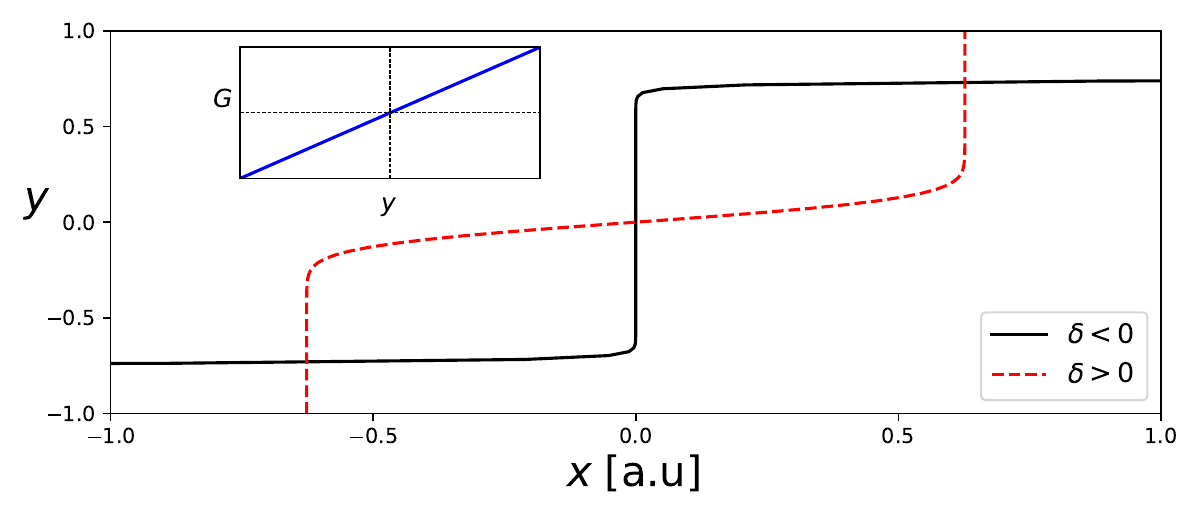}
        \put(9,9){\textbf{\small(b)}}
    \end{overpic}
    
    \begin{overpic}[width=0.7\linewidth, trim={4mm 4mm 4mm 4mm}, clip]{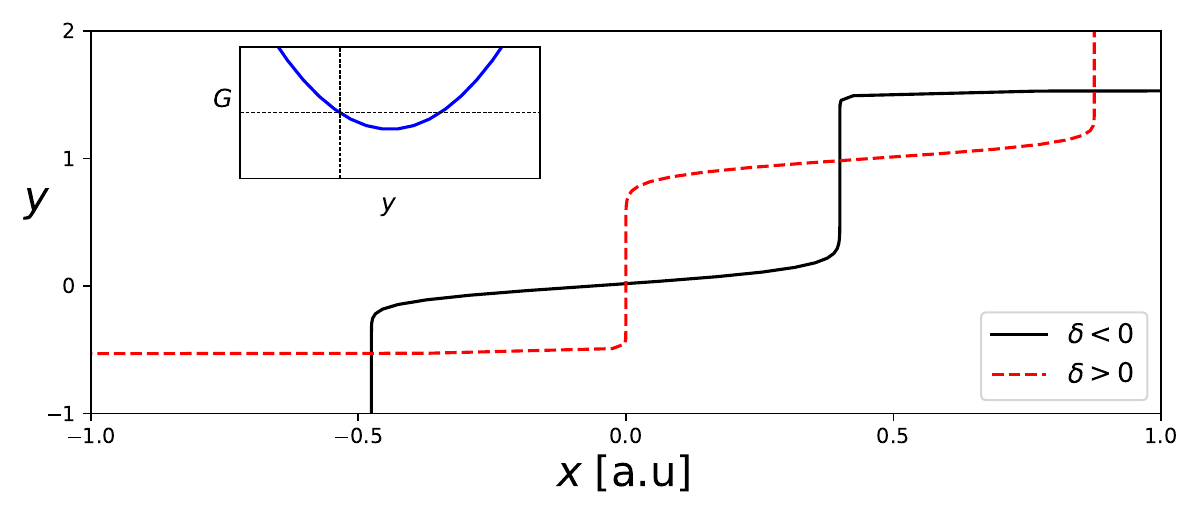}
        \put(9,9){\textbf{\small(c)}}
    \end{overpic}

    \begin{overpic}[width=0.7\linewidth, trim={4mm 4mm 4mm 4mm}, clip]{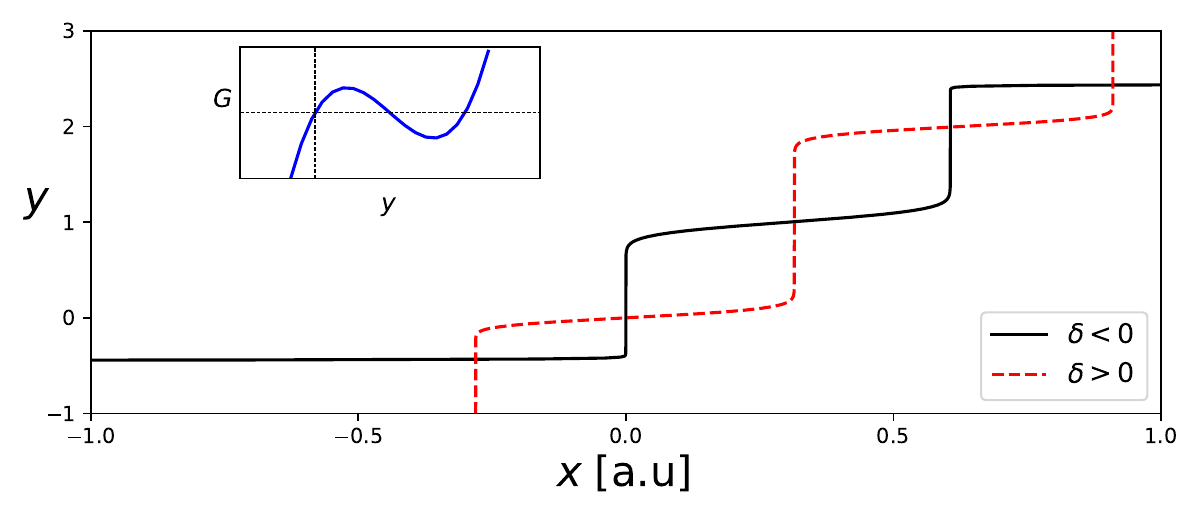}
        \put(9,9){\textbf{\small(d)}}
    \end{overpic}
    \caption{Solutions of \eq{eq:boundaryEQ} when $\gFUNCnorm(y)$ takes the form shown in each respective inset. All solutions have $|\delta| = 0.01$. The `arbitrary units' (a.u.) designation on the $x$-axis emphasizes that, due to affine symmetry, there is formally no scale to the $x$-dependence of the solutions. All solutions satisfy $y(0) = 0$, with the other boundary condition $y'(0)$, which simply controls the horizontal scale of the solution, adjusted for each case to fit the pertinent behaviour on the same axis. The functional forms for the plots shown in (a) and (b) can be derived analytically, as shown in \App{app:auxGfunc}.}
    \label{fig:staircaseG}
\end{figure}

\noindent The basis for this conjecture is demonstrated in \Fig{fig:staircaseG}, which shows solutions of \eq{eq:boundaryEQ} for a polynomial $\gFUNCnorm(y)$. The derivation of \eq{eq:stairWIDTH} and the analytical solution for certain special cases of $\gFUNCnorm(y)$ are presented in \App{app:auxGfunc}. Let us now demonstrate the role that these staircase solutions play in determining the globally stable temperature profiles for isobaric and constant-density plasmas.

\subsection{Isobaric density profile with Chapman--Enskog friction}
\label{sec:isoCHAP}

Let us consider a situation when the density profile is set by pressure balance. This means that $n(T)$ is given as
\begin{equation}
    n(T)
    = \frac{\plasmaBETA B_z^2}{8\pi T}
    .
    \label{eq:isobaricDEF}
\end{equation}

\noindent One therefore has
\begin{equation}
    \Hall(T)
    =
    \left(\frac{T}{\tau} \right)^{5/2}
    , \quad
    T(\Hall) = 
    \tau \Hall^{2/5}
    , \quad
    n(\Hall)
    =
    \frac{\plasmaBETA B_z^2}{8\pi \tau \Hall^{2/5}}
    ,
\end{equation}

\noindent where we have introduced the magnetization temperature
\begin{equation}
    \tau \doteq
    \left(
        \frac{ \sqrt{2} }{6 \sqrt{\pi}}Z e^2 m^{3/2} c^2 \plasmaBETA \cycloF
        \log \Lambda
    \right)^{2/5}
    .
\end{equation}

\noindent Consequently,
\begin{align}
    \pd{T} \Hall
    = \frac{5}{2} \frac{
        \Hall^{3/5}
    }{
        \tau
    }
    , \quad
    \pd{T}^2 \Hall
    = \frac{15}{4}
    \frac{
        \Hall^{1/5}
    }{
        \tau^2
    }
    , \quad
    n'(\Hall)
    =
    - \frac{2}{5} \frac{n}{\Hall}
    .
\end{align}

\noindent The auxiliary functions \eq{eq:auxG12} and \eq{eq:auxG} therefore take the following forms:
\begin{subequations}
    \label{eq:isobaricAUX}
    \begin{align}
        \gTpsCOEF
        &=
        \frac{5}{2 \tau \Hall^{2/5}}
        \left\{
            \plasmaBETA
            \Hall
            \frac{
                [\Delta_\beta(\Hall)]^2
            }{
                10 \alpha_\perp(\Hall)
            }
            + \gTppCOEF'(\Hall) \Hall
        \right\}
        , \\
        \gTppCOEF
        &=
        \beta_\wedge(\Hall) 
        + \frac{
            3 \alpha_\perp(\Hall) 
            - 5 \Hall \alpha_\perp'(\Hall)
        }{2 \plasmaBETA \Hall}
        , \\
        \gFUNC(\Hall)
        &=
        \frac{
            3 \gTppCOEF(\Hall)
            - 2 \Hall^{2/5} \tau \gTpsCOEF(\Hall)
        }{5 \gTppCOEF(\Hall) \Hall}
        .
    \end{align}
\end{subequations}

\begin{figure}
    \centering
    \includegraphics[width=0.7\linewidth,trim={20mm 6mm 4mm 2mm},clip]{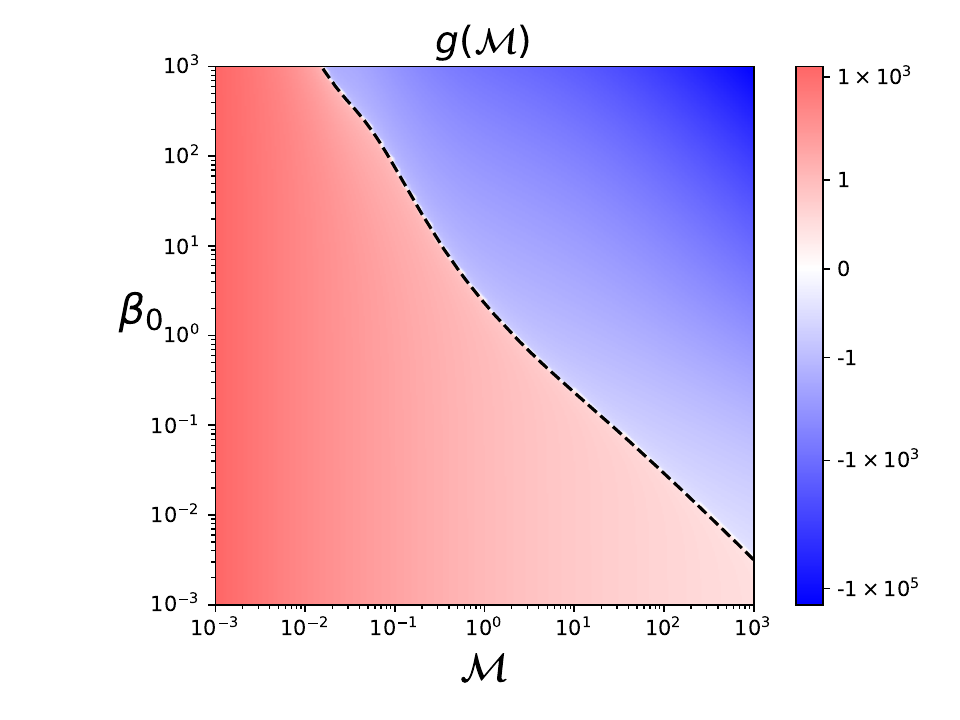}

    \hspace{-9mm}\includegraphics[width=0.6\linewidth,trim={4mm 8mm 4mm 14mm},clip]{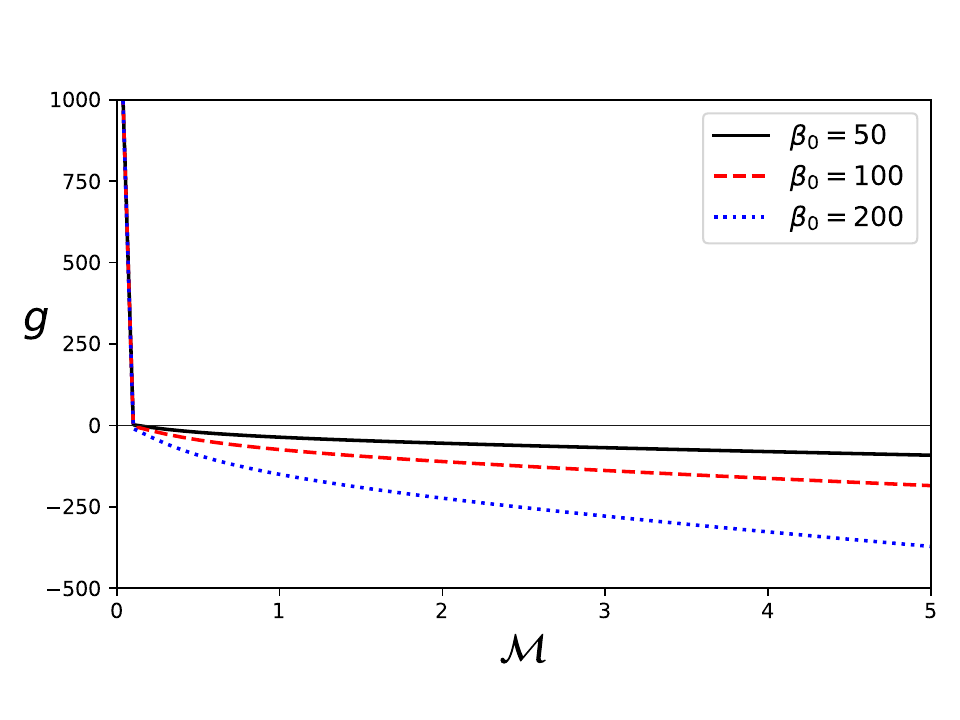}
    \caption{Contour plot (top) and lineouts at select $\plasmaBETA$ values (bottom) for $\gFUNC(\Hall)$ when the friction coefficients are obtained using the Lorentz collision operator. The dashed black contour in the top panel indicates the root set $\gFUNC = 0$ across which a staircase step is expected to form when $\plasmaBETA$ is large.}
    \label{fig:gCONTisobaric}
\end{figure}

\noindent A contour plot of $\gFUNC$ versus $\Hall$ and $\plasmaBETA$ is shown in \Fig{fig:gCONTisobaric}, along with lineouts along $\Hall$ for some values of~$\plasmaBETA$. It is clear that $\gFUNC(\Hall)$ becomes large for large~$\plasmaBETA$, as anticipated. Moreover, $\gFUNC(\Hall)$ has a single root corresponding to the single root of $\TpsCOEF$ (\Fig{fig:drives}), satisfying the criterion for a staircase to form.

Using known asymptotics \eq{eq:lim0} of the Lorentz friction coefficients, it is straightforward to show that $\Hall \gFUNC \to 8/5$ as $\Hall \to 0$. Hence, the temperature profile is guaranteed to be positive everywhere (\App{app:positivity}). To obtain a simple analytical approximation for the solution to \eq{eq:marginalZ}, it is reasonable to take
\begin{equation}
    \gFUNC(\Hall)
    \approx \frac{8}{5 \Hall} 
    \left(
        1 - \frac{\Hall}{\Hall_*}
    \right)
    ,
    \label{eq:rationalG}
\end{equation}

\noindent with $\Hall_*$ being the single root. Specifically, when $\plasmaBETA \gg 1$, $\Hall_*$ can be approximately calculated using the $\Hall \ll 1$ limit of the Lorentz friction coefficients to be
\begin{equation}
    \lim_{\plasmaBETA \to \infty}
    \Hall_* = 8 \sqrt{ \frac{\alpha_\parallel}{105 \plasmaBETA} }
    \approx \frac{0.4}{\sqrt{\plasmaBETA}}
    .
    \label{eq:isoMbetaLARGE}
\end{equation}

\noindent Note, importantly, that a geometrical-optics description of this parameter regime is not valid because \eq{eq:WKBvalid} predicts that $k_{z,\text{max}} L \sim \Hall_* \ll 1$ when $\Hall_*^2 \plasmaBETA \sim 1$ and $\Hall_* \ll 1$. The continuation of the root line to small $\plasmaBETA$ can be computed using the $\Hall \to \infty$ limit of the friction coefficients \eq{eq:limINFTY} to give 
\begin{equation}
    \lim_{\plasmaBETA \to 0}
    \Hall_* = \frac{2 \sqrt{6}}{\beta_\parallel \plasmaBETA} 
    \approx 
    \frac{3.3}{\plasmaBETA}
    .
\end{equation}

\noindent Since $\plasmaBETA \ll 1$, no staircase is expected to form in this parameter regime. For arbitrary~$\plasmaBETA$, a simple interpolation of the two limits can be used to obtain
\begin{equation}
    \Hall_* \approx
    \frac{3.3}{\plasmaBETA} + \frac{0.4}{\sqrt{\plasmaBETA}}
    .
    \label{eq:rootINTERP}
\end{equation}

As shown in \App{app:auxGfunc}, the solution to \eq{eq:marginalZ} can be computed analytically for $\gFUNC(\Hall)$ given by \eq{eq:rationalG}:
\begin{equation}
    \frac{
        z(\Hall) - z(\Hall_1)
    }{
        z(\Hall_2) - z(\Hall_1)
    }
    = 
    \frac{
        \gamma\left(
            - \frac{3}{5}
            ,
            - \frac{8 \Hall}{5 \Hall_*}
        \right)
        - \gamma\left(
            - \frac{3}{5}
            ,
            - \frac{8 \Hall_1}{5 \Hall_*}
        \right)
    }{
        \gamma\left(
            - \frac{3}{5}
            ,
            - \frac{8 \Hall_2}{5 \Hall_*}
        \right)
        - \gamma\left(
            - \frac{3}{5}
            ,
            - \frac{8 \Hall_1}{5 \Hall_*}
        \right)
    }
    ,
    \label{eq:exampleT}
\end{equation}

\begin{figure}
    \centering
    \includegraphics[width=0.7\linewidth,trim=4mm 14mm 4mm 24mm, clip]{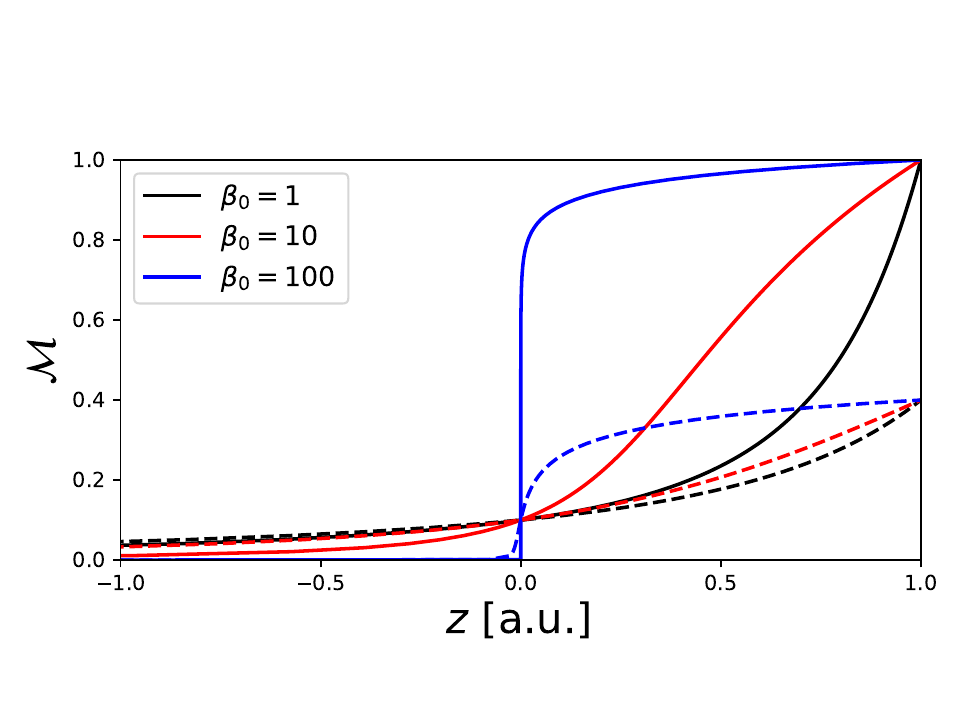}

    \includegraphics[width=0.7\linewidth,trim=4mm 14mm 4mm 24mm, clip]{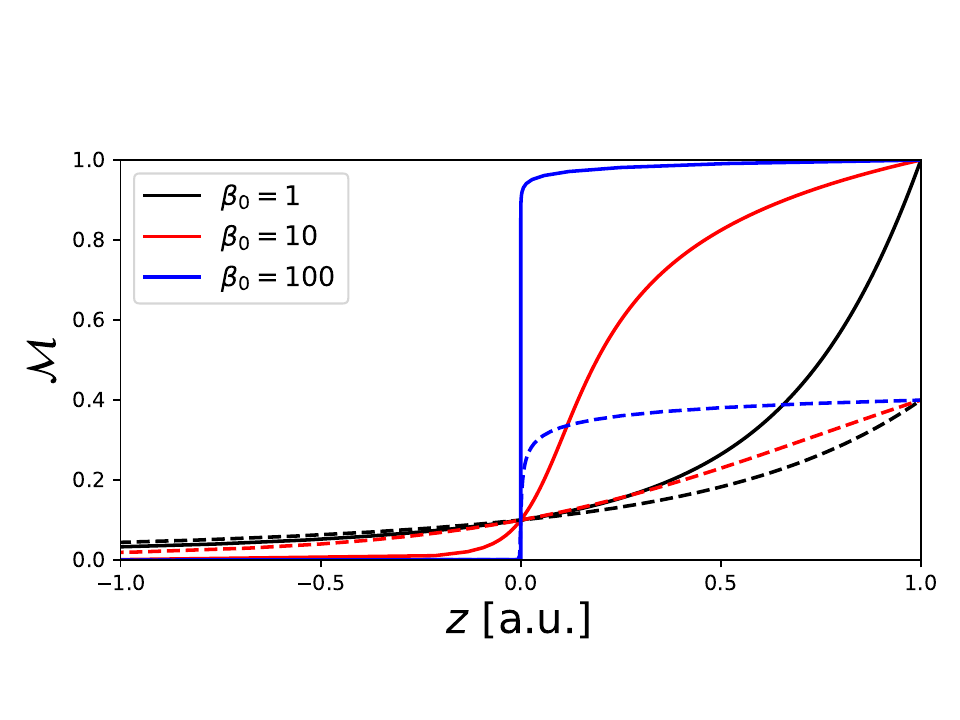}
    \caption{Solution \eq{eq:marginalZ} for the marginally stable magnetization $\Hall \propto T^{5/2}$ at various values of $\plasmaBETA$ for Lorentz friction coefficients \citep{Lopez24a} and isobaric plasma~\eq{eq:isobaricDEF}. The boundary conditions are $z(\Hall_1) =0$, $z(\Hall_2) = 1$, $\Hall_1 = 0.1$, and $\Hall_2 = 1$ (solid) or $\Hall_2 = 0.4$ (dashed). The top plot uses the analytical approximation presented in \eq{eq:exampleT} with $\Hall_*$ defined in \eq{eq:rootINTERP}, while the bottom plot is the numerically computed solution.}
    \label{fig:exampleSTEP}
\end{figure}

\noindent where $\gamma(s,z)$ is the lower incomplete Gamma function~\citep{Olver10a}. Importantly, $\gamma(s,0)$ is divergent when $s < 0$ so $\Hall(z)$ is positive-definite. Figure \ref{fig:exampleSTEP} shows the solution \eq{eq:exampleT} at different values of $\plasmaBETA$ for two different boundary conditions using the approximation for $\Hall_*$ provided in \eq{eq:rootINTERP}. A step-function profile clearly develops as $\plasmaBETA$ increases for both boundary conditions, demonstrating the robustness of the temperature staircase. For comparison, \Fig{fig:exampleSTEP} also presents numerically computed solutions of \eq{eq:marginalZ}. Overall, the analytical approximation is seen to capture all the salient features of the temperature profile, but underestimates the sharpness of the staircase step because the approximation \eq{eq:rationalG} does not reproduce the correct gradient across the root, \ie $\gFUNC'(\Hall_*)$.

\subsection{Constant density profile with Chapman--Enskog friction}
\label{sec:consCHAP}

Let us now consider the simpler case of constant density:
\begin{equation}
    n(\Hall)
    = n_0
    .
\end{equation}

\noindent One therefore has
\begin{equation}
    \Hall(T)
    =
    \left(\frac{T}{\tau_0} \right)^{3/2}
    , \quad
    T(\Hall) = 
    \tau_0 \Hall^{2/3}
    ,
\end{equation}

\noindent where the magnetization temperature now takes the form
\begin{equation}
    \tau_0 \doteq
    \left(
        \sqrt{\frac{2 m}{\pi}}
        \frac{
            Z e^2 \plasmaF^2 \log \Lambda
        }{
            3 \cycloF
        }
    \right)^{2/3}
    .
\end{equation}

\noindent Consequently,
\begin{align}
    \pd{T} \Hall
    = \frac{3}{2} \frac{
        \Hall^{1/3}
    }{
        \tau_0
    }
    , \quad
    \pd{T}^2 \Hall
    = \frac{3}{4}
    \frac{
        \Hall^{-1/3}
    }{
        \tau_0^2
    }
    ,
\end{align}

\noindent and the auxiliary functions \eq{eq:auxG12} and \eq{eq:auxG} take the following forms:
\begin{subequations}
    \label{eq:constantAUX}
    \begin{align}
        \gTpsCOEF
        &=
        \frac{1}{2 \Hall^{2/3} \tau_0}
        \left\{
            \betaEFFECT
            \Hall^{5/3}
            \frac{
                [\Delta_\beta(\Hall)]^2
            }{
                2 \alpha_\perp(\Hall)
            }
            + 3 \gTppCOEF'(\Hall) \Hall
        \right\}
        , \\
        \gTppCOEF
        &=
        \beta_\wedge(\Hall) 
        + \frac{3}{2}
        \frac{
            \alpha_\perp(\Hall) 
            - \Hall \alpha_\perp'(\Hall)
        }{\betaEFFECT \Hall^{5/3}} 
        , \\
        \label{eq:gFUNCcons}
        \gFUNC(\Hall)
        &=
        \frac{
            \gTppCOEF(\Hall)
            - 2 \Hall^{2/3} \tau_0 \gTpsCOEF(\Hall) 
        }{
            3 \gTppCOEF(\Hall) \Hall
        }
        ,
    \end{align}
\end{subequations}

\noindent where we have defined the effective plasma beta
\begin{equation}
    \betaEFFECT
    = \frac{8 \pi n_0 \tau_0}{B_z^2}
    \approx
    2.61 \times 10^{6}
    \left( Z \log \Lambda \right)^{2/3}
    \left( \frac{ n_0 }{ 10^{20}~\text{cm}^{-3} } \right)^{5/3}
    \left( \frac{B_z}{ 10^3 ~\text{G}}\right)^{-8/3}
    .
    \label{eq:betaEFF}
\end{equation}

\noindent Clearly, $\betaEFFECT$ can be made large for realistic plasma parameters, so, provided that a root to \eq{eq:gFUNCcons} exists, a sharp magnetization staircase is expected to form for constant density profiles as well.

\begin{figure}
    \centering
    \includegraphics[width=0.7\linewidth,trim={18mm 6mm 4mm 2mm},clip]{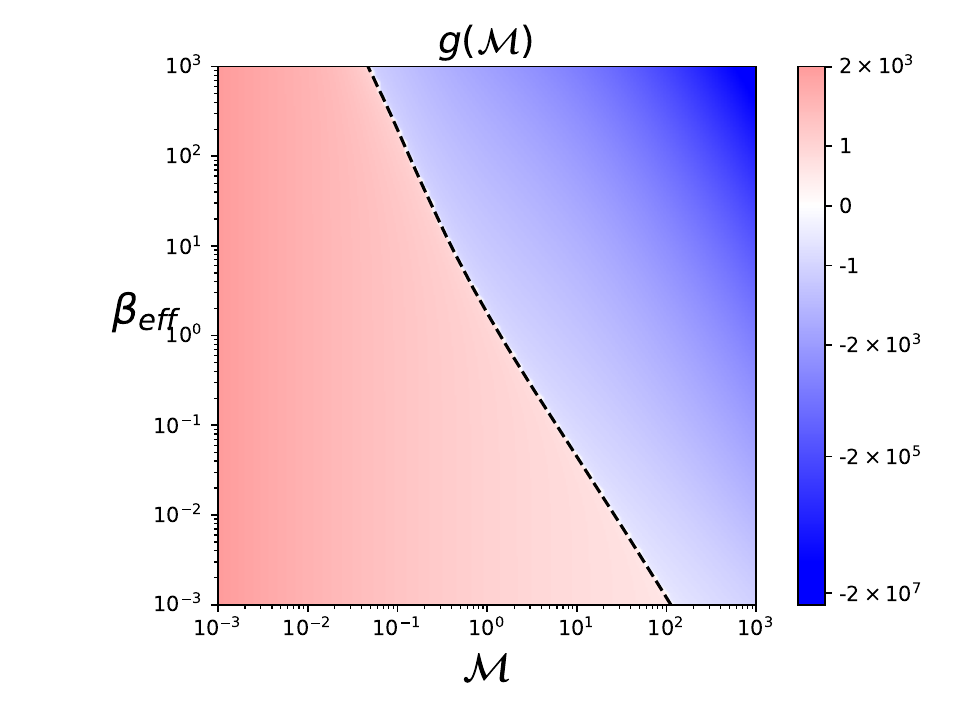}

    \hspace{-9mm}\includegraphics[width=0.6\linewidth,trim={4mm 8mm 4mm 14mm},clip]{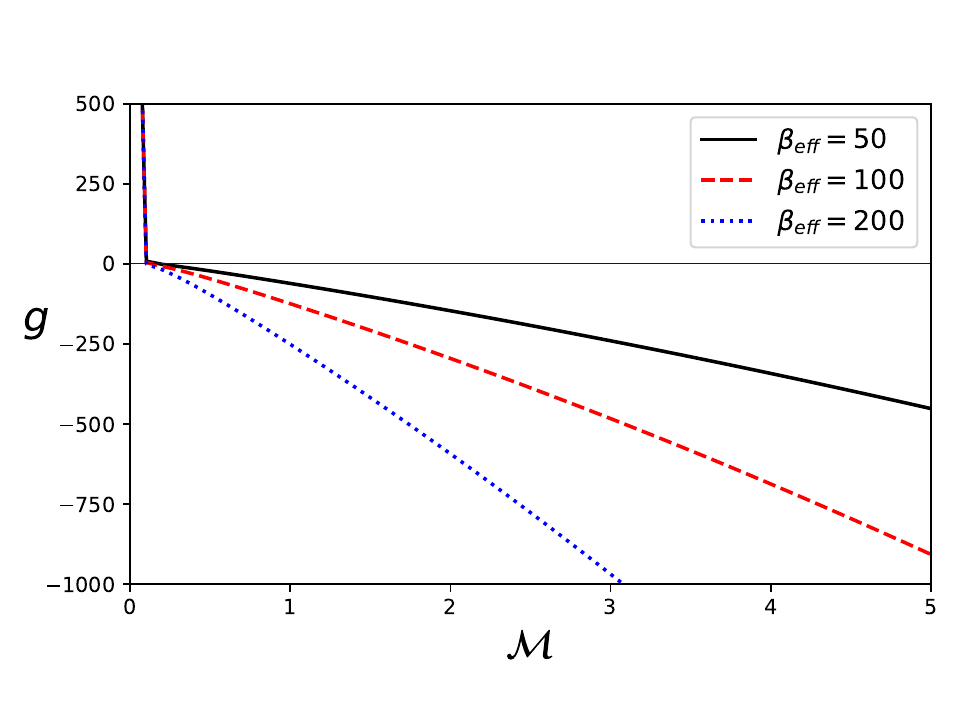}
    \caption{Same as \Fig{fig:gCONTisobaric}, but for constant density. The definition of $\betaEFFECT$ is~\eq{eq:betaEFF}.}
    \label{fig:gCONTconst}
\end{figure}

Figure \ref{fig:gCONTconst} shows a contour plot of $\gFUNC$ as a function of $\Hall$ and $\betaEFFECT$, along with lineouts along $\Hall$ for some values of $\betaEFFECT$. Analogously to \Fig{fig:gCONTisobaric}, $\gFUNC(\Hall)$ becomes large for large $\betaEFFECT$ and has a single root line. Therefore, the behavior of the solution \eq{eq:marginalZ} will have the same qualitative features as those seen in \Fig{fig:exampleSTEP}, namely, a positive-definite magnetization profile possessing a single staircase step across the root whose approximate interpolated form is
\begin{equation}
    \Hall_* \approx
    \frac{0.5}{\betaEFFECT^{3/8}}
    + \frac{1.9}{\betaEFFECT^{3/5}}
\end{equation}

\noindent (the two terms individually constitute the $\betaEFFECT \to \infty$ and the $\betaEFFECT \to 0$ limits of $\Hall_*$, respectively). By comparing Figs.~\ref{fig:gCONTconst} and \ref{fig:gCONTisobaric}, however, we see that $\gFUNC'(\Hall_*)$ is larger when the plasma density is constant instead of isobaric; hence, the staircase associated with \Fig{fig:gCONTconst} will be sharper than either the analytical approximation given by \eq{eq:exampleT} or the numerical solution presented in \Fig{fig:exampleSTEP}.


\section{Back-reaction on the background temperature profile}
\label{sec:quasilinear}

Having discussed at length the linear growth rate of the collisional whistler instability, let us now briefly investigate how the instability modifies the background temperature profile. To second order in $\epsilon$, \eq{eq:Tnonlin} can be written in two equivalent forms:
\begin{subequations}
	\label{eq:TevoFRIC}
    \begin{align}
		\label{eq:TevoADVECT}
		\frac{3}{2} n \, \pd{t} T
		&= \epsilon^2 B_z^2 \resSOURCE
		- \epsilon^2 n \advectV \pd{z} T
		- \pd{z} q_z
		, \\
		\label{eq:TevoENERGY}
		\frac{3}{2} n \, \pd{t} T
		&= - \pd{z} \modQ
		- \frac{\epsilon^2 B_z^2}{8 \pi} \ave{\Symb{D}_A} \inten
        .
	\end{align}
\end{subequations}

\noindent The first form \eq{eq:TevoADVECT} emphasizes the advection-diffusion dynamics involved in the temperature evolution, while the second form \eq{eq:TevoENERGY} emphasizes the flow of energy throughout space and the transfer of energy from plasma to waves. Here $\resSOURCE$ represents the heating source due to the work done by the resistive ($\alpha$) friction force on the perturbed flow, $\advectV$ the wave-driven advection velocity due to the work done by the thermoelectric ($\beta$) friction force on the perturbed flow, $\modQ$ the modified heat flux due to the additional wave-driven Poynting flux contribution, and $\ave{\Symb{D}_A} \inten$ the energy sink to excite fluctuations; their respective definitions are
\begin{subequations}
\label{eq:perturbQUANT}
\begin{align}
    \label{eq:resSOURCE}
    \resSOURCE
    &\doteq
    \frac{\resist}{8\pi} 
    \left(
        \inten \ave{k_z^2}
        + \frac{1}{4} \pd{z}^2 \inten
    \right)
    + \epsilon^2 \frac{ \Delta_\alpha c^2}{16 \pi \omega_p^2 \collT} \inten^2 \ave{k_z}^2
    \left(
        1 + \epsilon^2 \frac{\inten}{2}
    \right)^{-1}
    \nonumber\\
    &=
    \frac{\resist}{8\pi} 
    \left(
        \inten \ave{k_z^2}
        + \frac{1}{4} \pd{z}^2 \inten
    \right)
    + O(\epsilon^2)
    , \\
    \label{eq:thermoADVECT}
    \advectV
    &\doteq
    \frac{\cycloF c^2}{2 \plasmaF^2}
    \left(
        \frac{
            \Delta_\beta \ave{k_z} \inten
        }{
            1 + \epsilon^2 \inten/2
        } 
        -
        \frac{
            \beta_\wedge \, \pd{z} \inten
        }{
            2 \sqrt{ 1 + \epsilon^2 \inten/2 }
        }
    \right)
    \nonumber\\
    &=
    \frac{\cycloF c^2}{4 \plasmaF^2}
    \left(
        2 \Delta_\beta \ave{k_z} \inten
        - \beta_\wedge \, \pd{z} \inten
        \nullFrac
    \right)
    + O(\epsilon^2)
    ,
\end{align}
\begin{align}
    \modQ
    &\doteq
    q_z
    + \frac{\epsilon^2 B_z^2}{16 \pi} 
    \left(
        \nerstV \inten
        + \frac{\inten \pd{z} \resist - \resist \, \pd{z} \inten }{2}
    \right)
    , \\
    q_z
    &=
    - n T
    \left[
        \frac{\collT}{m}
        \left(
            \kappa_\parallel
            - \frac{\epsilon^2}{2} \Delta_\kappa \inten 
        \right)
        \pd{z} T
        + \epsilon^2 \frac{\cycloF c^2}{2 \plasmaF^2}
        \left(
            \Delta_\beta \ave{k_z} \inten
            + \frac{\beta_\wedge}{2}
            \, \pd{z} \inten
        \right)
    \right]
    + O(\epsilon^3)
    .
\end{align}
\end{subequations}

\noindent The appropriate lowest-order expressions for $\resist$, $\nerstV$, and $\Symb{D}_A$ are given in \eq{eq:hamCONSTIT} and \eq{eq:Daham}. It is important to note that, when combined with \eq{eq:intenEQ}, \eq{eq:TevoENERGY} manifestly conserves the total energy of the electron MHD system of equations (the appropriate expressions for energy conservation are presented in \App{app:energy}).

\subsection{Frictional cooling}
\label{sec:fricCOOL}

As required by energy conservation, the growth of whistler waves due to friction implies that the friction must be cooling the temperature profile at the same time (at least volumetrically, \ie neglecting fluxes). This is counterintuitive since friction is often considered a source of heating instead of cooling. Indeed, the frictional work due to resistivity ($\resSOURCE$) is positive definite and therefore always a heat source. However, the advection velocity $\advectV$ of the temperature profile due to friction is not sign-definite, and, depending on the signs of $\pd{z} \inten$ and $\ave{k_z}$, it can be aligned with $\pd{z} T$ and therefore be a cooling flow%
\footnote{Importantly, frictional cooling still produces entropy~\citep{Kolmes21c} and therefore does not violate any fundamental laws of thermodynamics.}.

More quantitatively, let us suppose that the wave profile is given by a quasi-monochromatic (and also quasi-eikonal) field of the form%
\footnote{This simple field profile is chosen to illustrate the key physics that might be at play as the instability tries to saturate. Since the geometrical-optics approximation is not generally satisfied, one does not expect a quasi-monochromatic field to remain such as time progresses.}
\begin{equation}
    \psi(z) = \sqrt{\inten(0)} \, \exp\left[ \frac{z}{2 \lenI} +i \int_0^z k_\text{max}(\zeta) \dd \zeta \right]
    ,
    \label{eq:psiQUASI}
\end{equation}

\noindent where $\lenI$ is the intensity gradient lengthscale. One can then calculate the instantaneous resistive heating rate \eq{eq:resSOURCE} and thermoelectric advection velocity \eq{eq:thermoADVECT} as
\begin{align}
    \resSOURCE
    =
    \left[
        \left(
            \frac{\growthV}{2\resist}
        \right)^2
        + \left(
            \frac{1}{2 \lenI}
        \right)^2
    \right]
    \frac{\resist \inten}{8\pi}
    , \quad
    \advectV
    =
    \left(
        \frac{\growthV^2}{\resist} 
        - \frac{\beta_\wedge}{m \cycloF} \frac{\pd{z}T}{\lenI}
        \nullFrac
    \right)
    \frac{m c^2 \cycloF^2}{4 \plasmaF^2} \frac{\inten}{\pd{z}T}
    .
\end{align}

\noindent Hence, we see that (i) the resistive heating is manifestly positive-definite, as required, and (ii) the thermoelectric advection velocity becomes a cooling flow, \ie $\text{sign}(\advectV) = \text{sign}(\pd{z} T)$, when wave intensity and temperature gradients oppose each other, viz., when $\lenI^{-1} < \growthV^2 m \cycloF \lenT/(\resist \beta_\wedge T)$. Furthermore, since the total frictional heating can be expressed as
\begin{align}
    \epsilon^2 B_z^2 \resSOURCE
    - \epsilon^2 n \advectV \pd{z} T
    &=
    \frac{\epsilon^2 B_z^2}{32 \pi} 
    \left(
        \resist \lenI^{-2}
        - 2 \nerstV \lenI^{-1}
        - \frac{\growthV^2}{\resist} 
    \right)
    \inten
    ,
    \label{eq:totalFRICheat}
\end{align}

\noindent the total frictional heating will be negative when
\begin{equation}
    \frac{
        \nerstV - \sqrt{\nerstV^2 + \growthV^2}
    }{\resist}
    < \lenI^{-1} < 
    \frac{
        \nerstV + \sqrt{\nerstV^2 + \growthV^2}
    }{\resist}
    .
    \label{eq:negFRICcond}
\end{equation}

\noindent If we take $\pd{z} T > 0$, then \eq{eq:negFRICcond} can be equivalently written as
\begin{equation}
    - \frac{\Hall \plasmaBETA}{2 \alpha_\perp}
    \left(
        \sqrt{\beta_\wedge^2 + \Delta_\beta^2}
        + \beta_\wedge
    \right)
    < \lenT \lenI^{-1} < 
    \frac{\Hall \plasmaBETA}{2 \alpha_\perp}
    \left(
        \sqrt{\beta_\wedge^2 + \Delta_\beta^2}
        - \beta_\wedge
    \right)
    .
\end{equation}

\noindent Interestingly, the condition \eq{eq:negFRICcond} is satisfied for a constant intensity profile $\lenI^{-1} = 0$ because the two endpoints of \eq{eq:negFRICcond} necessarily have opposite signs (but note that the interval is not symmetric about zero since its center is $\nerstV \neq 0$). This means that friction will cool the temperature profile in the early stages of the instability when whistlers grow from an initially homogeneous noise-level of fluctuations.

\subsection{Reduced heat flux}
\label{sec:suppressFLUX}

Next, let us consider how the heat flux gets modified by the collisional whistler instability. For the quasi-eikonal field given by \eq{eq:psiQUASI}, the heat flux takes the form
\begin{equation}
    - \frac{q_z}{q_0}
    =
    \kappa_\parallel
    + \frac{\epsilon^2}{2}
    \left(
        \frac{\Delta_\beta^2}{2 \alpha_\perp}
        + \frac{\beta_\wedge}{\Hall \plasmaBETA} \frac{\lenT}{\lenI}
        - \Delta_\kappa
    \right) \inten
    , \quad
    q_0 = \frac{n T \thermalV^2 \collT}{\lenT}
    .
    \label{eq:heatFLUXmanipulate}
\end{equation}

\noindent Hence, we see that the net effect of the instability on the heat flux results from the competition of three terms. The first two $O(\epsilon^2)$ terms are associated with the Ettingshausen effect
\footnote{For detailed discussions of the Ettingshausen effect, see, e.g., \citet{Chittenden93} and \citet{Kolmes21a}.},
which is the additional heat flux [beyond the standard enthalpy flux~\citep{Epperlein86}] carried by faster moving, less collisional electrons whose directional symmetry is broken with a mean flow. The first of these terms is always positive and therefore always enhances the heat flux; in contrast, the second term can reduce the heat flux when $\lenI$ and $\lenT$ are oppositely oriented and can even overcome the first term if $\lenT \lenI^{-1}$ is sufficiently negative (meaning that $\inten$ is sufficiently sharply peaked):
\begin{equation}
    \frac{\lenT}{\lenI}
    < - \frac{\Hall \plasmaBETA \Delta_\beta^2}{2 \alpha_\perp \beta_\wedge}
    \approx - 981 \plasmaBETA \Hall^4
    ,
    \label{eq:negETTINGS}
\end{equation}

\noindent where the final approximation is for $\Hall \ll 1$. This heat-flux-reduction mechanism can be readily achieved in the high-$\plasmaBETA$, low-$\Hall$ regime in which the temperature staircases discussed in \Sec{sec:global_stable} also form, since in this limit the right-hand side of \eq{eq:negETTINGS} goes to zero as $-25/\plasmaBETA$ [see \eq{eq:isoMbetaLARGE}]. The third $O(\epsilon^2)$ term in \eq{eq:heatFLUXmanipulate}, which is always negative, is the reduction of the effective conductivity due to the transverse magnetic field perturbations generated by the instability causing the temperature gradient and the total magnetic field to become misaligned. 

\subsection{Marginally stable heat flux}

Finally, let us suppose that the temperature profile is in a globally marginally stable state such that $\ave{\Symb{D}_A} = 0$ (\Sec{sec:global_stable}). Dynamically, one expects an arbitrary temperature profile to be driven towards such a state on the instability timescale, which can be faster than the conduction timescale (see \Fig{fig:whistleREL}). This is because, as shown in \Fig{fig:drives}, the instability growth rate $\Symb{D}_A$ is an increasing function of temperature (and conversely, the damping rate is a decreasing function of temperature). Energy conservation \eq{eq:TevoENERGY} then implies that higher temperatures are increasingly cooled by the instability (thereby decreasing $\Symb{D}_A$) while lower temperatures are increasingly heated (thereby decreasing $-\Symb{D}_A$) to create temperature plateaus separated by transition regions where the temperature profile has remained unchanged because $\Symb{D}_A \approx 0$ initially. The result is a profile that has $\Symb{D}_A = 0$ everywhere.

When $\ave{\Symb{D}_A} = 0$ globally, the total frictional heating can be written as an energy flux [see \eq{eq:TevoENERGY}], which combines with the heat flux to yield
\begin{equation}
    - \frac{\modQ}{\kappa_\parallel q_0}
    =
    1 
    - \epsilon^2 \fluxREDUCE \inten 
    ,
    \label{eq:modQreduce}
\end{equation}

\noindent where the flux-reduction factor is
\begin{align}
    \fluxREDUCE
    =
    \frac{1}{2 \kappa_\parallel}
    \left[
        \Delta_\kappa
        + \frac{5}{2} \frac{\alpha_\perp' }{ \Hall \plasmaBETA^2}
        - \frac{\beta_\wedge}{\Hall \plasmaBETA} 
        \left( \frac{\lenT}{\lenI} + 1 \right)
        - \frac{\alpha_\perp }{ \Hall^2 \plasmaBETA^2 } 
        \left( \frac{\lenT}{\lenI} + \frac{3}{2} \right)
        - \frac{\Delta_\beta^2}{2 \alpha_\perp}
    \right]
    .
    \label{eq:fluxREDUCEdef}
\end{align}

\begin{figure}
    \centering
    \includegraphics[width=0.7\linewidth,trim={16mm 6mm 30mm 2mm}, clip]{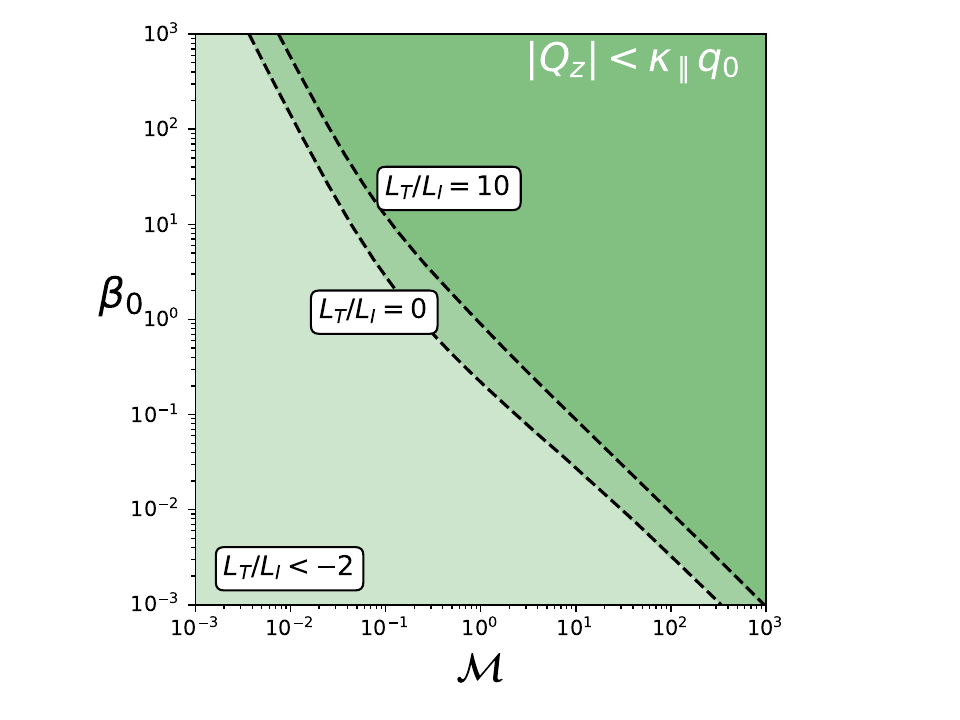}
    \caption{Regions (green) where the total heat flux $\modQ$ \eq{eq:modQreduce} is reduced due to the presence of whistler waves generated by the collisional whistler instability at global marginal stability. This reduction is controlled by the lengthscale ratio $\lenT \lenI^{-1}$ and occurs when $ \fluxREDUCE > 0$ \eq{eq:fluxREDUCEdef}; these regions are shown in green above the correspondingly labeled line. For $\lenT \lenI^{-1} < -2$, the entire plot range has a reduced total heat flux.}
    \label{fig:fluxREDUCE}
\end{figure}

\noindent Here we have imposed pressure balance for simplicity; the general expression is obtained by replacing $5/2$ with $\lenT \pd{z} \Hall /\Hall$ in the second term. As in \eq{eq:heatFLUXmanipulate}, we see that the intensity gradient is capable of reducing the total heat flux. In this case, the two mechanisms available are the Ettingshausen mechanism discussed in \Sec{sec:suppressFLUX} and the frictional cooling discussed in \Sec{sec:fricCOOL} (now coupled to the heat flux by imposing global marginal stability). Both are controlled by the lengthscale ratio $\lenT \lenI^{-1}$. As shown in \Fig{fig:fluxREDUCE}, making $\lenT \lenI^{-1}$ negative causes the heat flux to be reduced ($|\modQ| < \kappa_\parallel q_0$) over a large region of parameter space%
\footnote{ One can have reduced heat flux even with $\lenT \lenI^{-1} > 0$ provided that $\plasmaBETA$ and $\Hall$ are both sufficiently large ($\plasmaBETA \Hall \gtrsim \lenT \lenI^{-1}$) such that $\Delta_\kappa$ dominates \eq{eq:fluxREDUCEdef}.}.

Indeed, the flux-reduction factor $\fluxREDUCE$ can be made arbitrarily large by making $\lenT \lenI^{-1}$ increasingly negative. To make this more quantitative, let us consider the weakly magnetized (small-$\Hall$) limit and approximate all transport coefficients by their lowest-order asymptotic limits \eq{eq:lim0}. Then one has
\begin{align}
    \fluxREDUCE
    &\approx
    124 \Hall^2
    + \frac{1.34 }{ \plasmaBETA^2}
    - \frac{0.362}{ \plasmaBETA} 
    \left( \frac{\lenT}{\lenI} + 1 \right)
    - \frac{0.011 }{ \Hall^2 \plasmaBETA^2 } 
    \left( \frac{\lenT}{\lenI} + \frac{3}{2} \right)
    - 1200 \Hall^4
    \nonumber\\
    &\approx
    \frac{1}{\plasmaBETA}
    \left(
        21.7
        - 0.423 \frac{\lenT}{\lenI} 
    \right)
    \approx
    \frac{\Delta_\kappa}{2\kappa_\parallel}
    - \frac{0.423}{\plasmaBETA} \frac{\lenT}{\lenI} 
    ,
    \label{eq:approxFLUXreduce}
\end{align}

\noindent where in the second line, \eq{eq:isoMbetaLARGE} has been used to evaluate $\fluxREDUCE$ near the steepest point in the temperature staircase, taking $\plasmaBETA$ to be large as well. Using this, we can place a bound on the required value of $\lenT \lenI^{-1}$ to achieve strong heat-flux reduction as follows. By simple rearrangement, \eq{eq:modQreduce} can be written as
\begin{equation}
    \epsilon^2 \inten 
    =
    \frac{1}{\fluxREDUCE}
    \left(
        1 
        + \frac{\modQ}{\kappa_\parallel q_0}
    \right)
    \le 1
    ,
\end{equation}

\noindent where the inequality ensures that the small-amplitude expansion is not grossly violated. If the heat flux is strongly reduced, then $\modQ \approx 0$ and one would have
\begin{equation}
    \fluxREDUCE
    \ge 1
    ,
\end{equation}

\noindent or, equivalently, using \eq{eq:approxFLUXreduce}
\begin{equation}
    - \frac{\lenT}{\lenI}
    \ge 
    \frac{\plasmaBETA}{0.423} 
    \left(
        \frac{\kappa_\parallel + \kappa_\perp}{2\kappa_\parallel}
    \right)
    \gtrsim
    2 \plasmaBETA
    ,
    \label{eq:boundFLUXreduce}
\end{equation} 

\noindent since $\kappa_\parallel \le \kappa_\parallel + \kappa_\perp \le 2 \kappa_\parallel$.

Thus, the collisional whistler instability is capable of reducing the electron heat flux in principle, somewhat similar to the collisionless whistler instability~\citep{Levinson92,Pistinner98,Komarov18,RobergClark18a}. The persistence time $\tau$ of the temperature profile is then nominally lengthened by the same factor $\tau \sim \diffuseT/|1 - \epsilon^2 \fluxREDUCE \inten|$ until it is ultimately set by the persistence time of the intensity profile itself, which in turn is set by advection and refraction (\ie evolving the $z$ and the $k_z$ dependence of $\Symb{W}$, respectively). The advection-limited persistence time is expected to be comparable to $\tau$ (and thus not a limiting factor) since the total Poynting flux that accounts for the intensity advection [the term in square brackets in \eq{eq:energyFRIC}] and the modified heat flux $\modQ$ used to estimate $\tau$ only differ by subdominant terms. The refraction timescale, however, is difficult to estimate given that it inherently involves breaking the quasi-eikonal ansatz \eq{eq:psiQUASI}, requiring one to reconsider the full phase-space dynamics of $\Symb{W}$. This is beyond the scope of the present work.

Let us conclude this section with a brief word of caution regarding the bound obtained in \eq{eq:boundFLUXreduce}. The inequality \eq{eq:boundFLUXreduce} represents a stricter requirement on $\lenI$ compared to \eq{eq:negETTINGS} because \eq{eq:boundFLUXreduce} requires second-order effects to become comparable to the lowest-order heat flux, whereas \eq{eq:negETTINGS} results from comparing two second-order terms. Clearly, such an occurrence would also indicate that the perturbative approach underlying \eq{eq:perturbQUANT} has potentially become invalid; if $\lenT \lenI^{-1}$ is too large, even an infinitesimal intensity will grow quickly in space and become relatively large elsewhere. Therefore, our analysis here is merely a simple first attempt to understand what is required of $\inten(z)$ to greatly reduce the heat flux assuming quasilinear theory holds for all $\lenT \lenI^{-1}$. Future investigations can be conducted to see how nonlinear physics modifies this constraint. Conversely, it is likely that the heat-flux suppression by the collisional whistler instability will be a small effect for high-beta plasmas within the strict validity of our fluid slab model.


\vspace{2mm}

\section{Conclusion}
\label{sec:concl}

In this work, we have shown that the electron MHD equations with Braginskii friction in a 1-D slab geometry are unstable with respect to transverse magnetic perturbations. We call this instability the collisional whistler instability, since the dispersion relation contains the usual group-velocity dispersion of whistler waves. We show that for a large region of parameter space, the fastest-growing/least-damped whistler waves do not satisfy the geometrical-optics approximation. This necessitates using the Wigner--Moyal formalism to describe their dynamics, which we derive (\Sec{sec:main}). Extra terms are found in the instability growth rate involving gradients of the background plasma that would not be present had the geometrical-optics approximation been applied. The physical origin of these terms and their impact on the instability threshold are discussed in \Sec{sec:stable}. 

In particular, we show that the extra stabilization provided by the new terms allows for non-constant temperature profiles to emerge and persist (\Sec{sec:global_stable}). These stable temperature profiles are expected to be established quickly, on the instability timescale, since the quasilinear damping of the instability on the background temperature drives the system to marginal stability ($\Symb{D}_A = 0$ globally). In the high-beta limit, these stable temperature profiles generically take the form of a staircase with affine symmetry (shifts and rescalings of the spatial coordinate). For simple density profiles, viz., constant or isobaric profiles, the staircase has a single step that occurs at low temperature where the plasma is effectively unmagnetized. More exotic density profiles can yield multi-step staircases: e.g., choosing a power-law density profile ($n \propto \Hall^\sigma$) gives a temperature profile with multiple steps but only when the plasma beta is small, and as a consequence the multi-step staircase is not `sharp'.

Finally, we discuss the back-reaction of the collisional whistler instability on the plasma temperature profile (\Sec{sec:quasilinear}). The instability is able to modify the temperature profile via frictional heating and Ettingshausen heat flux so that total energy is conserved. Interestingly, there exists a regime in which the instability cools the plasma via friction rather than heats it; this regime necessarily occurs in the initial stages of the instability. The Ettingshausen heat flux is also capable of canceling a portion of the conductive heat flux when the intensity gradient of the collisional whistler instability is anti-aligned with the temperature gradient. In principle, the collisional whistler instability might be capable of strongly reducing the heat flux through these mechanisms, but for high-beta plasmas, a strong reduction is unlikely to occur in the manner envisioned here as this would require the wave-intensity profile to be essentially delta-shaped. Non-geometrical-optics behaviour, nonlinear effects, or even synergistic interplay with kinetic microinstabilities (as these quickly modify fluid transport coefficients away from the standard Braginskii expressions used here) might relax this conclusion, but dedicated simulations are required to investigate this further. In this sense, our work here should be considered an initial investigation in which a number of simplifications have been made to elucidate the basic physics at play and to obtain initial estimates of the various parameter dependencies. More detailed followup investigations can now be performed in which these simplifications are sequentially relaxed.

\section*{Acknowledgements}

The authors would like to thank Elijah Kolmes for insightful discussions. 

\section*{Funding}

The work of NAL and, in part, of AAS was supported by STFC (grant number ST/W000903/1). The work of AAS was also supported in part by EPSRC (grant number EP/R034737/1) and the Simons Foundation via the Simons Investigator award. AFAB was supported by UKRI (grant number MR/W006723/1).

\section*{Declaration of interests}

The authors report no conflict of interest.


\appendix


\section{Conditions for no density or temperature fluctuations}
\label{app:decouple}


In what follows, fluctuating and mean components are denoted respectively as $\fluct{f}$ and~$\mean{f}$.

\subsection{No density fluctuations}

The density equation \eq{eq:densMHD} demands that $n$ remain constant in time. Hence, there can be no fluctuating component to the density since that would require a nonzero time derivative.


\subsection{No temperature fluctuations}

Suppose that $\fluct{T} = 0$. For this to be a possible solution of the linearized fluid equations, the magnetic-field perturbations must satisfy
\begin{align}
	0 
	&=
	\frac{c}{4 \pi n e} (\nabla \times \fluct{\Vect{B}}) \cdot 
	\left(
		\frac{3}{2} n \nabla T 
		- T \nabla n
		+ \mean{\Vect{R}}
	\right)
	+
	\frac{c}{4 \pi n e} (\nabla \times \mean{\Vect{B}}) \cdot \fluct{\Vect{R}}
	- \nabla \cdot \fluct{\Vect{q}} 
 .
\end{align}

\noindent Clearly, this can be satisfied if the following three conditions are met: the no-mean-flow condition:
\begin{equation}
	\nabla \times \mean{\Vect{B}}  = \Vect{0}
    ,
    \label{eq:cond1}
\end{equation}

\noindent the solenoidal condition for the heat-flux perturbations:
\begin{equation}
	\nabla \cdot \fluct{\Vect{q}} = 0
    ,
    \label{eq:cond2}
\end{equation}

\noindent and the transversality condition for the perturbed flows:
\begin{equation}
	(\nabla \times \fluct{\Vect{B}}) \cdot 
	\left(
		\frac{3}{2} n \nabla T 
		- T \nabla n
		+ \mean{\Vect{R}}
	\right)
	= 0
    .
    \label{eq:cond3}
\end{equation}

\noindent When these are satisfied, an eigenmode involving only magnetic-field fluctuations may exist.


\subsection{Verification of conditions for slab model}

Let us now verify that the above three conditions for the absence of temperature fluctuations are satisfied for the slab model used in the main text. First note that
\begin{equation}
    \mean{\Vect{B}} = 
	\begin{pmatrix}
		0 \\
		0 \\
		B_z
	\end{pmatrix}
	, \quad
	\fluct{\Vect{B}}
	= 
	\begin{pmatrix}
		\fluct{B}_x \\
		\fluct{B}_y \\
		0
	\end{pmatrix}
    ,
\end{equation}

\noindent whence
\begin{equation}
    \nabla \times \mean{\Vect{B}} = 0
	, \quad
	\nabla \times \fluct{\Vect{B}}
	=
	\begin{pmatrix}
		- \fluct{B}_y' \\
		\fluct{B}_x ' \\
		0
	\end{pmatrix}
    .
\end{equation}

\noindent The condition \eq{eq:cond1} is manifestly satisfied. 

Next, the fluctuating component of the Chapman--Enskog heat flux~\eq{eq:CEfrictionQ} takes the form
\begin{equation}
    \fluct{\Vect{q}}
	=
	- n \collT \thermalV^2
	\left(
		\Delta_\kappa \frac{\fluct{\Vect{B}} \mean{\Vect{B}} + \mean{\Vect{B}} \fluct{\Vect{B}} }{|B|^2}
		+ \kappa_\wedge \frac{\fluct{\hatmap{B}} }{|B|} 
	\right) \cdot \nabla T
	- \frac{\cycloF c^2}{\plasmaF^2} \frac{n T}{B_z} \, 
	\left[
		\Delta_\beta \frac{\mean{\Vect{B}} \mean{\Vect{B}} }{|B|^2}
		+ \beta_\perp \IMat{3}
		+ \beta_\wedge \frac{\mean{\hatmap{B}} }{|B|} 
	\right] \cdot \nabla \times \fluct{\Vect{B}}
    ,
\end{equation}

\noindent where we have truncated at quadratic order in the fluctuation amplitude. Using the fact that $\nabla T$ is parallel to $\unit{z}$ (and thereby parallel to $\mean{\Vect{B}}$ and orthogonal to $\fluct{\Vect{B}}$), that $\nabla \times \fluct{\Vect{B}}$ is perpendicular to $\mean{\Vect{B}}$, and that
\begin{equation}
    \unit{z} \cdot \fluct{\Vect{B}} = 0
	, \quad
	\unit{z} \cdot \fluct{\hatmap{B}} \cdot \unit{z} = 0
	, \quad
	\unit{z} \cdot \nabla \times \fluct{\Vect{B}} = 0
	, \quad
	\unit{z} \cdot \mean{\hatmap{B}} = \Vect{0}
    ,
\end{equation}

\noindent (where the second relation follows from the antisymmetry of hat-map matrices), one sees that
\begin{equation}
    \nabla \cdot \fluct{\Vect{q}}
	= \pd{z} \left( \unit{z} \cdot \fluct{\Vect{q}} \right)
	= 0
    .
\end{equation}

\noindent The condition \eq{eq:cond2} is therefore satisfied as well. 

Lastly, note that for the Chapman--Enskog friction~\eq{eq:CEfrictionQ}, to lowest order in the fluctuation amplitude, one has $\mean{\Vect{R}} = - n \beta_\parallel \nabla T$. It therefore follows that
\begin{equation}
    (\nabla \times \fluct{\Vect{B}}) \cdot 
	\left(
		\frac{3}{2} n \nabla T 
		- T \nabla n
		+ \mean{\Vect{R}}
	\right)
	\propto
    (\nabla \times \fluct{\Vect{B}}) \cdot \unit{z}
    = 0
    .
\end{equation}

\noindent Thus, the condition \eq{eq:cond3} is also satisfied.

\section{Limiting forms of the Chapman--Enksog friction coefficients for the Lorentz collision operator}
\label{app:transport}

Here we list the limiting forms of the Lorentz transport coefficients in the large- and small-magnetization limits, as these expressions are used to develop various analytical approximations presented in the main text. These expressions are repeated from \citet{Lopez24a}. 

As $\Hall \to 0$, one has
\begin{subequations}
    \label{eq:lim0}
    \begin{align}
        \lim_{\Hall \to 0} \alpha_\perp
        &=
        0.295
        + 7.30 \Hall^2
        , \quad
        \lim_{\Hall \to 0} \alpha_\wedge
        =
        0.933 \Hall
        , \\
        \lim_{\Hall \to 0} \beta_\perp
        &=
        1.50 - 139 \Hall^2
        , \hspace{6mm}
        \lim_{\Hall \to 0} \beta_\wedge
        =
        9.85 \Hall
        , \\
        \lim_{\Hall \to 0} \kappa_\perp
        &=
        13.6 - 3360 \Hall^2
        , \quad
        \lim_{\Hall \to 0} \kappa_\wedge
        =
        173 \Hall
        .
    \end{align}
\end{subequations}

\noindent Importantly, all perpendicular coefficients are equal to their respective parallel component at $\Hall = 0$, \ie $\alpha_\perp(\Hall = 0) = \alpha_\parallel$, etc. Finally, as $\Hall \to \infty$, one has
\begin{subequations}
    \label{eq:limINFTY}
    \begin{align}
        \lim_{\Hall \to \infty} \alpha_\perp
        &=
        1 - 1.43 \Hall^{-2/3}
        , \quad
        \lim_{\Hall \to \infty}
        \alpha_\wedge
        =
        2.53 \Hall^{-2/3}
        , \\
        \lim_{\Hall \to \infty}
        \beta_\perp
        &=
        6.33 \Hall^{-5/3}
        , \hspace{10mm}
        \lim_{\Hall \to \infty}
        \beta_\wedge
        =
        1.50 \Hall^{-1}
        , \\
        \lim_{\Hall \to \infty}
        \kappa_\perp
        &=
        3.25 \Hall^{-2}
        , \hspace{13mm}
        \lim_{\Hall \to \infty}
        \kappa_\wedge
        =
        2.50 \Hall^{-1}
        .
    \end{align}
\end{subequations}

\section{Overview of Wigner--Weyl transform}
\label{app:Weyl}

Here we summarize the main definitions and identities for the Wigner--Weyl transform (WWT) and associated operator calculus that are necessary to derive the results presented in this work (see \citealt{Case08} for a gentle introduction, or \citealt{Tracy14}, \citealt{Dodin19}, and \citealt{Dodin22} for more detailed discussions and generalizations). The WWT, denoted $\Weyl$, maps a given operator $\oper{A}$ to a corresponding phase-space function~$\Symb{A}$ (called the Weyl symbol of $\oper{A}$):
\begin{align}
	\Symb{A}(z, k_z)
	&= \Weyl\left[ \oper{A}(\oper{z}, \oper{k}_z) \right]
	\doteq
	\int \dd s \,
	\exp\left(
		i k_z s
	\right)
	\bra{z - s/2}
	\oper{A}
	\ket{z + s/2}
	.
	\label{eq:wigner}
\end{align}

\noindent As a corollary, one has
\begin{equation}
	\int \frac{\dd k_z }{2\pi} \Symb{A}(z, k_z)
	= \bra{z} \oper{A} \ket{z}
    .
	\label{eq:symbINT}
\end{equation}

\noindent The relevant applications of this result are as follows:
\begin{subequations}%
    \label{eq:traceEQS}%
    \begin{align}
        \psi^* \psi 
        &= 
        \braket{z}{\psi} \braket{\psi}{z}
        = \bra{z} \oper{W} \ket{z}
        = \int \frac{\dd k_z}{2\pi} \Symb{W}
    	, \\
        \psi^* \pd{z} \psi 
        &=
        i \bra{z} \oper{k}_z \ket{\psi} \braket{\psi}{z}
        = i \bra{z} \oper{k}_z \oper{W} \ket{z}
        = i \int \frac{\dd k_z}{2\pi} \Weyl\left[ \oper{k}_z \oper{W} \right]
        , \\
        (\pd{z} \psi)^* \pd{z} \psi
        &= \bra{z} \oper{k}_z \ket{\psi} \bra{\psi} \oper{k}_z \ket{z}
        = \bra{z} \oper{k}_z \oper{W} \oper{k}_z \ket{z}
        = \int \frac{\dd k_z}{2\pi} \Weyl\left[ \oper{k}_z \oper{W} \oper{k}_z \right]
        ,
    \end{align}
\end{subequations}

\noindent where all symbols are defined in the main text.

The WWT is invertible, although we shall not quote the inverse transform here as it is not needed for our purposes. The WWT also preserves hermiticity,
\begin{equation}
    \Weyl\left[\oper{A}^\dagger \right] = \Symb{A}^*
	,
\end{equation}

\noindent so that a Hermitian operator maps to a real-valued function. This, combined with the linearity of the WWT, means that the Hermitian and anti-Hermitian parts of a general operator and its associated symbol are in exact correspondence. We make use of this property in \Sec{sec:main} when identifying the instability growth rate without appealing to geometrical optics.

The WWT of the product of two operators can be concisely represented as the Moyal product $\star$ of their symbols:
\begin{equation}
	\Weyl[\oper{A}\oper{B}] = \Symb{A}(z, k_z) \star \Symb{B}(z, k_z) .
	\label{eq:weylMOYAL}
\end{equation}

\noindent This non-commutative product is given explicitly in the integral form
\begin{subequations}
	\label{eq:moyalDEF}%
	\begin{equation}
		\Symb{A} \star \Symb{B}
		=
		\int \frac{\dd u \, \dd v \, \dd \kappa \, \dd K}{(2\pi)^2}
		\exp\left[
			i (k_z - \kappa) u
			+ i (k_z - K) v
		\right]
		\Symb{A}\left( z - \frac{v}{2}, \kappa \right)
		\Symb{B} \left( z + \frac{u}{2}, K \right)
        ,
	\end{equation}
	
	\noindent which follows from the definition \eq{eq:wigner}, or equivalently via the pseudo-differential representation
	\begin{equation}
	\Symb{A} \star \Symb{B}
	= \left. 
		\sum_{s = 0}^\infty
		\left( 
			i \frac{
				\pd{z} \pd{\kappa} - \pd{k_z} \pd{\zeta}
			}{2}
		\right)^s
		\frac{\Symb{A}(z, k_z) \Symb{B}(\zeta, \kappa) }{s!}  
	\right|_{\zeta = z, \kappa = k_z}
	.
\end{equation}
\end{subequations}

\noindent Using this, one can show that
\begin{equation}
    \left( \Symb{A} \star \Symb{B} \right)^*
    = \Symb{B}^* \star \Symb{A}^*
    ,
\end{equation}

\noindent which also follows from the result $(\oper{A} \oper{B})^\dagger = \oper{B}^\dagger \oper{A}^\dagger$. As further corollaries, one has the integral identities
\begin{subequations}
	\begin{align}
		\label{eq:intKZweyl}
        \int \dd k_z \,
		\Symb{A} \star \Symb{B}
		&=
		\int \frac{\dd \zeta \, \dd \kappa \, \dd K}{\pi}
		\Symb{A}(\zeta, \kappa)
		\Symb{B}(\zeta, K)
		\exp\left[
			2 i (\kappa - K) (z - \zeta)
		\right]
		, \\
		\int \dd z \, \dd k_z \, 
		\Symb{A} \star \Symb{B}
		&=
		\int \dd z \, \dd k_z \, 
		\Symb{A}(z, k_z) \Symb{B}(z, k_z)
        .
	\end{align}
\end{subequations}

\noindent One can thus compute the following relevant WWT pairs:
\begin{subequations}
    \begin{align}
        \Weyl \left[f(\oper{z}) \right] &= f(z) 
        , \\
        \label{eq:WWTkz}
        \Weyl \left[ \oper{k}_z \oper{G}(\oper{z}, \oper{k}_z) \right] 
        \equiv k_z \star \Symb{G}(z, k_z)
        &= k_z \Symb{G}(z, k_z) - \frac{i}{2} \pd{z} \Symb{G}(z, k_z)
        , \\ 
        \Weyl \left[ \oper{G}(\oper{z}, \oper{k}_z) \oper{k}_z \right] 
        \equiv \Symb{G}(z, k_z) \star k_z
        &= k_z \Symb{G}(z, k_z) + \frac{i}{2} \pd{z} \Symb{G}(z, k_z)
        , \\
        \Weyl \left[ \oper{k}_z \oper{G}(\oper{z},\oper{k}_z) \oper{k}_z \right] 
        \equiv k_z \star \Symb{G}(z, k_z) \star k_z
        &= k_z^2 \Symb{G}(z, k_z) + \frac{1}{4} \pd{z}^2 \Symb{G}(z, k_z)
        ,
    \end{align}
\end{subequations}

\noindent where $\Symb{G}$ is the corresponding symbol of $\oper{G}$.


\section{Derivation of collisional whistler dispersion relation and growth rate via polar decomposition}
\label{app:polar}

As an alternative means of deriving the Hamiltonian \eq{eq:hamiltonian}, let us represent $\psi$ by its polar decomposition%
\footnote{Note that this is traditionally the first step in the short-wavelength WKB approximation, whereafter one would introduce the additional assumption that $\theta$ varies more rapidly than $\inten$. We shall not impose this latter constraint here, and thereby maintain an exact treatment.}
\begin{equation}
    \psi = \sqrt{\inten} \, \exp(i \theta)
    .
    \label{eq:polarDECOMP}
\end{equation}

\noindent In terms of $\inten$ and $\theta$, the field-evolution equation \eq{eq:psiNONLIN} becomes
\begin{align}
    \psi \pd{t} \theta
    - i \psi \frac{\pd{t} \inten }{2 \inten }
    &=
    \left(
        \groupDISP
        - i \resist
    \right)
    \left[
        - (\pd{z} \theta)^2
        - \frac{(\pd{z} \inten)^2}{4 \inten^2}
        + i \pd{z}^2\theta
        + \frac{\pd{z}^2\inten}{2 \inten}
        + i \frac{(\pd{z} \theta) (\pd{z} \inten)}{\inten}
    \right]\psi
    \nonumber\\
    &
    + \left(
        \pd{z} \groupDISP
        - \growthV
        + i \nerstV
        - i \pd{z} \resist
    \right)
    \left(i \pd{z} \theta + \frac{\pd{z} \inten}{2 \inten} \right)\psi
    - \pd{z}\left(
        \growthV
        - i \nerstV
    \right) \psi
    .
\end{align}

\noindent Dividing by $\psi$ and then collecting real and imaginary parts gives separate evolution equations for $\theta$ and $\inten$:
\begin{align}
    \pd{t} \theta
    &=
    - \Symb{D}_H\left(\pd{z} \theta, z \right)
    - \frac{
        \pd{z} \left[
            \pd{k} \Symb{D}_A\left(\pd{z} \theta, z \right) \inten 
            - \frac{1}{2} \pd{z}( \groupDISP \inten) 
        \right]
    }{2\inten}
    + \frac{1}{4} \groupDISP
    \left[ 
        \pd{z}^2\inten - \frac{(\pd{z}\inten)^2}{\inten}
    \right]
    ,\\
    \pd{t} \inten
    &=
    2 \Symb{D}_A\left(\pd{z} \theta, z \right) \inten
    - \pd{z}
    \left[
        \pd{k} \Symb{D}_H\left(\pd{z} \theta, z \right) \inten
        - \frac{1}{2} \pd{z} (\resist \inten)
    \right]
    + \frac{1}{2} \resist 
    \left[ 
        \pd{z}^2 \inten - \frac{(\pd{z} \inten)^2}{\inten}
    \right]
    .
    \label{eq:polarIeq}
\end{align}

\noindent Thus, we see that the Hermitian and anti-Hermitian parts of the Hamiltonian identified in \eq{eq:hamiltonian} via WWT-based methods emerges from the traditional approach as well. In fact, the evolution equation for the polar amplitude \eq{eq:polarIeq} is identical to that for the wave intensity given by \eq{eq:intenEQ}. This is because, as discussed further in \App{app:quasiEikonal}, the final set of gradient terms in \eq{eq:polarIeq} encode the `non-eikonal' bandwidth $\ave{k_z^2} - \ave{k_z}^2$ that can be combined with the term $\Symb{D}_A\left(\pd{z} \theta, z \right)$ to yield the Wigner-averaged growth rate~$\ave{\Symb{D}_A}$. The non-eikonal bandwidth being automatically contained in the Wigner-based formalism, rather than being a separate term that must be included in the evolution equations, is a theoretical advantage of that approach.


\section{Wigner function bandwidth and quasi-eikonal fields}
\label{app:quasiEikonal}

In the standard geometrical-optics (eikonal) limit, the Wigner function for a quasi-monochromatic wave is often approximated as a delta function along a given level curve of $\Symb{D}_H(k_z,z)$%
\footnote{See, \eg the discussion and cited literature in \citet{Donnelly21} and \citet{Dodin22}.}. %
Hence, one would have $\ave{f(k_z)} = f(\ave{k_z})$ for any function subjected to the averaging operator $\ave{}$ defined in \eq{eq:aveDEF}. Generally speaking, however, non-eikonal deviations of $\Symb{W}$ provide a bandwidth that makes this equality no longer hold. Let us consider this explicitly for $\ave{k_z^2}$.

By definition, the Wigner function for the polar-decomposed wavefield given by \eq{eq:polarDECOMP}~is
\begin{equation}
    \Symb{W}
    = \int \dd s \,
    \sqrt{
        \inten(z + s/2) \inten(z - s/2)
    }
    \exp\left[
        i k_z s
        + i \theta(z - s/2)
        - i \theta(z + s/2)
    \right]
    .
\end{equation}

\noindent It is straightforward to show that
\begin{align}
    \int \frac{\dd k_z}{2\pi}
    \Symb{W}(k_z,z)
    &=
    \int \dd s \,
    \sqrt{
        \inten(z + s/2) \inten(z - s/2)
    }
    \exp\left[
        i \theta(z - s/2)
        - i \theta(z + s/2)
    \right]
    \delta(s)
    \nonumber\\
    &= \inten(z)
    ,
\end{align}

\noindent and also that
\begin{align}
    \ave{k_z} &= \frac{1}{\inten(z)}
    \int \frac{\dd k_z}{2\pi}
    k \, \Symb{W}(k_z,z)
    \nonumber\\
    &=
    \frac{i}{\inten(z)}
    \int \dd s \,
    \delta(s)
    \pd{s}
    \left\{
        \sqrt{
            \inten(z + s/2) \inten(z - s/2)
        }
        \exp\left[
            i \theta(z - s/2)
            - i \theta(z + s/2)
        \right]
    \right\}
    \nonumber\\
    &=
    \theta'(z)
    .
\end{align}

\noindent Hence, the lowest two moments of $\Symb{W}$ for a polar-decomposed field behave identically to what would be expected for eikonal fields. However, let us compute the second moment:
\begin{align}
    \ave{k_z^2} &= \frac{1}{\inten(z)}
    \int \frac{\dd k_z}{2\pi}
    k_z^2 \, \Symb{W}(k_z,z)
    \nonumber\\
    &=
    - \frac{1}{\inten(z)}
    \int \dd s \,
    \delta(s)
    \pd{s}^2
    \left\{
        \sqrt{
            \inten(z + s/2) \inten(z - s/2)
        }
        \exp\left[
            i \theta(z - s/2)
            - i \theta(z + s/2)
        \right]
    \right\}
    \nonumber\\
    &=
    \ave{k_z}^2
    + \frac{
        [\inten'(z)]^2
        - \inten(z) \inten''(z)
    }{4 [\inten(z)]^2}
    .
\end{align}

\noindent The bandwidth of a non-eikonal field is therefore given by
\begin{equation}
    \ave{k_z^2} - \ave{k_z}^2
    = \frac{
        [\inten'(z)]^2
        - \inten(z) \inten''(z)
    }{4 [\inten(z)]^2}
    .
\end{equation}

We can then define a `quasi-eikonal' wavefield as a non-eikonal wavefield that nevertheless exhibits no bandwidth for a desired set of moments. Since we are only concerned with moments up to $\ave{k_z^2}$, for our purposes a quasi-eikonal field corresponds to the constraint
\begin{equation}
    [\inten'(z)]^2
    = \inten(z) \inten''(z)
    ,
\end{equation}

\noindent which is satisfied for any exponential intensity profile, viz.,
\begin{equation}
    \inten(z) = c_1 \exp(c_2 z)
    ,
    \label{eq:quasieikonal}
\end{equation}

\noindent with $c_1$ and $c_2$ arbitrary constants. With no bandwidth, intensity profiles given by \eq{eq:quasieikonal} can now be described rigorously with concepts normally restricted to geometrical optics, such as intensity profiles being advected by a well-defined group velocity and being amplified or damped by a well-defined growth rate. This latter property is crucial for the conclusions drawn in the main text.


\section{Sufficient condition for positive temperature profile}
\label{app:positivity}

For \eq{eq:marginalZ} to correspond to a physical profile, one must have $\Hall(z) \ge 0$ everywhere. For this to occur, $\Hall = 0$ must be an impassable boundary for the flow of the governing differential equation. One possible mechanism for this to occur is if $\Hall = 0$ is an asymptote. This would imply that
\begin{equation}
    \lim_{\Hall \to 0^+}
    z(\Hall) \to \pm \infty
    , \quad
    \lim_{\Hall \to 0^+}
    z'(\Hall) \to \mp \infty
    .
    \label{eq:asymptote}
\end{equation}

\noindent Consider that
\begin{equation}
    z'(\Hall)
    = 
    z'(\Hall_2)
    \exp\left[
        \int_{\Hall}^{\Hall_2}
        \dd m \, \gFUNC(m)
    \right]
    .
\end{equation}

\noindent Importantly, since the exponential function is always positive, $z'$ can never change sign. Hence, for \eq{eq:asymptote} to hold, one must have
\begin{equation}
    \lim_{\Hall \to 0^+}
    \int_{\Hall}^{\Hall_2}
    \dd m \, \gFUNC(m)
    \to + \infty
    .
\end{equation}

One class of divergent integrals is obtained when 
\begin{equation}
    \gFUNC(\Hall) = \frac{A}{\Hall}
\end{equation}

\noindent for some $A$. Then one has
\begin{equation}
    \int_{\Hall}^{\Hall_2}
    \dd m \, \gFUNC(m)
    =
    \int_{\Hall}^{\Hall_2}
    \dd m \, \frac{A}{m}
    = A \log \left( \frac{\Hall_2}{\Hall}\right)
    .
\end{equation}
    
\noindent Clearly, if $A > 0$, the integral diverges logarithmically. But this is not enough to ensure positivity of $\Hall(z)$: we also require that $z(\Hall)$ diverges. Straightforward calculation gives
\begin{equation}
    z(\Hall) = z(\Hall_1) + 
    z'(\Hall_2) \Hall_2^A
    \frac{\Hall^{1-A} - \Hall_1^{1-A}}{1 - A}
    .
\end{equation}

\noindent If this is to diverge as well, then one must have
\begin{equation}
    A > 1
    .
\end{equation}

\noindent This condition is sufficient to ensure that $\Hall$ remains positive.


\section{Calculations pertaining to magnetization staircases}
\label{app:auxGfunc}

Here we derive solutions to the differential equation \eq{eq:boundaryEQ} governing globally marginally stable temperature profiles in certain simple cases where analytical treatment is possible.

\subsection{Constant G(y)}

Consider
\begin{equation}
    \gFUNCnorm(y) = \paramONE
\end{equation}

\noindent for a constant $\paramONE$. Using \eq{eq:marginalZ} with appropriate boundary conditions, we compute
\begin{align}
    z(y) 
    &= 
    z'(y_0)
    \int_{y_0}^y
    \dd \mu \,
    \exp\left(
        -
        \int_{y_0}^\mu
        \dd m \, \frac{\paramONE}{\delta}
    \right)
    \nonumber\\
    &=
    z'(y_0)
    \int_{y_0}^y
    \dd \mu \,
    \exp\left[
        \frac{\paramONE(y_0 - \mu)}{\delta}
    \right]
    \nonumber\\
    &=
    \delta z'(y_0)
    \frac{
        1
        - \exp\left[
            \paramONE(y_0 - y)/\delta
        \right]
    }{\paramONE}
    .
\end{align}

\noindent This can be inverted to obtain the solution
\begin{equation}
    y(z)
    = 
    y_0
    - \frac{\delta}{\paramONE}
    \log\left(
        1
        - \frac{\paramONE}{\delta} y'_0 z
    \right)
    .
\end{equation}

\noindent This solution is shown in \Fig{fig:staircaseG} (a).

\subsection{Linear G(y)}

Next consider a linear function
\begin{equation}
    \gFUNCnorm = A (y - y_*)
    ,
\end{equation}

\noindent with a root occurring at $y_*$. Then
\begin{align}
    z(y)
    &=
    z'(y_0)
    \int_{y_0}^y
    \dd \mu \,
    \exp\left[
        \frac{\paramONE}{\delta}
        \int_{y_0}^\mu
        \dd m \,
        \left( y_* - m \right)
    \right]
    \nonumber\\
    &=
    z'(y_0)
    \int_{y_0}^y
    \dd \mu \,
    \exp\left[
        - \frac{\paramONE}{2 \delta}
        \left( \mu - y_*\right)^2
        + \frac{\paramONE}{2\delta}
        \left( y_0 - y_*\right)^2
    \right]
    \nonumber\\
    &=
    \frac{
        \exp\left[
            \frac{\paramONE}{2 \delta}
            \left( y_0 - y_*\right)^2
        \right]
    }{y'_0}
    \sqrt{\frac{\pi \delta}{2 \paramONE}}
    \left\{
        \textrm{erf}
        \left[
            \sqrt{\frac{\paramONE}{2 \delta}}
            \left(
                y - y_*
            \right)
        \right]
        - \textrm{erf}
        \left[
            \sqrt{\frac{\paramONE}{2 \delta}}
            \left(
                y_0 - y_*
            \right)
        \right]
    \right\}
    .
    \label{eq:linGsol}
\end{align}

\noindent This solution is shown in \Fig{fig:staircaseG} (b). Note that the continuation from positive to negative $\paramONE/\delta$ requires the identity
\begin{equation}
    \textrm{erfi}(z) = - i \textrm{erf}(i z)
    .
\end{equation}

\subsection{Rational G(y)}

Finally, let us consider the class of rational functions given by
\begin{equation}
    \gFUNCnorm(y)
    = \frac{\paramONE}{y} 
    \left(
        1 - \frac{y}{y_*}
    \right)
    ,
\end{equation}

\noindent where $y_*$ can be either positive or negative. Then
\begin{align}
    z(y)
    &=
    z'(y_0)
    \int_{y_0}^y
    \dd \mu \,
    \exp\left[
        \frac{\paramONE}{\delta}
        \int_{y_0}^\mu
        \dd m \,
        \left(
            \frac{1}{y_*} 
            - \frac{1}{m}
        \right)
    \right]
    \nonumber\\
    &=
    z'(y_0)
    \int_{y_0}^y
    \dd \mu \,
    \left( \frac{\mu}{y_0} \right)^{-\paramONE/\delta}
    \exp\left(
        \frac{\paramONE}{\delta}
        \frac{\mu - y_0 }{y_*}
    \right)
    .
\end{align}

\noindent To compute the remaining integral, we first make the variable substitution
\begin{equation}
    \mu
    = \left|\frac{\delta y_*}{\paramONE} \right| u e^{i\pi - i \varphi}
    , \quad
    \mu^{-\paramONE/\delta} \dd \mu
    = 
    \left(
        \left|\frac{\delta y_*}{\paramONE} \right| 
        e^{i\pi - i \varphi} 
    \right)^{(\delta - \paramONE)/\delta} u^{-\paramONE/\delta} \dd u
    ,
\end{equation}

\noindent where $\paramONE/\delta y_*= \left| \paramONE/\delta y_* \right| e^{i \varphi}$. We then obtain
\begin{align}
    z(y)
    &=
    z'(y_0)
    y_0
    \left(
        \left|\frac{\delta y_*}{\paramONE y_0} \right| 
        e^{i\pi - i \varphi}  
    \right)^{(\delta - \paramONE)/\delta}
    \exp\left(
        - \frac{\paramONE y_0}{\delta y_*}
    \right)
    \int_{\left| \paramONE y_0/\delta y_* \right| e^{i \varphi - i\pi}}^{\left| \paramONE y/\delta y_* \right| e^{i \varphi - i\pi}}
    \dd u \,
    u^{-\paramONE/\delta}
    e^{- u}
    \nonumber\\
    &=
    z'(y_0)
    y_0
    \frac{
        \gamma\left(
            \frac{\delta - \paramONE}{\delta}
            ,
            \left| \frac{\paramONE y}{\delta y_*} \right| e^{i \varphi - i\pi}
        \right)
        - \gamma\left(
            \frac{\delta - \paramONE}{\delta}
            ,
            \left| \frac{\paramONE y_0}{\delta y_*} \right| e^{i \varphi - i\pi}
        \right)
    }{
        \left(
            \left|\frac{\delta y_*}{\paramONE y_0} \right| 
            e^{i\pi - i \varphi}  
        \right)^{(\paramONE - \delta)/\delta}
        \exp\left(
            \frac{\paramONE y_0}{\delta y_*}
        \right)
    }
    ,
\end{align}

\noindent where $\gamma(a,z) = \int_0^z \dd t \, t^{a - 1} e^{-t}$ is the lower incomplete Gamma function \citep{Olver10a}.

\subsection{Derivation of \eq{eq:stairWIDTH}}

The width of the boundary layer can be estimated by the gradient at the steepest location, which occurs at the root of $\gFUNCnorm$. Specifically, the width of the staircase at a root~$y_*$ of $\gFUNCnorm$ is given by \eq{eq:marginalZ} as
\begin{equation}
    W
    = y_* z'(y_*)
    = y_*
    z'(y_2)
    \left\{
        \exp\left[
            -
            \int_{y_2}^{y_*}
            \dd Y \, \gFUNCnorm(Y)
        \right]
    \right\}^{1/\delta}
    .
\end{equation}

\noindent Suppose that the root is simple, so that we can approximate the local behavior of $G(y)$ with a linear profile
\begin{equation}
    \gFUNCnorm = \gFUNCnorm'(y_*) (y - y_*)
    .
\end{equation}

\noindent Then one readily computes
\begin{equation}
    W \approx
    y_*
    z'(y_2)
    \left\{
        \exp\left[
            \frac{(y_2 - y_*)^2}{2}
        \right]
    \right\}^{\gFUNCnorm'(y_*)/\delta}
    ,
\end{equation}

\noindent whence \eq{eq:stairWIDTH} follows.


\section{Energy-conservation relations}
\label{app:energy}

Nominally, the total energy in a wave-plasma system is given by the sum of the particle kinetic and thermal energies along with the energy of the electromagnetic field. However, as we have neglected the electron inertia and the displacement current (and, of course, the ion motion entirely), the energy invariant for the electron MHD equations consists solely of thermal and magnetic contributions and satisfies the local conservation law
\begin{equation}
	\pd{t}
	\left(
		\frac{3}{2} n T 
		+ \frac{|\Vect{B}|^2 }{8 \pi}
	\right)
	+ \nabla \cdot
	\left(
		\Vect{q}
		+ \frac{5}{2} n T \Vect{u}
		+ \frac{c}{4 \pi} \Vect{E} \times \Vect{B}
	\right)
	= 0
    ,
    \label{eq:energyGEN}
\end{equation}

\noindent where $\Vect{u}$ and $\Vect{E}$ are given by the expressions
\begin{equation}
	\Vect{u} = - \frac{\cycloF c^2 }{\plasmaF^2 B_z} \nabla \times \Vect{B}
	, \quad
	\Vect{E} = - \frac{\Vect{u} \times \Vect{B}}{c} - \frac{\nabla(n T)}{n e} + \frac{\Vect{R}}{ne}
    .
\end{equation}

\noindent For the slab geometry considered here, it can be shown that \eq{eq:energyGEN} takes the $1$-D form
\begin{equation}
	\pd{t}\left(
		\frac{3}{2} n T
		+ \frac{|\fluct{\Vect{B}}_\perp|^2 }{8 \pi}
	\right)
	+ \pd{z}
	\left(
		q_z
		+ \frac{\cycloF c^2 }{\plasmaF^2 B_z}
		\Vect{R}_\perp \cdot \Mat{J} \cdot \fluct{\Vect{B}}_\perp
		- \frac{\cycloF c^2}{4 \pi \plasmaF^2 }
		\fluct{\Vect{B}}_\perp \cdot
		\Mat{J} \cdot
		\pd{z} \fluct{\Vect{B}}_\perp
	\right)
	= 0
    ,
\end{equation}

\noindent or equivalently in terms of the complex wavefunctions $\psi$ and $\xi$:
\begin{equation}
	\pd{t}\left(
		\frac{3}{2} n T
		+ \epsilon^2 B_z^2 \frac{\psi^* \psi }{16 \pi}
	\right)
	+ \pd{z}
	\left[
		q_z
		+ \epsilon \frac{\cycloF c^2}{\plasmaF^2}
		\frac{\Im\left( \psi^* \xi \right)}{2}
		+ \epsilon^2 B_z^2 \frac{\cycloF c^2}{\plasmaF^2}
		\frac{\Im\left( \psi^* \pd{z} \psi \right)}{8 \pi}
	\right]
	= 0
    .
    \label{eq:energyPSI}
\end{equation}

\noindent Finally, for the specific case when the friction is given by the Chapman--Enskog expression~\eq{eq:CEfrictionQ}, one can show that \eq{eq:energyPSI} takes the form
\begin{align}
	&\pd{t}
	\left(
		\frac{3}{2} n T
		+ \frac{ \epsilon^2 B_z^2 }{16 \pi} \inten
	\right)
	+ \pd{z}
	\left[
		q_z
		+ \frac{\epsilon^2 B_z^2}{8 \pi}
        \left(
            \groupDISP \ave{k_z} \inten
        	+ \nerstV \inten
        	- \frac{1}{2}\resist \,
        	\pd{z} \inten
        \right)
	\right]
	= 0
    .
    \label{eq:energyFRIC}
\end{align}

\bibliography{Biblio.bib}
\bibliographystyle{jpp}

\end{document}